\documentclass[final,3p,times, twocolumn, authoryear]{elsarticle}

\usepackage{amssymb}
\usepackage{amsmath}

\usepackage{newtxtext,newtxmath}
\usepackage{graphicx}
\usepackage{subfigure}
\usepackage[export]{adjustbox}
\usepackage{float}
\usepackage{hyperref} 

\makeatletter
\def\ps@pprintTitle{%
   \let\@oddhead\@empty
   \let\@evenhead\@empty
   \let\@oddfoot\@empty
   \let\@evenfoot\@empty}
\makeatother

\begin{document}

\begin{frontmatter}



\title{Bridging simulations of kink instability in relativistic magnetized jets with radio emission and polarisation} 


\author[1,2]{Nikita Upreti}
\author[1]{Bhargav Vaidya}
\author[1]{Amit Shukla}

\address[1]{Discipline of Astronomy, Astrophysics and Space Engineering, Indian Institute of Technology Indore, Khandwa Road, Simrol, 453552, India}
\address[2]{Max Planck Institute for Plasma Physics, Boltzmannstraße 2, D-85748 Garching, Germany}

\begin{abstract}
Relativistic outflows emanating from active galactic nuclei can extend up to kiloparsec scales in length, displaying a variety of complex morphologies. This study explores the intricate morphologies of such relativistic jets, mainly focusing on creating a bridge between magnetic instabilities in jets with observational signatures from complex radio galaxies. In particular, we aim to study the role of dynamical instabilities in forming distinctive morphological features by employing 3D relativistic magnetohydrodynamic (RMHD) simulations of rotating jets. Our simulations have further used the hybrid Eulerian-Lagrangian framework of the \texttt{PLUTO} code and generated the synthetic synchrotron emission and polarisation maps to compare with the observed signatures.
Our analysis based on simulations of a continuously injected jet suggests that current-driven instabilities, notably the $|m|=1$ mode, generate ribs-like structures that are seen in some of the recent radio galaxies using  MeerKat, e.g. \textit{MysTail}.
In our contrasting simulations of the restarted jet, the kink-instability driven ribs-like structures were formed relatively near the nozzle. In both cases, the jet dissipates its pre-existing magnetic energy through these instabilities, transitioning to a more kinetic energy dominant state. The turbulent structures resulting from this dissipation phase are filamentary and resemble the tethers as observed for the case of \textit{MysTail}. 
This pilot study essentially provides a plausible qualitative explanation by bridging simulations of kink instability to produce synthetic radio features resembling the observed complex radio morphology of \textit{MysTail}.
\end{abstract}




\end{frontmatter}



\section{Introduction}

Active Galactic Nuclei (AGN) are compact, highly luminous regions at the centers of galaxies that outshine the rest of the galaxy across the entire electromagnetic spectrum. These powerful astrophysical objects emit energy ranging from low-energy radio waves to high-energy (TeV) gamma rays, making them some of the most visible and intriguing subjects for study \citep[see][and references therein]{Blandford_review}. Additionally, some of these active galaxies feature long collimated, magnetized, relativistic jets which can extend up to very large distances, ranging from hundreds of kilo-parsec (kpc) to a few megaparsecs (Mpc) from their nucleus \citep{leahy1984bridges,carilli1996cygnus,10.1093/mnras/stw1468,2020A&A...642A.153D}. These relativistic jets, which are believed to be the result of matter accreting onto the central black hole, are electromagnetically powered by drawing rotational energy from either the black hole or the accretion disk transferred through the threaded magnetic fields \citep{blandford1977electromagnetic,payne}. The rotation of the black hole or accretion disk causes these magnetic fields to twist, which naturally results in the creation of toroidal fields. This process ultimately leads to forming a helical field configuration \citep{Tchekhovskoy2015}. During the propagation of these jets, the ratio between the poloidal and toroidal magnetic fields is critical. The determination of this pitch angle is affected by the density of the medium the jet is traversing through, with denser environments resulting in a reduced pitch angle, indicative of a more intense toroidal magnetic field \citep{bromberg_tchekov}. Thus, the configuration of the magnetic field and the surrounding ambient medium plays a crucial role in both the collimation and the propagation of jets across cluster scales in terms of dynamical instabilities, thereby influencing the morphology of radio jets \citep{guan2014relativistic, yi-hao_chen}.

Radio Galaxies have traditionally been classified based on morphology as FRI and FRII types \citep{fanaroff1974morphology}, depending upon the presence or absence of the radio lobes. 
Observations have also revealed that some radio galaxies also exhibit systematic changes in the jet propagation direction, resulting in tailed or winged sources \citep{hardcastle2019,cotton2020,muller2021,pandge2022}. Relative motion between the host galaxy and the ambient medium is believed to be an underlying cause of tailed radio galaxies \citep{O_Neill_2019}. This causes the jets to bend in the same direction, resulting in mirror symmetry. The winged morphology, on the other hand, has a unique inversion symmetry in which the jets bend in opposite directions to form X-, S-, or Z-shaped structures \citep{ekers,Gopal-Krishna_2003,hodges-kluck,hardcastle2019}. The formation of such jet structures has been discussed following the back-flow model \citep{leahy1984bridges,capetti,rossi2017}, re-orientation model that involves sudden flip of black hole spin axis \citep{dennett-thorpe,rottmann2001jet, Hodges-Kluck_2010, Giri_2023} or the presence of dual AGNs producing a precessing or curved jet \citep{gower1982,krause2019}.


The advent of low-frequency radio observations has resulted in producing some interesting and complex radio galaxies \citep{MIGHTEE,LoTSS,gmrt,knowles,erosita, Velovic}. 
One such recent example is of RG2 (as defined by \citet{riseley}) or \emph{MysTail} (as defined by \citet{rudnick}), which lies at the northwest of the galaxy cluster Abell 3266. 
This unique and complex radio galaxy was first identified in a shallow narrow-band observation by \citet{Murphy1999} and was suggested to represent a dying radio galaxy tail with a possible origin from the elliptical galaxy to the immediate southwest. 
This source is part of the MeerKAT GCLS (\citep{knowles}) and studied in detail by \citet{rudnick}. The complex nature of this source is evident from the presence of a prominent \textit{ribbed tail} and \textit{tethers} as seen in the 1.28 GHz band. \citet{riseley} also observed these interesting features using the 9 arcsec resolution ATCA map with 2.1 GHz. 
Their study could also identify a compact radio counterpart to the cluster-member galaxy J043045.37-612335.8 ($z$= 0.0628; \citet{dehghan2017}) as a possible source in line with the suggestion made by \citet{Murphy1999}. 
Further, based on the estimate of core prominence at 2.1 GHz, \citet{riseley} have proposed this source would rather belong in the category of remnant radio galaxies. 
Additionally, polarisation measurements were also carried out to study the structure of magnetic fields for the source both within the ribbed tail and in tethers (see \citet{rudnick}). 
Several plausible mechanisms that could potentially manifest such a complex morphology have been discussed, which has motivated the present study to provide physical constraints on the potential origin of complex morphological signatures. 
In this regard, we have developed 3D RMHD simulations using the hybrid approach to create synthetic observations for qualitative comparison with the unique features like ribbed tail and tethers as observed in \textit{MysTail} source.

This paper is organized as follows. In section \ref{sec2}, we elaborate on the numerical approach adopted for this study along with the initial conditions used for the spectral modeling in our hybrid approach. The results of various simulations carried out are mentioned in section \ref{sec3}. Finally, the discussion associated with the obtained results is detailed and summarized in section \ref{sec4}. 

\section{Numerical Approach}
\label{sec2}
In this section, we shall describe in detail the numerical approach adopted for this work. In particular, we shall emphasize on the computational code used, the initial and boundary conditions adopted, and also detail the methodology used for calculating synthetic emission from our 3D RMHD simulations. 

\subsection{Numerical Code and Scaling}
We have performed 3D simulations of a relativistic, rotating, and magnetized jet using the \texttt{PLUTO} code \citep{pluto2007}. The code essentially solves the set of special RMHD equations as follows:
\begin{equation}
    \frac{\partial}{\partial t}
    \begin{pmatrix}
        D \\
        \mathbf{m}\\
        E_t \\
        \mathbf{B}
    \end{pmatrix}
    + \nabla \cdot
    \begin{pmatrix}
        D\mathbf{v} \\
        w \gamma^2 \mathbf{v}\mathbf{v} - \mathbf{b}\mathbf{b} + \mathbf{I}p_t \\
        \mathbf{m} \\
        \mathbf{vB} - \mathbf{Bv}
    \end{pmatrix}^T
    = \mathbf{0},
\end{equation}

where density in the observer's frame is represented as $ D = \gamma \rho $, with $ \rho $ being the rest mass density and $ \gamma $ denotes the Lorentz factor of the fluid. The three-dimensional velocity vector is represented by $ \mathbf{v}$ and the covariant magnetic field is defined as,
\begin{equation}
     [b^0, \mathbf{b}] = [\gamma \mathbf{v} \cdot \mathbf{B}, \mathbf{B}/\gamma + \gamma(\mathbf{v} \cdot \mathbf{B})\mathbf{v}] 
\end{equation}
in which \textbf{B} denotes the magnetic field in the lab frame. The total enthalpy, denoted as $ w_t $, is defined as $ w_t = \rho h + \frac{\mathbf{B}^2}{\gamma^2} + (\mathbf{v} \cdot \mathbf{B})^2 $, incorporating the specific enthalpy $ h $. The momentum density is expressed as $ \mathbf{m} = w_t \gamma^2 \mathbf{v} - b^0 \mathbf{b} $. The total pressure of the system, $ p_t $, is defined by $ p_t = p_g + \frac{\mathbf{B}^2}{2\gamma^2} + \frac{(\mathbf{v} \cdot \mathbf{B})^2}{2} $, where $ p_g $ represents the gas pressure \citep{del2003, mignone_taub}. The total energy (including the rest mass energy), $ E_t $, is given by $ E_t = w_t \gamma^2 - b^0b^0 - p_t $. Lastly, $ I $ represents the identity matrix.

For our simulation, we implemented a 3D Cartesian grid with coordinates in the range of $ x \in \left[-\frac{L_x}{2}, \frac{L_x}{2}\right], y \in \left[-\frac{L_y}{2}, \frac{L_y}{2}\right]  \text{ and } z \in \left[0, L_z\right]$, where $L_x =  L_y = 44 l_0$ and $L_z = 135 l_0$. The numerical domain is discretized using a uniform grid with $N_{x}, N_y$, and $N_z$ cells, respectively.  
The domain and grid dimensions adopted for the present work are showcased in Figure~\ref{fig:comp_domain} and Table~\ref{tab:Table 1}. 
We utilized a five-wave HLLD Riemann solver \citep{mignone2009five} in conjunction with a piecewise parabolic scheme for reconstruction \citep{colella1984, Mignone_2005} and a second-order Runge-Kutta time-stepping method. To implement the solenoidal condition of the magnetic field $\nabla \cdot \mathbf{B} = 0$ within our domain, we utilized the divergence cleaning technique \citep{dedner2002}. We further employ the Taub-Matthews equation of state as given by \citep{mignone_taub}, which features a variable adiabatic index whose value is 4/3 for hot gas and 5/3 in the cold gas limit.

In general, RMHD simulations are scale-independent; however, the present work involves the calculation of synthetic emission maps associated with radiative cooling in physical units. It is, therefore, necessary to select an appropriate set of scaling factors to transform all simulated quantities from code units into physical units. These scaling factors are set based on the specifications of the chosen reference source. 
In \texttt{PLUTO} code, the variables are standardized based on three fundamental scale parameters: the unit length $l_0$, the unit speed $v_0$, and the unit density $\rho_0$. The remaining quantities are then normalized based on these scales. Considering our study is centered on \textit{MysTail}, a radio galaxy extending across hundreds of kilo-parsecs within a cluster \citep{rudnick,riseley}, we have defined the unit length as the jet radius, $l_0$ = 1\, kpc. The unit velocity is set as the speed of light, $v_0=c$, and the unit density is chosen as $\rho_0= 1.661 \times 10^{-30}$\,g\,$cm^{-3}$ (see Fig.~\ref{fig:comp_domain}). Other variables are then normalized accordingly as $t_0=l_0/v_0$, $B_0=v_0\sqrt{4\pi \rho_0}$, $p_0=\rho_0 v_0^2$, $T_0=\mu m_{\mu}v_0^2/2k_B$, where $t_0$, $B_0$,$p_0$ and $T_0$ represent the unit time, unit magnetic field, unit pressure, and unit temperature respectively.

\subsection{Initial and Boundary conditions}
\begin{figure*}[t]
    \centering
    \includegraphics[width=\linewidth,keepaspectratio]{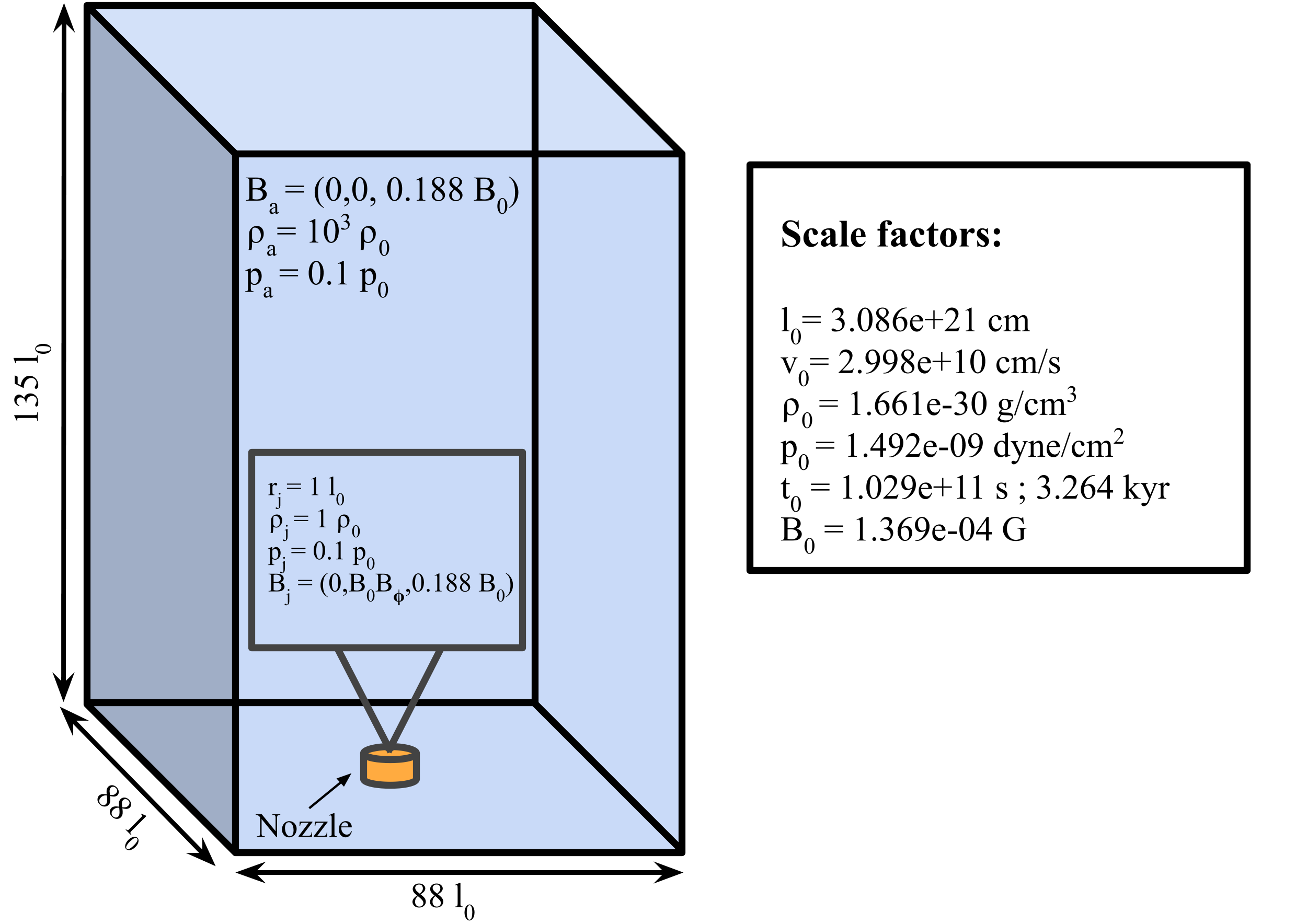}
    \caption{ A cartoon representation of computational domain along with scale factors used for the simulation runs. It illustrates key simulation parameters, including density ($\rho$), pressure (p), and magnetic fields within both the ambient medium and the injection nozzle. $r
    _j$ denotes the radius of the jet at the time of injection. The ambient medium is characterized by a constant density profile and a purely vertical magnetic field, denoted as $B_{\rm a}$. However, the jet is injected through the cylindrical nozzle, featuring a helical magnetic field configuration. The vertical component of the magnetic field of the jet is identical to that of the ambient, while the radial profile for the toroidal component ($B_{\phi}$) is provided in the appendix (see Equation \ref{b_phi profile}).}
    \label{fig:comp_domain}
\end{figure*}
The jet is launched in the positive z-direction from an injection nozzle located at the origin, which has a radius of $r_j = 1\,l_0$ and a nozzle height of $z_j = 1\,l_0$ (see the orange cylinder in Fig.~\ref{fig:comp_domain}). Surrounding the nozzle, a stationary ambient medium is initialized with a constant density $\rho_a = 1000 \rho_0$. This results in the ambient density being of the order $10^{-27}$ g cm$^{-3}$ with the chosen scale and in agreement with typical cluster values \citep{sarazin}. With this choice, a jet-to-ambient density contrast is set as $\eta = 1 \times 10^{-3}$, indicating that the jet is less dense compared to the ambient medium. 

 We further employ a stable jet nozzle as an internal boundary to inject the magnetized, relativistic, and rotating jet. Such a nozzle is initialized so as to maintain a radial equilibrium configuration, following the model described by \cite{Bodo_2019}. 
For details about the governing equation and to view the radial profiles of the key variables in the jet, please refer to \ref{appendixA}. The initial gas pressure is set at $p_{g0} = 0.1 \rho_0 v_0^2$ throughout the computational domain. As the injected jet is pressure matched, the injection nozzle also has the same pressure as the ambient initially. 
Furthermore, the ambient magnetic field, $B_{\rm a}$, is chosen to be purely vertical (along $\mathbf{\hat{z}}$) and uniform across the domain. Its magnitude is set to $B_{\rm a} = 0.188 B_0$ and matches the constant vertical component of the magnetic field within the jet nozzle to ensure the domain remains divergence-free at the outset. 

As the focus of this work is to study the formation of ribs and tethers in a particular radio galaxy \textit{MysTail}, we have carried out a limited parameter survey and focused on two different runs, i.e., Steady and Restarted, where the restarted jet is interrupted at time $t_{\rm stop}$ and again restarted at time $t_{\rm restart}$ and steady jet is never interrupted till the final time. 
Moreover, the dynamics of our jet is determined by five main parameters: $\gamma_z$, the Lorentz factor in the $\mathbf{\hat{z}}$ direction; $P_c = rB_{\rm j,z}/B_{\rm j, \phi}$, which is the magnetic pitch parameter; $M_a^2=\rho \gamma_z^2/B_{\rm j}^2$, the ratio of matter's energy density to the magnetic energy density; $\alpha$, a parameter for the strength of rotation; and $\omega$, the jet's angular velocity, all specified at the jet's central axis. The initial values for these parameters, which define our jet injection, are detailed in table~\ref{tab:Table 1}. We have also carried the steady run using two different grid resolutions ($N_x, N_y, N_z$) for a resolution study. 

Using the parameters adopted in our study, we computed the average injected jet power without considering the rest mass energy as follows\citep[e.g.,][]{mignone_taub, hardcastle, mukherjee2020}:

\begin{multline}
    Q_{\text{RMHD}} = \pi r^2_j v_z \Bigg[\gamma (\gamma - 1) \rho c^2 + ((h - 1) \gamma^2 - \Theta)\rho + \frac{\mathbf{B^2}}{2} \\
    + \frac{|\mathbf{v}|^2 |\mathbf{B}|^2 - (\mathbf{v} \cdot \mathbf{B})^2}{2}\Bigg]
\end{multline}

Here, $c$ represents the speed of light, $v$ is the jet velocity, $\gamma$ is the total Lorentz factor of the jet, $h$ is the specific enthalpy of the jet, and $\Theta$ is the temperature of the jet. Utilizing the initial parameters and configurations, we estimated the average jet power to be $2.25 \times 10^{44}$ erg s$^{-1}$. 
We also estimated the radio power of the injected jet at $1.285$ GHz to be  $6.96 \times 10^{40}$ erg s$^{-1}$ ($\approx 5.41 \times 10^{24}$ W Hz$^{-1}$), using empirical relation from \citet{birzan2008} and \citet{cavagnolo}.

\begin{table*}[t]
    \centering
    \resizebox{\linewidth}{!}{%
    \begin{tabular}{cccccccccccccc}
    
         \textbf{Run type}& \textbf{$\gamma_{zc}$}& \textbf{$P_c$}&
         \textbf{$M_a$}&
         \textbf{$\alpha$}& \textbf{$\Omega_c$}& 
         \textbf{$\overline{\sigma}$}&
         \textbf{$\overline{\beta}$}&
         \textbf{$N_x$}&
         \textbf{$N_y$}&
         \textbf{$N_z$}& 
         \textbf{$t_{stop}$}& \textbf{$t_{restart}$}& \textbf{$t_{final}$}\\
         (1)& (2)& (3)& (4)&(5)&(6)&(7)&(8)&(9)&(10)&(11)&(12)&(13)&(14)\\
         \hline
         Steady & 5 & 0.01 & 1  & 1  & 1.377&19.6&0.012& 528  & 528  & 810  & -- & -- & 1200  \\
         \hline
         Steady\_HR & 5 & 0.01 & 1  & 1  & 1.377&19.6&0.012  & 704  & 704  & 1080  & -- & -- & 1200  \\
        \hline
        Restarted & 5 & 0.01  & 1  & 1  & 1.377&19.6&0.012  & 528  & 528  & 810 & 500 & 900 & 1200 \\
        \hline
    \end{tabular}%
    }
    \caption{Input parameters for the injection nozzle(in code units). The parameters detailed in columns 2 through 8 include the Lorentz factor in the z-direction at the central axis ($\gamma_{zc}$), the pitch parameter at the central axis ($P_c$), the Alfvénic Mach number ($M_a$), the rotational parameter ($\alpha$), the angular velocity of the jet at the axis ($\Omega_c$), the average standard magnetization in the injection nozzle ($\overline{\sigma}$) and the average plasma beta value in the injection nozzle ($\overline{\beta}$). Columns 9 through 11 detail the grid resolution for the respective simulation runs. Also note that columns 12 and 13 record the times at which jet injection was paused and subsequently restarted for the restarted jet simulation, respectively, designated as $t_{\text{stop}}$ and $t_{\text{restart}}$. Finally, column 14 shows the final time up to which the simulation was run.  }
    \label{tab:Table 1}
\end{table*}

Additionally, to ensure a smooth transition and avoid sharp discontinuities at the jet-ambient boundary ($r=r_j$), the quantities injected at the nozzle are smoothed.
In addition to the internal boundary for the nozzle, we've applied an outflow boundary condition on all remaining boundaries in all directions. We ran the simulation for a total time of $t/t_0$=1200, which amounts to $\approx 3.92$ million years. 
This choice of time span and domain size ensures that the jet remains well within the boundary zones to prevent any boundary effects on jet dynamics.

\subsection{Spectral modelling}
For modeling the non-thermal spectral signatures of the relativistic jets, particularly synchrotron emission, we utilized the Lagrangian particle module of the \texttt{PLUTO} code \citep{Vaidya_2018, Nikhil2021, dipanjan2021}.
This approach models an ensemble of microparticles (i.e., electrons) as a single Lagrangian particle (macro-particle) that follows the fluid in the Eulerian grid. 
It incorporates physical processes, particle acceleration, and radiative losses at a finer, sub-grid level. 
Initially, these micro-particles within each Lagrangian macro-particle follow a power-law energy distribution. 
Subsequently, it consistently evolves the energy distribution of these particles across time and momentum space by solving the Fokker-Planck equation, which also accounts for the acceleration caused by diffuse shock acceleration (DSA) and different cooling processes. 
The spectral update is done at each simulation step depending on the underlying fluid quantity that governs the terms of the Fokker-Planck equation. 
For example, the radiative processes include energy loss due to adiabatic expansion, synchrotron radiation, and inverse Compton scattering of particles by the surrounding Cosmic Microwave Background (CMB) photons. 
These processes depend on the local velocity and magnetic field distribution (see \cite{Vaidya_2018} for more details). 

The Lagrangian particles are injected into the system through the nozzle along with the fluid. To inject these macro-particles, we specified a circular area with a radius of $r_j$ located a pixel height above the jet nozzle. This area is divided into 20 radial zones. The 20 zones are then subdivided into 36 angular divisions, which amounts to injecting 720 macro-particles per single injection episode. 
We optimized the injection rate so that, by the end of the simulation, we would have roughly $10^7$ Lagrangian particles in the domain, which would be sufficient to sample the jet and better represent its emission properties.
The ensemble of micro-particles (i.e., electrons) within each macro-particle is initialized with an energy distribution following a power law distribution as:
\begin{equation}
N(\gamma_{\rm e}) = N_0 \gamma_{\rm e}^{-\alpha} \label{eq:n_0}
\end{equation}.

Here, $N(\gamma_{\rm e})$ represents the number of particles per unit volume with a Lorentz factor $\gamma_e$, and $\alpha$ represents the initial power-law index, which is initially set to very steep ($\alpha = 9$). Such an initial distribution will be subsequently updated based on the various micro-physical processes encountered by the macro-particle as it traverses within the fluid. We have used the initial bounds on electron Lorentz factor as $\gamma_{e, min}=10^2$ and $\gamma_{e, max}=10^8$. Evidently, as the jet propagates into the medium, the acceleration of particles and the non-thermal cooling processes will alter these bounds and lead to the distribution of particles deviating from the initial power-law. The initial value of $N_0$ is set by estimating non-thermal electron number density $n_{\mu}$.

\begin{equation}
   \int_{\gamma_{\text{e, min}}}^{\gamma_{\text{e, max}}} N_0 \gamma_{\rm e}^{-\alpha} \, d\gamma_{\rm e} = n_{\mu}
\end{equation}

To quantify $n_{micro}$, we first assume an equipartition stage between the magnetic energy density at the jet base and the radiating electron's energy density given as \citep{sayan2022, Giri_2022}:

\begin{equation}
U_e = \frac{m_ec^2}{2} \int_{\gamma_{\text{e, min}}}^{\gamma_{\text{e,max}}} \gamma N(\gamma_{\rm e}) \, d\gamma_{\rm e} \Bigl(= \frac{B^2_{\rm eq}}{8\pi}\Bigr) = \epsilon \frac{B^2_{\rm dyn}}{8\pi}
\label{eq:equipartition}
\end{equation}  

where $U_{\rm e}$ represents the energy density of electrons, with $m_ec^2$ denoting the rest mass energy of an electron, and $B_{\rm eq}$ representing the strength of the equipartition magnetic field. It is noteworthy that $B_{\rm eq}$ is used as a substitute for the magnetic field, defined so that its magnetic energy density equals the energy of radiating electrons; thus, it cannot be regarded as the physical magnetic field. We define $\epsilon$ as the ratio of equipartition energy to the energy of dynamical magnetic field $B_{\rm dyn}$.  
In all of our simulation runs, we opted for $\epsilon=0.01$, ensuring that the particle energy initially remains in sub-equipartition with the energy present in the magnetic field, i.e., $B_{\rm eq}<B_{\rm dyn}$. 
Additionally, as the magnetic field in our injection nozzle varies according to the profiles outlined in Appendix A, we defined $B_{\rm dyn}$ as the average magnetic field strength within the injection nozzle, resulting in $B_{\rm dyn} = 5.09$ in code units. With these parameter selections, we obtained $n_{\mu}=2.3 \times 10^{-9}$ cm$^{-3}$.

This initial condition undergoes subsequent updates in the presence of radiative cooling and diffusive shock acceleration. The considered radiative losses encompass synchrotron and inverse Compton losses due to the surrounding cosmic microwave background photons for which the galaxy's redshift is set at $z = 0.0628$ \citep{riseley}. 
Synchrotron emissivities are computed for two frequencies ($\nu$): 0.9 GHz and 1.285 GHz, along a line of sight with angle ($\theta, \phi$) = ($90^{\circ},0^{\circ}$). Intensity maps are then obtained by integrating emissivity along the specified line of sight for each frequency. Additionally, the particle module yields Stokes parameters ($Q_{\nu}$ and $U_{\nu}$) as output, representing linearly polarised synchrotron radiation. Utilizing these parameters, we also investigate the polarised state of the jet at a frequency of $\nu=1.285$ GHz.
In this work, for simplicity, we have only included particle acceleration due to shocks and have not considered effects due to turbulent acceleration \citep[e.g,][]{sayan2022} and magnetic re-connection. Further, we also do not incorporate any feedback on the underlying fluid. This implies that any update in the spectral energy of the macro-particle does not modify the local fluid conditions. However, for the runs presented, we have observed that our choice of free parameters to extract thermal energy from the fluid at shock acceleration has a very small effect on fluid energetics during the course of our simulations.

\section{Results} \label{sec3}
In this study, we explore the intricate morphologies that emerge in jets spanning several thousand parsecs. This is accomplished by introducing a rotating, relativistic, and magnetized jet into a comparatively denser ambient medium. We are particularly interested in understanding how the distinctive structures known as ribs and tethers form within AGN jets at the kilo-parsec scale. We start with a reference case, continuously injecting the jet into the environment, to thoroughly examine how the jet evolves and propagates. Following this, we present a case where the jet injection is temporarily halted and then restarted, serving as a model for intermittent jet activity. Finally, qualitatively compare the structures formed in our simulations with those observed in the \textit{Mystail} galaxy.

\subsection{Steady Run}
\begin{figure*}
   \centering
    \subfigure[2-Dimensional slices of density distribution in log scale.]{
        \includegraphics[ width=\linewidth]{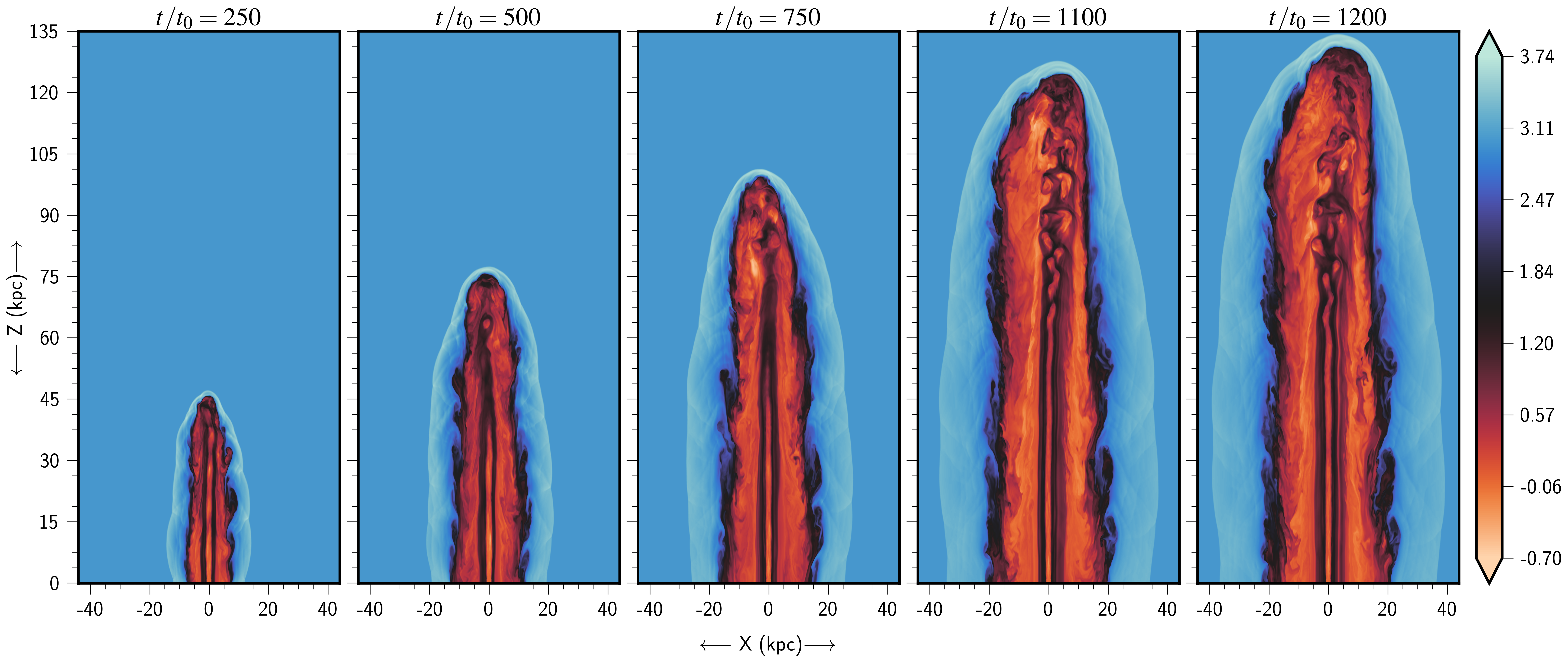}\label{fig:rho}}

   \vspace{1cm}

    \subfigure[2-Dimensional slices of magnetization parameter ($\sigma$) distribution in log scale.]{
        \includegraphics[width=1\linewidth]{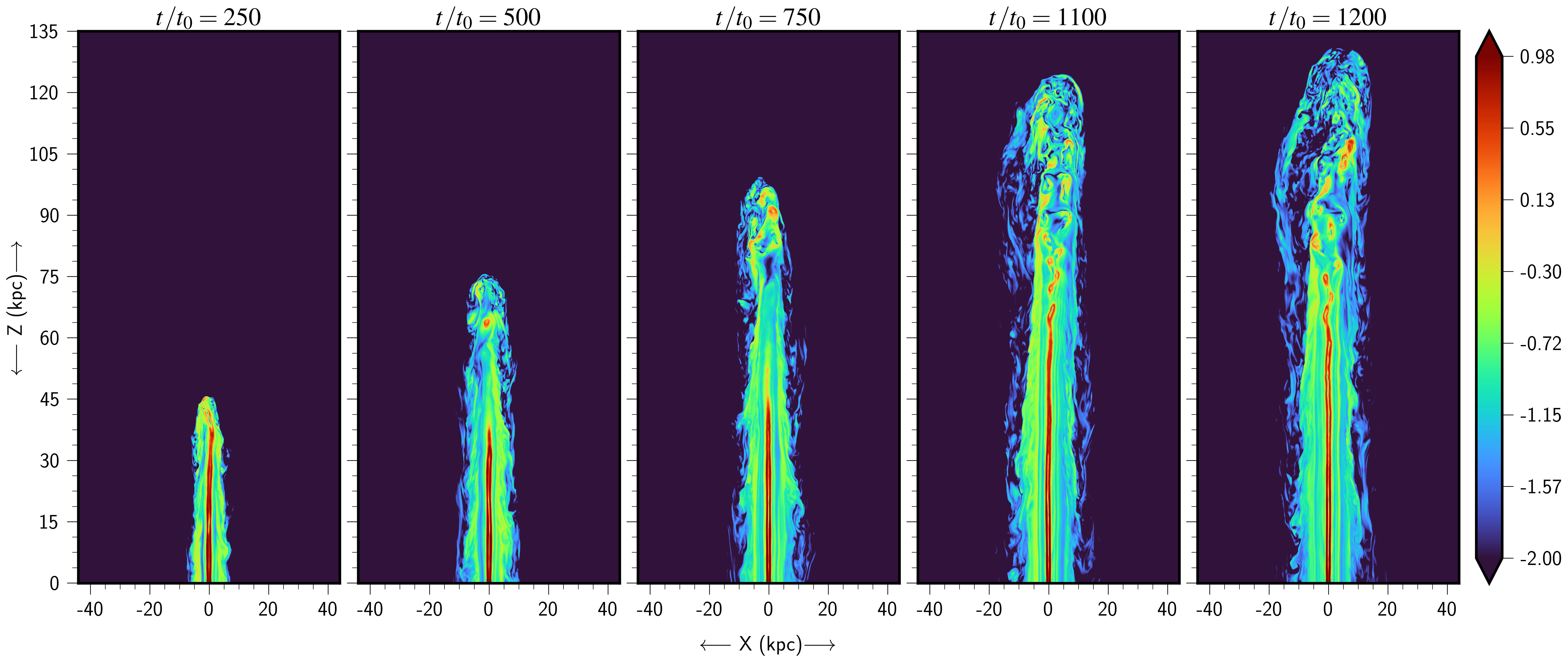}\label{fig:sigma}}
        \caption{This figure presents snapshots of density and the magnetization parameter ($\sigma$) for the \textit{steady run} in the y=0 plane at various time steps: $t/t_0 = 250$, $500$, $750$, $1100$, and $1200$, expressed in code time units ($t_0$, equivalent to 3.264 kyr). Rendered in logarithmic scale to accentuate the detailed variations within the distributions, these images portray the jet's evolution in the medium. In subplot (a), the density distribution draws attention to the underdense spine of the jet, which continues to propagate despite the growing instabilities. The subplot (b) displays the distribution of the magnetization parameter, where sections of the jet that are magnetically dominant are indicated by high $\sigma$ values, clearly visible in the color red. } 
    \label{fig:combined} 
\end{figure*}

\begin{figure*}
    \centering
    \includegraphics[width=1\linewidth]{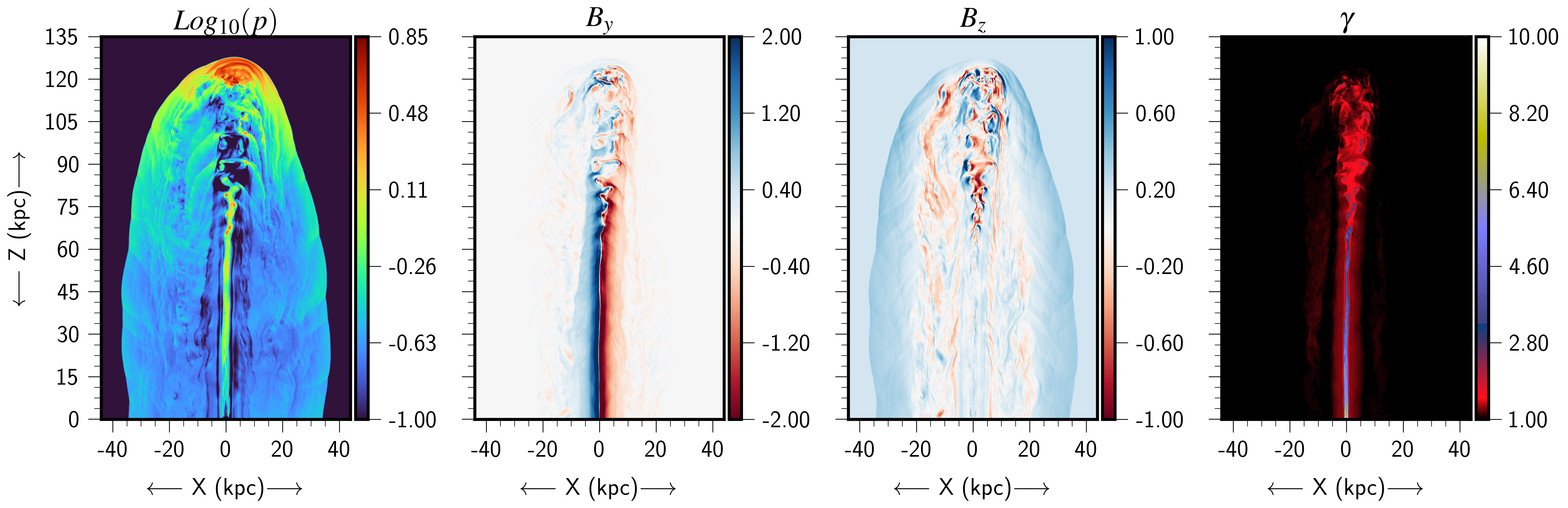}
    \caption{\textit{From left to right:} This figure illustrates the 2-D distribution of pressure (p) rendered in log scale, y-component of magnetic field ($B_y$),  z-component of magnetic field ($B_z$) and bulk Lorentz factor ($\gamma$) at time $t/t_0 = 1100$ (in code units, $t_0$) in the y=0 plane for the \textit{steady run}.  }
    \label{fig:110}
\end{figure*}

In this case, a rotating relativistic MHD jet is constantly injected into the medium via a nozzle, adhering to the initially set parameters (see Table~\ref{tab:Table 1}).
We will first discuss the propagation dynamics of the magnetized relativistic jet as it moves through an ambient medium, illustrated in figures (\ref{fig:combined}) and (\ref{fig:110}). 
These figures reveal the dynamical evolution of the jet starting from its point of injection. 

Initially, the jet travels at relativistic speeds in the z-direction, dispersing the low-density jet material into the surrounding medium, eventually terminating at the bow shock. This action results in the displacement of the uniform denser gas present in the ambient, leading to the formation of a cocoon of lower density around the jet. Figure (\ref{fig:rho}) presents 2-D slices of logarithmic density distribution in the y-z plane at various stages of the jet's evolution. By the end of the simulation run ($t/t_0$ = $1200$), the jet extends to a length of roughly 135 kpc. The density distribution images highlight the formation of an internal cavity within the jet, which is enclosed by a forward shock and a contact discontinuity. This is a characteristic feature shown to be present in several RMHD jet simulations \citep{kaiser,komissarov1998large,gaiblerr_krause,mignone_2010,guan2014relativistic,tchekov_2016, marti,massaglia,Perucho_López-Miralles_2023}.

A magnetized central spine, which is roughly the size of the injection region and inside which the majority of the current density is concentrated, remains present throughout the simulation run. Figure \ref{fig:sigma} illustrates the 2-D slices of the temporal evolution of the logarithmic distribution of magnetization parameter $\sigma$ defined as $B^2/\rho h$. We see that the jet has a distinct region (closer to injection) with $\sigma>> 1$, which denotes a magnetically dominated spine, and this extends into a region with $\sigma<<1$ (away from the nozzle). This denotes that the jet gradually transitions from being magnetically dominated to a more kinetically dominated state along its axis.

Additionally, the pressure slice in figure (\ref{fig:110}) at time $t/t_0 = 1100$ highlights areas of high pressure at both the termination point of the jet and along its spine, signaling the presence of strong shocks and recollimation shocks. \citep{norman1982structure,nawaz,Fuentes_2018,bodo_tavechio,marti,dipanjan2021,L_pez_Miralles_2022, Dubey2023}. Figure \ref{fig:110} also presents 2-dimensional slices illustrating the $y$ and $z$ components of the magnetic field, $B_y$ and $B_z$, respectively. The $B_y$ slice reveals the axisymmetric nature of the magnetic field and indicates that the jet possesses a magnetically dominated core. The strength of the $B_y$ component diminishes in the distant regions (along z) as instabilities develop in the jet spine.

Furthermore, in the vicinity of the jet head, the vertical magnetic field's strength, $B_z$, is amplified due to the compression of the field lines by the jet's forward shock (see figure \ref{fig:110}).
The jet also exhibits bulk relativistic velocities near the central axis, as evidenced by the 2D slice of the Lorentz factor at time $t/t_0 = 1100$ (Figure \ref{fig:110}). Closer to the injection nozzle, the jet's Lorentz factor is notably higher. However, at greater distances from the nozzle, the Lorentz factor reduces, and the flow transitions to being nearly non-relativistic near the jet head. This deceleration is attributed to the forward shock, which compels the jet material in its proximity to slow down. As a result, there is a backflow of jet material, contributing to the expansion of the cocoon. Over time, this process leads to a broadening of the jet.
As the rotating-magnetized jet propagates into the medium, instabilities start to set in the jet spine. When this jet gets farther away from the nozzle, these instabilities grow and give rise to non-linear structures. Over time, this results in the jet having more of a turbulent and disordered field configuration. These instabilities lead to the dissipation of the magnetic energy of the jet into kinetic energy and heat. 
\begin{figure*}
    \centering
    \includegraphics[width=\linewidth]{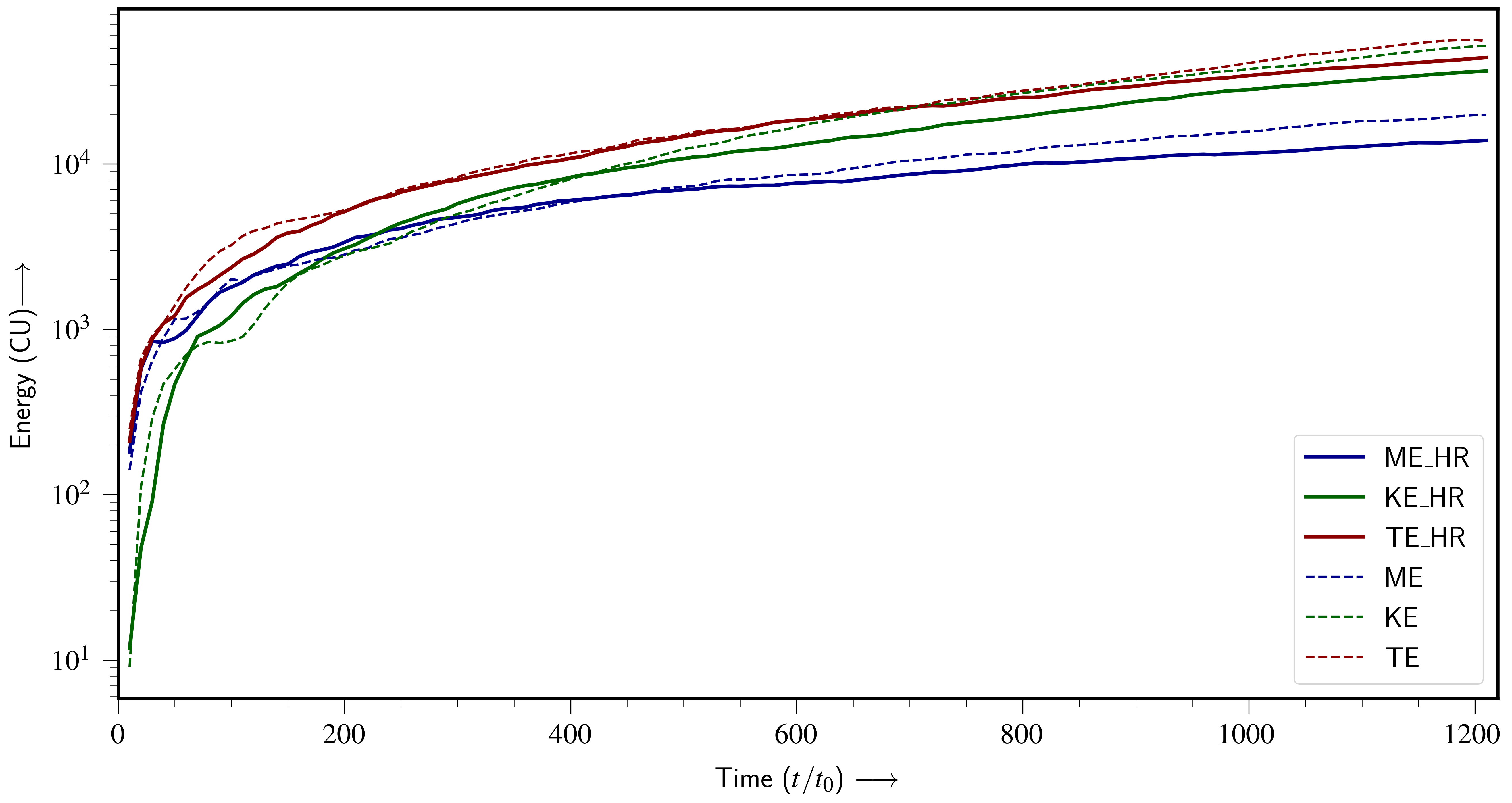}
    \caption{A plot of the time evolution of various energy components in the \textit{steady run}, comparing high-resolution (HR ) and low-resolution simulations with 8 and 6 cells per jet radius, respectively. Displayed are the volume-integrated kinetic energy density (KE), magnetic energy density (ME), and thermal energy density (TE), all scaled to code units. Although the initially injected energy is mostly magnetic in nature, it is converted into kinetic and thermal energy at later stages as the jet evolves.}
    \label{Energetics}
\end{figure*}
Such an exchange of energy is more clearly depicted in Figure \ref{Energetics}. The figure shows the evolution of the total magnetic energy density defined as $E_B = B^2/2 + \left[ |v|^2 |B|^2 - (v \cdot B)^2 \right]/2$, total kinetic energy density defined as $E_K = (\gamma - 1)\gamma\rho$ without including the rest mass energy, and total thermal energy density defined as $E_th= ((h - 1)\gamma^2 - \Theta )\rho$, all integrated across the jet's entire volume. For our simulation, the volume-averaged quantities are weighted by a passive scalar tracer to isolate jet material from the ambient medium, ensuring accurate computation of the jet's energy components. Additionally, we integrated above $z = 2$ to exclude the injection nozzle and focus on the jet's dynamics post-injection. It's observed that despite the overall increase in the jet's energies over time, as a result of continuous injection, the growth of magnetic energy density appears subdued. The magnetic energy of the jet gets dissipated in the form of kinetic energy and heat.

We conducted steady simulation runs at both higher and lower resolutions, using eight and six cells per jet radius, respectively. In Figure \ref{Energetics}, we compare the volume-averaged kinetic, magnetic, and thermal energies of the high-resolution (HR) run (solid line) with those of the low-resolution run (dotted line), finding that the differences in all energy components are within $(15-20\%)$. Despite these differences in global energy, we have seen that the properties of the kinks developed are qualitatively similar in both cases. So, given the computational cost and relatively smaller deviation observed with higher resolution, we have adopted 6 cells per jet radius as the standard for the restarted jet simulations.

\subsubsection{Growth rate of Current Driven Instabilities}
\begin{figure*}
    \centering
    \includegraphics[width=1\linewidth]{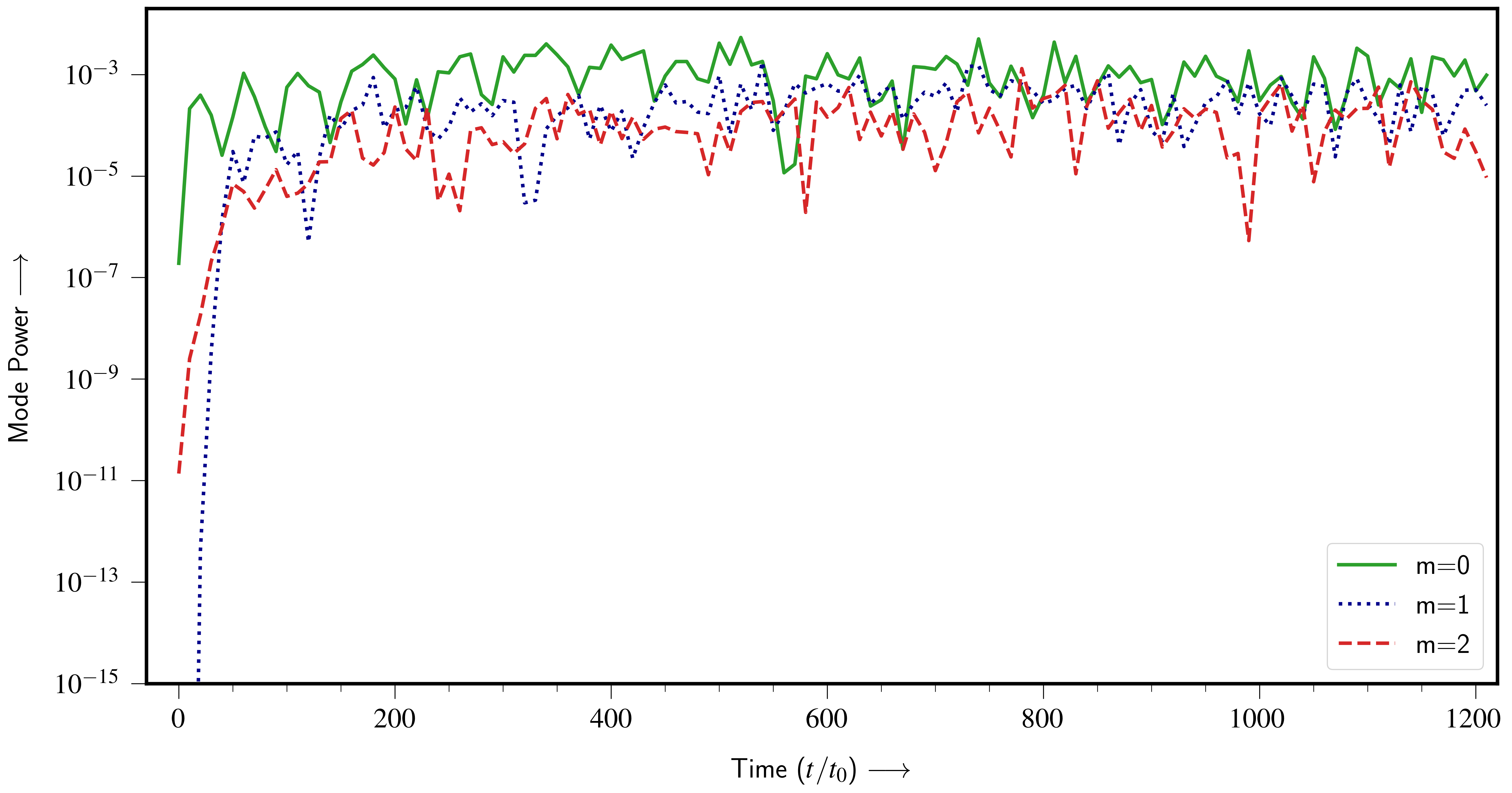}
    \caption{This figure depicts the temporal evolution of the power of various modes (in code units) in the current density distribution for the \textit{steady run}, highlighting that $|m|=1$ mode dominates amongst the non-axisymmetric modes.}
    \label{fig:power}
\end{figure*}

Since our model features a highly magnetized rotating jet, it is more prone to current-driven instabilities (CDIs). The growth of these current-driven instabilities leads to the deceleration of the jet and advances the transition of energy within it. A primary non-axisymmetric mode of CDIs is the $|m|=1$ mode, which is unique as it allows the fluid perturbations to interact across the axis, resulting in the jet's bodily shift across the axis. In these $|m|=1$ modes, the jet's center of mass transitions into a spiral configuration, indicative of the kink or screw instability mode \citep{1978mit..book.....B}. For other $|m|$ values, which are 0 and 2, the jet's center of mass remains aligned with the axis, and perturbations are concentrated mainly on the exterior. Consequently, Figure \ref{fig:power} illustrates the growth rate of various non-axisymmetric modes by calculating the power in the current density by using the Fourier transform. For each mode $m$, we define Fourier amplitude as follows:

\begin{equation}
    f(m, k) = \frac{\int_{r_{\text{min}}}^{r_{\text{max}}} \int_{0}^{2\pi} \int_{z_{\text{min}}}^{z_{\text{max}}} |J|e^{i(m\phi+kz)} r \, dr \, d\phi \, dz}{\int_{r_{\text{min}}}^{r_{\text{max}}} \int_{0}^{2\pi} \int_{z_{\text{min}}}^{z_{\text{max}}} r \, dr \, d\phi \, dz}
\end{equation}
where we integrated the magnitude of current density over a cylindrical volume in which the current is enclosed. 
In this calculation, we have used $r_{\text{min}} = 0$, $r_{\text{max}} = 6$, $z_{\text{min}} = 0$, and $z_{\text{max}} = 135$; $k = 2\pi / \lambda$ is the vertical wave-number, where $\lambda$ is the characteristic wavelength which we have taken to be 10 (in code length units). $m$ is the azimuthal wavenumber, for which we have considered three components, $m = 0, 1, 2$. The choice of the characteristic wavelength here is rather arbitrary and just to demonstrate the growth of the different modes. However, other wavenumbers also experience a similar sharp increase in the power of the kink mode.
The volume-averaged mode power in the current density $|J|$ is then calculated as  \citep{guan2014relativistic}:

\begin{equation}
    P(m, k) = |f(m, k)|^2 = \left\{\Re[f(m, k)]\right\}^2 + \left\{\Im[f(m, k)]\right\}^2
\end{equation}

The figure \ref{fig:power} reveals that the $m=0$ component is the dominant mode throughout the run. However, the $|m|=1$ mode experiences a sharp increase till $t/t_0 \approx 200$ (see blue dotted line) and predominates among non-axisymmetric modes over the full run. Besides, at later times, the growth rate of these non-axisymmetric modes is much slower. This dominance of $|m|=1$ component among the non-axisymmetrical modes in our jet model, which is characterized by a low magnetic pitch parameter (0.01), aligns with the similar findings from prior numerical simulations and linear stability analysis of magnetized jets \citep[e.g.,][]{Nakamura_2007,mignone2010,mizuno2012, bodo2013__,guan2014relativistic,bromberg_tchekov,barniolduran,Bodo_2019,bodo2021,meenakshi2023polarization,yi-hao_chen, Acharya2023}. The development of this kink mode leads to deformation and disruption of the jet's spine, resulting in significant wiggles and bends. 

We additionally study the Kruskal-Shafranov criterion \citep{kadomstev, shafranov1970hydromagnetic} as an alternative manner to verify the presence of current driven instabilities for $|m|=1$ mode. For a cylindrical MHD plasma with a constant current density and confined within a specific radius, the jet is unstable to kink modes when 
\begin{equation}
    q = (2\pi rB_p)/(L_z B_{\phi}) < q_{\text{crit}},
\end{equation}
here, $r$ represents the cylindrical radius, $B_p$ and $B_{\phi}$ are the poloidal and toroidal magnetic field components, respectively in the lab frame, and $L_z$ is the length of the plasma column. 

In an ideal MHD force-free regime, the critical value of $q$, $q_{\text{crit}}$, is 1; however, in relativistic MHD (RMHD), this threshold is lower \citep{narayan2009}.

\begin{figure*}[ht]
    \centering
    \includegraphics[width=\linewidth]{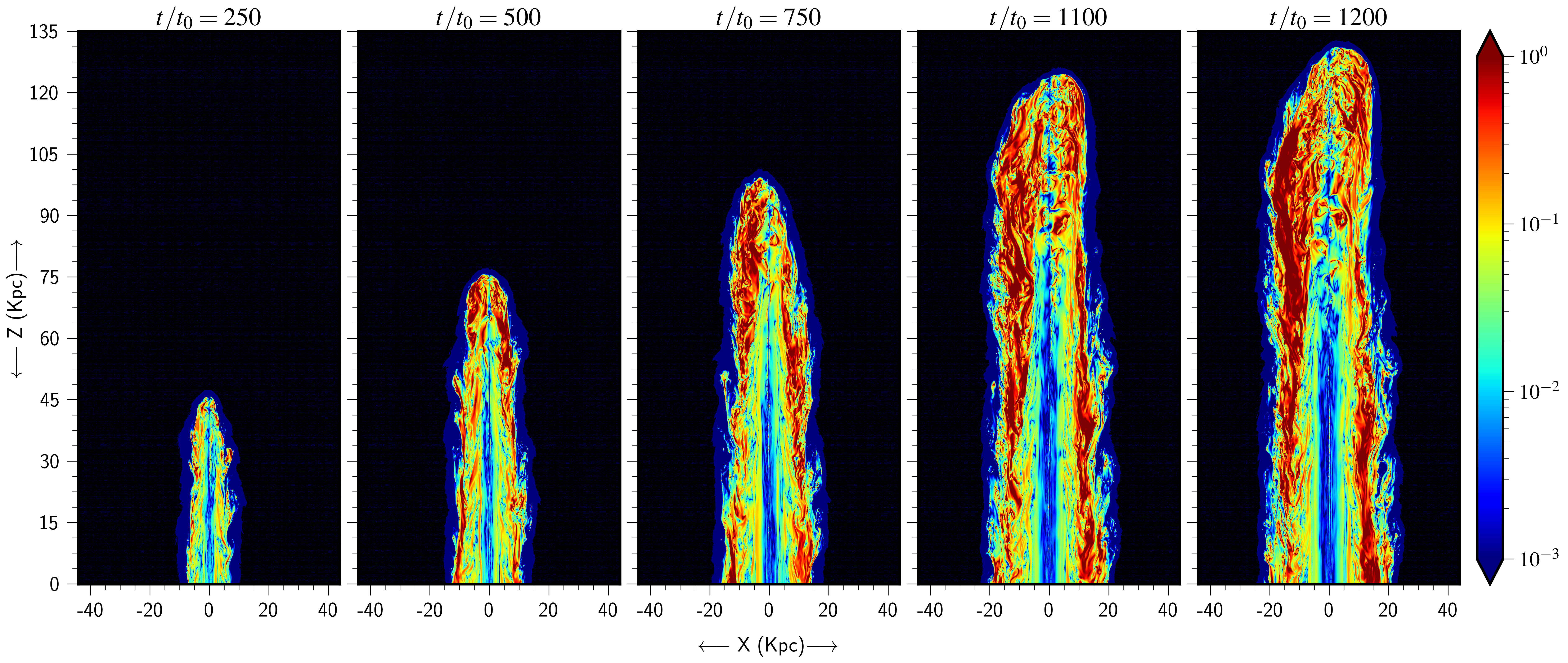}
    \caption{ This figure presents the distribution of $q=(2\pi \gamma rB_p)/(L_z B_{\phi})$ for the \textit{steady run} in the y=0 plane at various time steps: $t/t_0 = 250$, $500$, $750$, $1100$, and $1200$, with time expressed in code units ($t_0$, equivalent to 3.264 kyrs). The distribution is rendered using a logarithmic colormap to highlight the detailed variations within the jet. The jet's stability against kink modes is governed by the Kruskal-Shafranov criterion, which states that for $q < q_{\text{crit}} = 1 $, the jet is susceptible to kink instabilities. Since this criteria is derived for ideal MHD cylindrical plasma, so we keep our focus on the near-axis region where large currents are confined. In this region, we observe that $q$ is significantly less than 1, indicating that the jet spine is particularly susceptible to destabilization by the $|m|=1$ kink mode.}
    \label{fig:q_plot}
\end{figure*}

Figure \ref{fig:q_plot} displays the variation of $q$ at different time steps for the \textit{steady run}, with $L_z$ chosen as the length of the jet at that respective time. Since the Kruskal-Shafranov criterion is derived under highly idealistic assumptions, so we should focus on the near-axis region of the simulation where most of the current density is confined. We can see in figure \ref{fig:q_plot} that the near-axis region has $q<<1$ for $L_z<70$ where the large current is confined. Therefore, the jet is prone to destabilization and displays characteristics indicative of current-driven instabilities. The jet exhibits maximum instability to the non-axisymmetric $|m|=1$ mode, which initially experiences a sharp increase (see Figure~\ref{fig:power}). 

 In summary, our simulations of continuously injected 3D relativistic MHD jets show clear evidence of the presence of kink mode. As a result of this instability, the spine of the jet deforms but does not disrupt, and the kinetic energy of the jet grows at the expense of magnetic energy that is dissipated, resulting in a kinetically dominated jet far away from the injection nozzle. In the next section, we will attribute these dynamical features to spectral signatures for qualitative comparison with observations.

\subsubsection{Spectral signatures}

\begin{figure*}
    \centering
    \includegraphics[width=\linewidth]{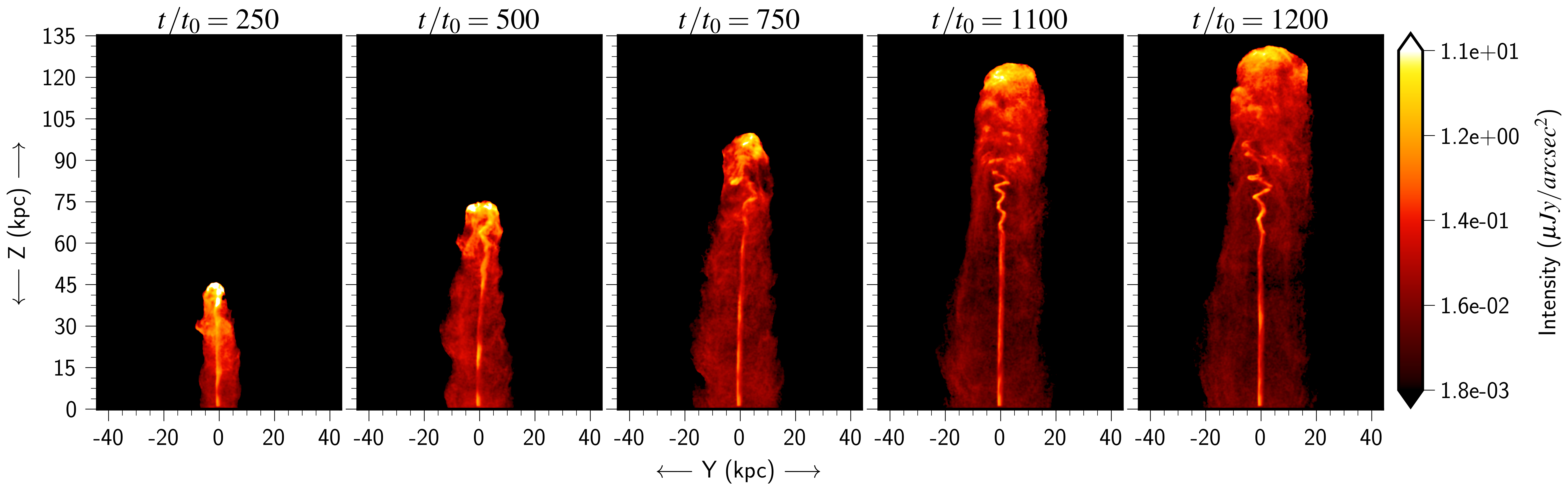}
    \caption{The figure presents a series of 2-D synchrotron intensity maps at 1.285 GHz, rendered in logarithmic scale and integrated along the line of sight ($\theta, \phi$) =($90^\circ, 0^\circ$), at time steps $t/t_0 = 250$, $500$, $750$, $1100$, and $1200$, measured in code time units ($t_0$, corresponding to 3.264 kyr). It showcases the evolution of the jet's spectral morphology over time. {The spectral evolution for the steady simulation run can be seen through this \href{https://drive.google.com/file/d/1wB_cRa8HVzj-1HJrF5MZ5uQInVeivIHE/view?usp=sharing}{movie link}.
 }}
    \label{fig:intensity}
\end{figure*}
In this section, we will discuss the spectral signatures of our modeled global jet. We computed the synthetic synchrotron emissivity at 1.285 GHz within the 3D domain by utilizing the Lagrangian particle module of the \texttt{PLUTO} code. Subsequently, we integrated this emissivity along the line of sight defined by $(\theta, \phi) = (90^{\circ},0^{\circ})$, which allows for an edge-on view of the jet, to generate 2D intensity maps. 
Figure \ref{fig:intensity} presents snapshots of the intensity maps at times $t/t_0 = 250$, $500$, $750$, $1100$, and $1200$. These maps illustrate that the synchrotron emission in this jet is concentrated along the main axis of the jet, with the intensity gradually decreasing towards the outer edges. As seen earlier from the figure \ref{fig:power}, the jet suffers from the sharp growth in the power of the kink mode till $t \approx 200 $. Notably, the snapshot captured at $t/t_0 = 250$ reveals the jet spine beginning to exhibit early indications of instabilities near the jet head yet largely retains a straight orientation. Snapshots at later times, $t/t_0 = 500$ and $t/t_0 = 750$, reveal significant wiggles and bends along the jet spine, which, nevertheless, still remains intact up to its full vertical reach. In subsequent timesteps, particularly at $t/t_0 = 1100$ and $1200$ as depicted in Figure \ref{fig:intensity}, there is a significant bending and twisting of the jet spine around its axis followed by its complete disruption in the vicinity of the jet head. The presence of turbulent backflow from the bow shock leads to this disintegration of the already highly unstable jet spine. However, even though our jet model has significant growth of instability modes along its spine, the jet does not stop progressing into the medium or gets completely destroyed along its axis. This is partially due to the continuous injection of the magnetic energy that drives the jet as can be seen in similar examples of numerical simulations \citep{keppens,huarte,guan2014relativistic,tchekov_2016,Massaglia_2019,L_pez_Miralles_2022,massaglia,meenakshi2023polarization}. The observed dominance of m=0 in figure \ref{fig:power} throughout the run, even in the non-linear stages, is in line with the non-disruption of the jet.

Such deformation of the jet spine results in the formation of peculiar morphological features within the jet. In the following section, we will compare our simulated model jet with the specific case that is the focus of our study, \textit{MysTail} and its unique characteristics referred to as "ribs and tethers."

\subsubsection{Qualitative Comparison with the \textit{MysTail}}
\label{subsec:comparison}
Since current-driven instabilities (CDIs) cause the jet to twist and bend at multiple points along its axis they lead to the formation of structures extending transversely to the jet's direction. This effect becomes particularly noticeable in the snapshots from $t/t_0 = 1100$ and $1200$ within the region of $65<z<90$, as can be seen in Figure \ref{fig:intensity}. These transverse structures bear resemblance to the rib-like features observed in \textit{MysTail}, as depicted in Figure 3 of \citep{rudnick}. 
Furthermore, at the forefront of the jet, where it interacts directly with the ambient medium, intense back-flow from the bow shock causes the jet spine to deform completely. This interaction gives rise to hot spots and turbulent features. Distinctly, filament-like structures can be observed near the jet head. For further comparative analysis, we chose a particular snapshot ($t/t_0$ = 1100) where we can see both ribs and tethers clearly due to prominent helical bending of the jet spine and filaments near the jet head (see Fig.~\ref{fig:intensity}). 

\begin{figure*}[t]
  \centering
    \includegraphics[width=\linewidth, keepaspectratio]{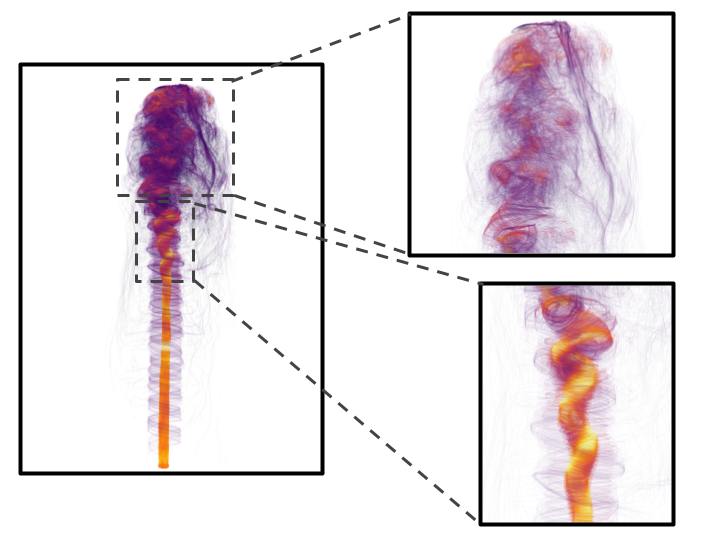}
   \caption{The figure illustrates the configuration of magnetic field lines at the time $t/t_0 = 1100$ ($\sim$ 3.59 Myr), focusing on specific features through zoomed-in panels. Here, the yellow and purple colors signify stronger and weaker magnetic field strengths, respectively. \textit{Top zoomed-in panel:} This section highlights the turbulent region close to the forward shock, featuring a chaotic magnetic field configuration, and filamentary structures which will lead to the formation of tethers-like structures.\textit{Bottom zoomed-in panels:} This section illustrates the intricate twists and bends in magnetic field configuration along the jet's spine, which manifest as transversely stretched, ribs-like structures reminiscent of the detailed morphology seen in \textit{MysTail}.}
  \label{fig:test}
\end{figure*}

The ribs and tethers seen in our simulations also show characteristic magnetic field profiles. Figure \ref{fig:test} showcases the 3-D magnetic field line configuration in our global jet simulation at $t/t_0 = 1100$ with zoomed-in panels for the regions containing ribs and tethers. 
Here, the yellow and purple colors signify stronger and weaker magnetic field strengths, respectively. In the upper section of the jet near the jet head, a relatively long magnetic filament is visible on the right side of the jet, extending up to $40-50\ \text{kpc}$. Similar long magnetic filaments have been observed in several cases, with spatial lengths that extends upto $220\ \text{kpc}$ in the radio galaxy 3C40B and few other sources\citep{owen, rajpurohit2022, brienza, magnetic_filament}. However, in our simulation, the long magnetic filament is not clearly visible in the intensity image due to the very low density in that region.

In the lower and middle sections of the jet, the magnetic field retains its tightly coiled helical structure as initially injected. We can see that the ribs have brighter patches in between, particularly around the bends, revealing denser and more ordered magnetic field lines. This is indicative of kink instability in the jet spine. Furthermore, shocks generated near the jet head from the interaction between the jet spine and strong back-flow result in the formation of bright hot spots. In between these hot spots, we can observe filaments with fainter emission (as seen in figure \ref{fig:intensity}) and are surrounded by turbulent magnetic fields. 
Additionally, the filamentary structures obtained from the simulation do not extend very far, as the ongoing energy injection continuously propels the jet further upwards. 

We further analyzed the spectral characteristics of these features in comparison to the observed spectral signatures of ribs and tethers. To determine the spectral index profile, we first derived the spectral index values based on the principle that synchrotron radiation intensity, $S_{\nu}$, scales with observation frequency $\nu$ to the power of $\alpha$ ($S_{\nu} \propto \nu^{\alpha}$), where $\alpha$ is the spectral index. Thus, the spectral index $\alpha$ is calculated using the formula:
\begin{equation}
    \alpha = \frac{\log_{10} (S_{\nu_1}/S_{\nu_2})}{\log_{10} (\nu_1/\nu_2)}
    \label{eq:spec_alpha}
\end{equation}
We've selected $\nu_1 = 0.9$ GHz and $\nu_2 = 1.285$ GHz as our frequencies of interest to compute the spectral index.

\begin{figure*}[t]
         \centering
        \includegraphics[width=\linewidth, keepaspectratio]{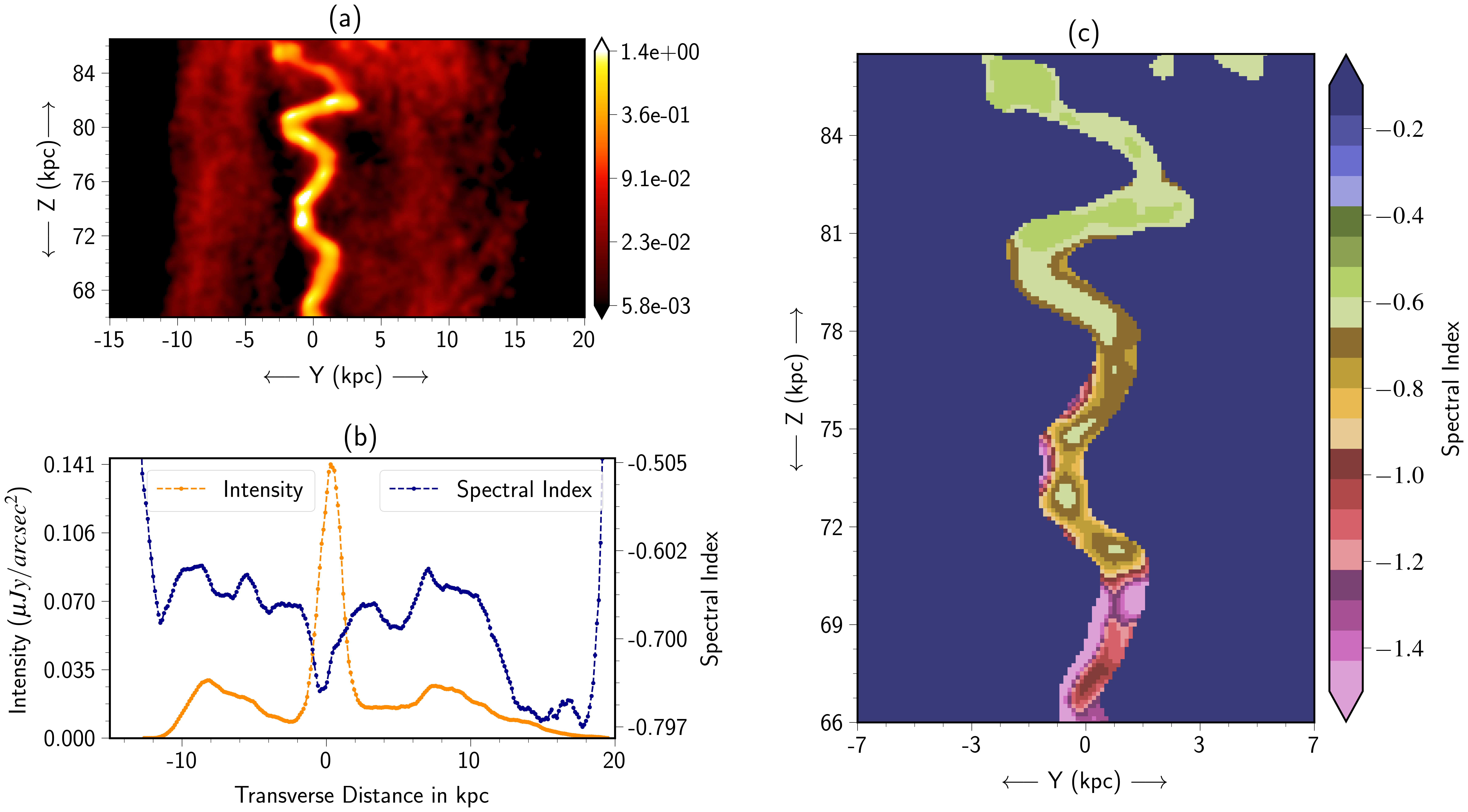}
        \caption{This figure presents the spectral signatures of simulated ribs observed at a frequency of 1.285 GHz during the \textit{steady run} at time $t/t_0 = 1100$. In the left panel, subfigure (a) showcases a region within the jet's intensity map characterized by transverse extensions resembling ribs, stretching up to 21.5 kpc in length. Subfigure (b) depicts a profile cut made in total intensity and spectral index across this patch of ribs, achieved by computing the median along the jet axis (Z-axis). The right panel, subfigure (c), illustrates the spectral index distribution within the ribs. Notably, cocoon emission has been omitted to focus specifically on the spectral index distribution within the ribs. } 
        \label{fig:ribsplot}
\end{figure*}

For the ribs, our methodology mirrors that used by \cite{rudnick}, where a profile cut was made across the particular section of the jet containing ribs in both intensity and spectral index, calculating the median values in the direction of the jet's propagation. While this method may not emphasize minor details, it effectively outlines the general trends in intensity and spectral index variations within the ribs. Figure \ref{fig:ribsplot} presents profile cuts of total intensity (in orange) and spectral index (in blue) across the ribs. The intensity profile exhibits a Gaussian distribution within the ribs region, aligning with observations noted in figure 6 of \cite{rudnick}.

The overall trend observed in the spectral index profile cut reveals that it is relatively flatter near the ribs, becomes steeper towards the edges of the jet's emission, and then flattens again as one moves further out. Notably, a distinct dip is observed in the profile cut around the left side of the ribs. This occurs because, despite the kink giving rise to transverse extensions, they are not closely stacked, leading to areas between them with fainter to absent cocoon emission, as can be seen in the figure \ref{fig:ribsplot}. Additionally, we mapped the spectral index distribution across the ribs, deliberately excluding cocoon emission to focus on the ribs' structure alone. Furthermore, regions of the ribs that exhibit higher brightness also display a flatter spectrum compared to adjacent areas. We observe that the spectral index tends to become flatter in the ribs as we move upward, which is in contrast to the observations. Therefore, while the intensity profile cuts resemble the observed features, the spectral index profile obtained from our simulations does not match the smoother profile obtained through the observations of \textit{MysTail}. 

\begin{figure*}
        \centering
        \includegraphics[width=\linewidth, keepaspectratio]{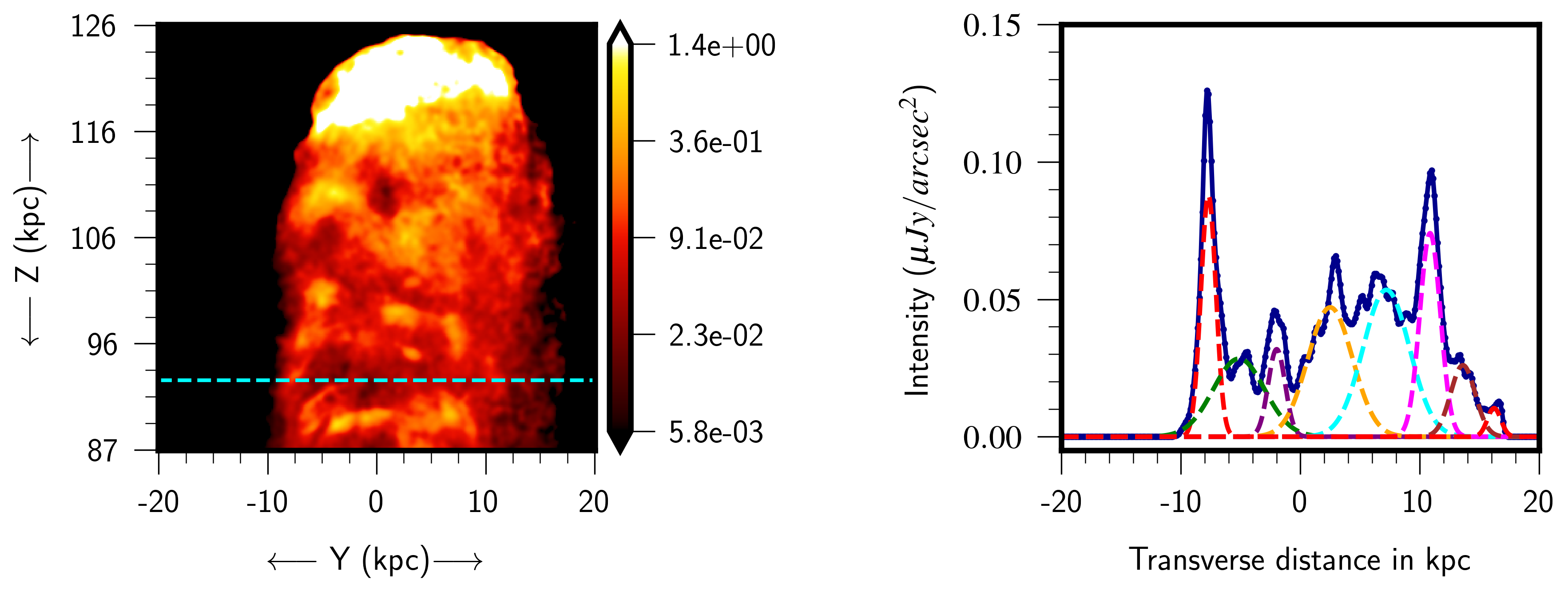}
        \caption{ This figure depicts the spectral signatures of tethers within the simulated jet, observed at a frequency of 1.285 GHz for the \textit{steady run} at time $t/t_0 = 1100$. The left panel showcases a region within the jet intensity map characterized by turbulent backflow, along with numerous hotspots and filamentary structures. The right panel displays a cross-sectional cut in total intensity across tethers made along the cyan-colored line shown in the left panel. The distinct peaks observed in total intensity (depicted in blue) are subsequently fitted with independent Gaussian functions, providing evidence for the presence of filamentary structures.}
        \label{fig:try2}
 \end{figure*}

As for tethers, we conducted a cross-sectional cut in total intensity across the area marked by a cyan line in Figure \ref{fig:try2}, identifying several distinct peaks. These peaks were modeled using eight separate Gaussian functions, analogous to the observational approach applied to tethers by \cite{rudnick}, where tethers were resolved into three separate Gaussians (see Figure 7 of \cite{rudnick}).

In summary, while kink instability and turbulent backflow contribute to the formation of structures somewhat similar to ribs and tethers, there are notable differences, such as the dip in the spectral index profile cut, the ribs' formation at a distance from the nozzle, and the tethers not being sufficiently elongated. These discrepancies led us to explore another scenario that might mitigate these anomalies while still attributing the formation of such structures to instabilities. Consequently, we conducted a simulation run featuring intermittent or restarted jet activity.

\subsection{Restarted Jet Activity}
\cite{riseley} studied this peculiar galaxy (\textit{MysTail}) and calculated its core prominence, and asserted that it aligns with the characteristics of known remnant radio galaxies. Therefore, it was suggested that this could indicate a phase of renewed jet activity. Furthermore, as previously discussed, the rib-like structures identified in the initial case were situated significantly distant from the nozzle and were not closely stacked. To check the feasibility of a remnant radio galaxy, we carried out a simulation featuring restarted jet activity. In this simulation,   the jet was halted at $t/t_0 = 500$ and then resumed injection at $t/t_0 = 900$. This pause at $t/t_0 = 500$ was strategically chosen to provide sufficient time for the current driven instability to evolve along the jet spine while also ensuring that the jet remained within the computational domain until the simulation concluded. 

\begin{figure*}
    \centering
    \subfigure[2-Dimensional slices of logarithmic density distribution]{
        \includegraphics[width=\linewidth]{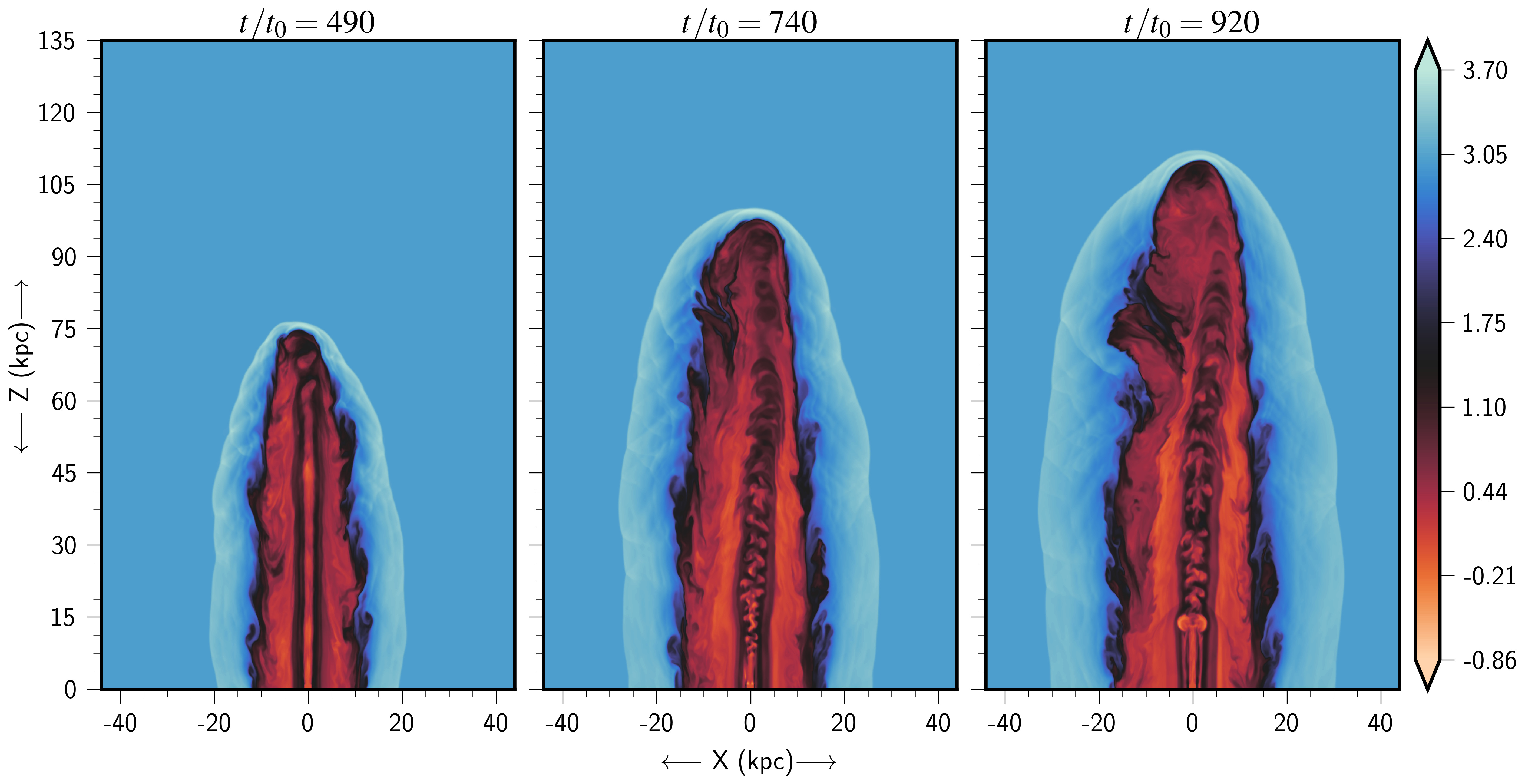}\label{fig:stopped_rho}}
     \subfigure[2-Dimensional slices of logarithmic pressure distribution]{
        \includegraphics[width=\linewidth]{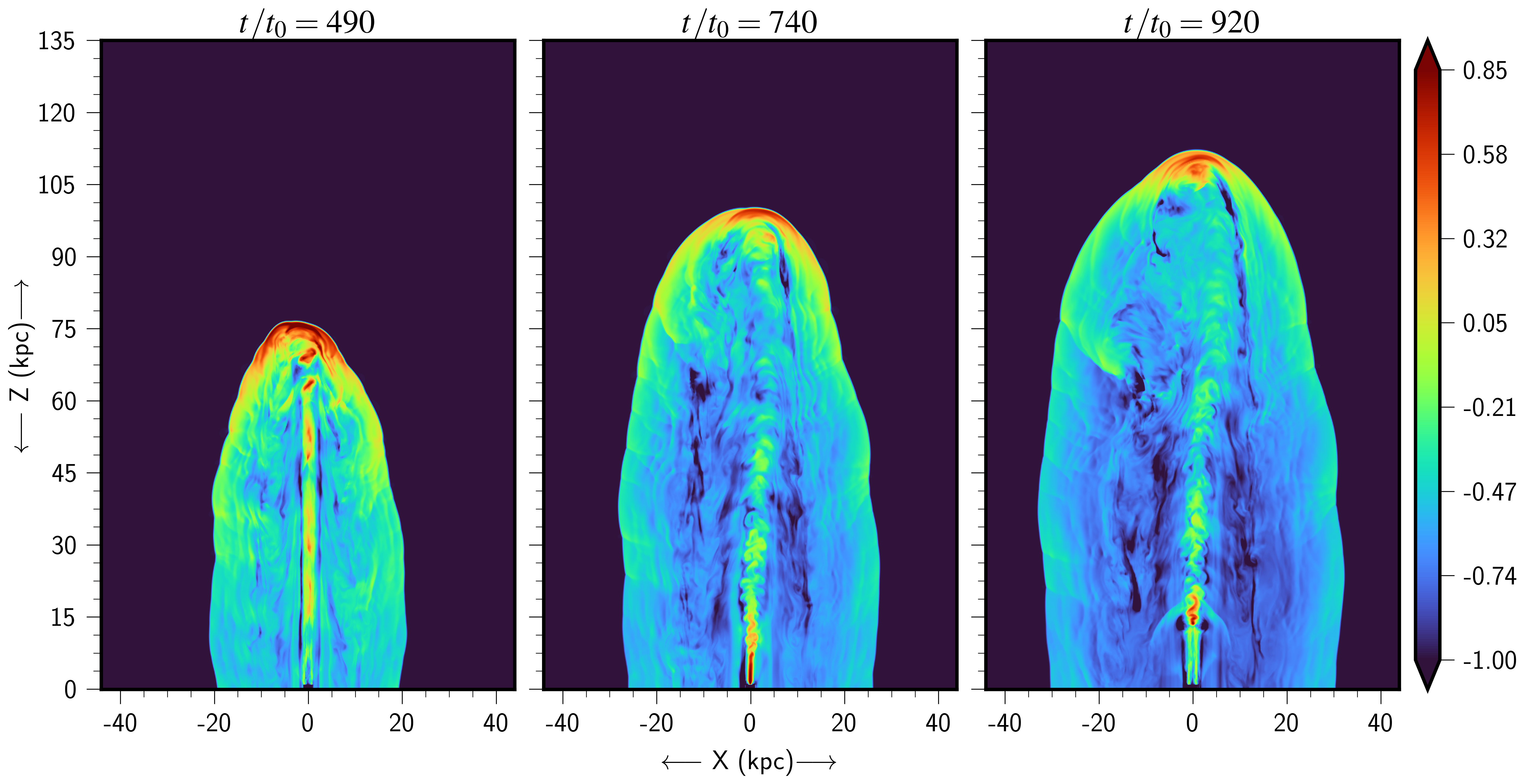}\label{fig:stopped_prs}}
    \caption{This figure illustrates three distinct phases of jet activity at times $t/t_0=490$, $t/t_0=740$, and $t/t_0=920$, representing the periods when the jet was undergoing continuous injection, when jet injection was paused for approximately 783.4 kyr, and when jet injection activity resumed for about 6.528 kyr, respectively (from left to right). The top panel displays the logarithmic density distribution, depicting the propagation of a kink towards the nozzle, causing the entire jet spine to become unstable. In the bottom panel, the logarithmic pressure distribution is presented, revealing a high-pressure region near the nozzle subsequent to the halting of jet injection.}
    \label{fig:stopped}
\end{figure*}

\subsubsection{Dynamical evolution}
Figure \ref{fig:stopped} presents the 2D slices of the logarithmic density and pressure distribution on the y-z plane, capturing three key moments: $t/t_0 = 490$, just before the jet injection was halted; $t/t_0 = 740$, a significant period after the jet injection was stopped; and $t/t_0 = 920$, shortly after the jet re-injection commenced. 
From the 2D slice plots of density and pressure at $t/t_0 = 490$, it is evident that the jet maintained collimation along its entire length, exhibiting slight wiggles near the jet head. Additionally, areas surrounding these wiggles and at the bow shock displayed elevated pressure values. Up to this point, the jet's dynamics were similar to those observed for the steady case. 
However, by halting the injection, we effectively stopped driving the jet forward, and instead of the jet expanding transversely and its spine disintegrating, it continued to advance at the expense of the magnetic energy. 
In our density and pressure distribution slices at $t/t_0 = 740$, a jet spine is still visible. Nonetheless, what was once a minor wiggle has now fully destabilized the jet spine while still keeping the spine structure intact to some extent. 
The patchy distribution of the density structure along the jet spine is attributed to the nonlinear interaction of various modes, including the $m=0$ mode after the jet injection has been stopped. Moreover, the bends observed in the patchy structure are due to the relaxation of the kink mode, which had already been influencing the spine prior to the injection being switched off.
As the instability progresses downwards, it leads to the formation of a high-pressure area near the nozzle, as illustrated in Figure \ref{fig:stopped_prs}. Upon resuming the injection at $t/t_0 = 920$ ($\approx$ 6.53 kyr after the jet restarts), an intriguing case of new jet interacting with remnant material becomes apparent. A less intense bow shock forms within the jet, attributed to the internal density of the pre-existing jet not being significantly higher than that of the jet material, leading to a subdued bow shock. Nonetheless, a high-pressure region is observed where the newly injected jet interacts with the pre-existing structure. As the injection continues, it begins to stabilize the jet spine once more, pushing instabilities away from the nozzle. Therefore, selecting such a time step was strategic, as at a later stage, the jet would have naturally stabilized itself. In the next sub-sections, we qualitatively compare the synthetic spectral and intensity maps from such complex dynamics of the restarted jet simulation run.

\subsubsection{Spectral Signatures and comparison with \textit{MysTail}}
\begin{figure*}[t]
    \centering
    \includegraphics[width=\linewidth]{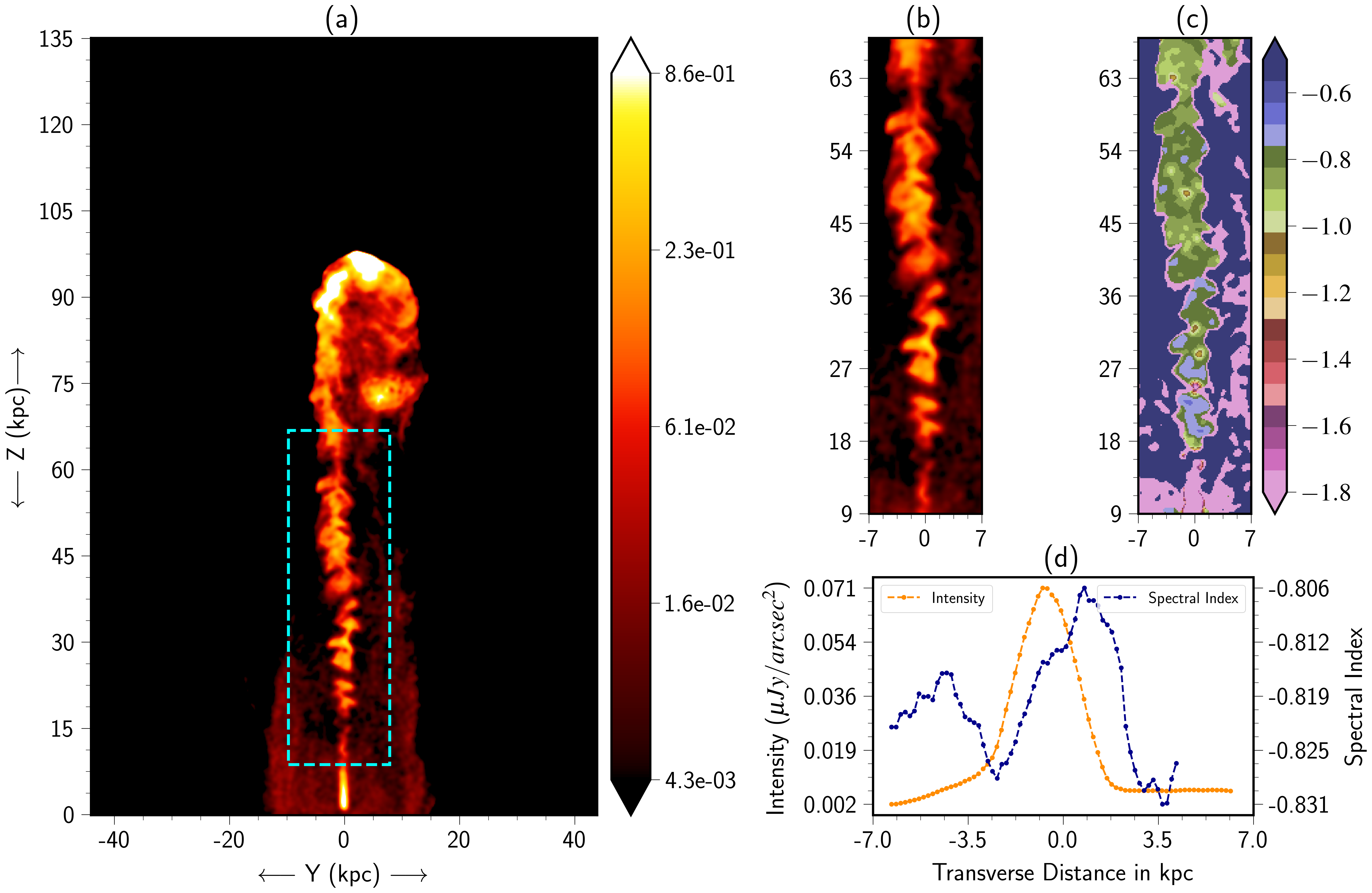}
    \caption{This figure illustrates the spectral morphology and signatures of the jet at time $t/t_0 = 740$, corresponding to 783.4 kyr after the cessation of jet injection. In the left panel, Subfigure (a) displays the integrated intensity map at a frequency of 1.285 GHz, with a cyan-colored rectangle highlighting a region within the jet exhibiting transverse extensions resembling a ribs-like structure. The right panel, Subfigure (b), presents a zoomed-in view of the total intensity map, showcasing ribs extending up to 58 kpc in length. Subfigure (c) illustrates the distribution of spectral indices within the ribs, while sub-figure (d) depicts a profile cut in total intensity and spectral index across the ribs patch, obtained by taking the median along the jet axis (Z-axis).}
    \label{fig:74}
\end{figure*}

\begin{figure*}[t]
    \centering
    \includegraphics[width=\linewidth]{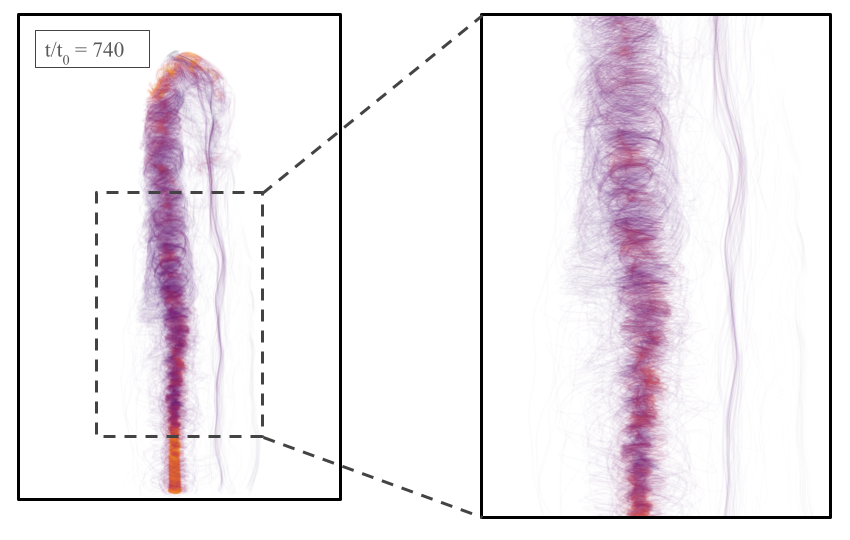}
    \caption{The figure illustrates the configuration of magnetic field lines at the time $t/t_0 = 740$ ($\sim$ 2.4 Myr). Here, the yellow and purple colors signify stronger and weaker magnetic field strengths, respectively. \textit{Zoomed-in panel:} This section illustrates the intricate twists and bends in magnetic field configuration along the jet's spine arising due to the relaxation of kink mode during the off-state, which manifest as transversely stretched, ribs-like structures reminiscent of the detailed morphology seen in \textit{MysTail}.}
    \label{fig:mag_740}
\end{figure*}

The 2-D intensity map was generated by integrating the synthetic synchrotron emissivity at 1.285 GHz along a line of sight of $(\theta, \phi)=(90^{\circ},0^{\circ})$. In this discussion, we specifically highlighted two timesteps of particular interest: $t/t_0 = 740$, marking a period when the jet had been halted for quite some time, and $t/t_0 = 920$, when the jet injection was recently resumed. Figure \ref{fig:74} illustrates the intensity map and spectral analysis focusing on the ribs at $t/t_0 = 740$, which is 783.4 kyrs after halting the jet injection. In the intensity map (panel (a)), we can distinctly observe transverse extensions featuring bright patches within the jet spine, highlighted by a blue box, which mirrors the rib-like features noted in \textit{MysTail}. We further computed the spectral index $\alpha$ following equation~\ref{eq:spec_alpha} and adopted the same two frequencies, $\nu_1 = 0.9$ GHz and $\nu_2 = 1.285$ GHz, as discussed in the section \ref{subsec:comparison}.
A zoomed-in picture of the intensity and the spectral index $\alpha$ within the blue box is shown in panels (b) and (c) of the figure \ref{fig:74}. Vertical profile cuts in intensity and spectral index (panel (d)) revealed a Gaussian profile that aligns qualitatively with the observations made by \cite{rudnick} (See their figure 6). In examining the spectral index map of the ribs, we noted that areas with brighter ribs exhibited a flatter spectral index $> -0.85$, suggesting the jet material underwent significant shocks. Furthermore, the spectral index distribution closely mirrors the trend in the observational data, characterized by a steeper spectrum encircling a region of flatter spectrum at the ribs' center. As one transitions towards the outer edges of the ribs, the emission spectrum steepens, especially near its edges. Further away from the jet axis, where there is no emission, the spectrum flattens. Moreover, Figure \ref{fig:mag_740} illustrates the magnetic field configuration at the timestep $t/t_0 = 740$, with a zoomed-in panel highlighting the region containing the ribs. The purple and yellow colors represent areas of weaker and stronger magnetic field strength, respectively. Notably, we observe bends and twists in the magnetic field within the ribs, which is attributed to the relaxation of the pre-existing kink mode that relaxed during the jet's off-state. The region near the jet head, which contains the backflow, exhibits a chaotic magnetic field configuration. We also obtained a long magnetic filamentary structure on the right side of the jet, similar to the one seen in Figure \ref{fig:test}, but significantly larger in scale.

\begin{figure*}[t]
    \centering
    \includegraphics[ width=\linewidth]{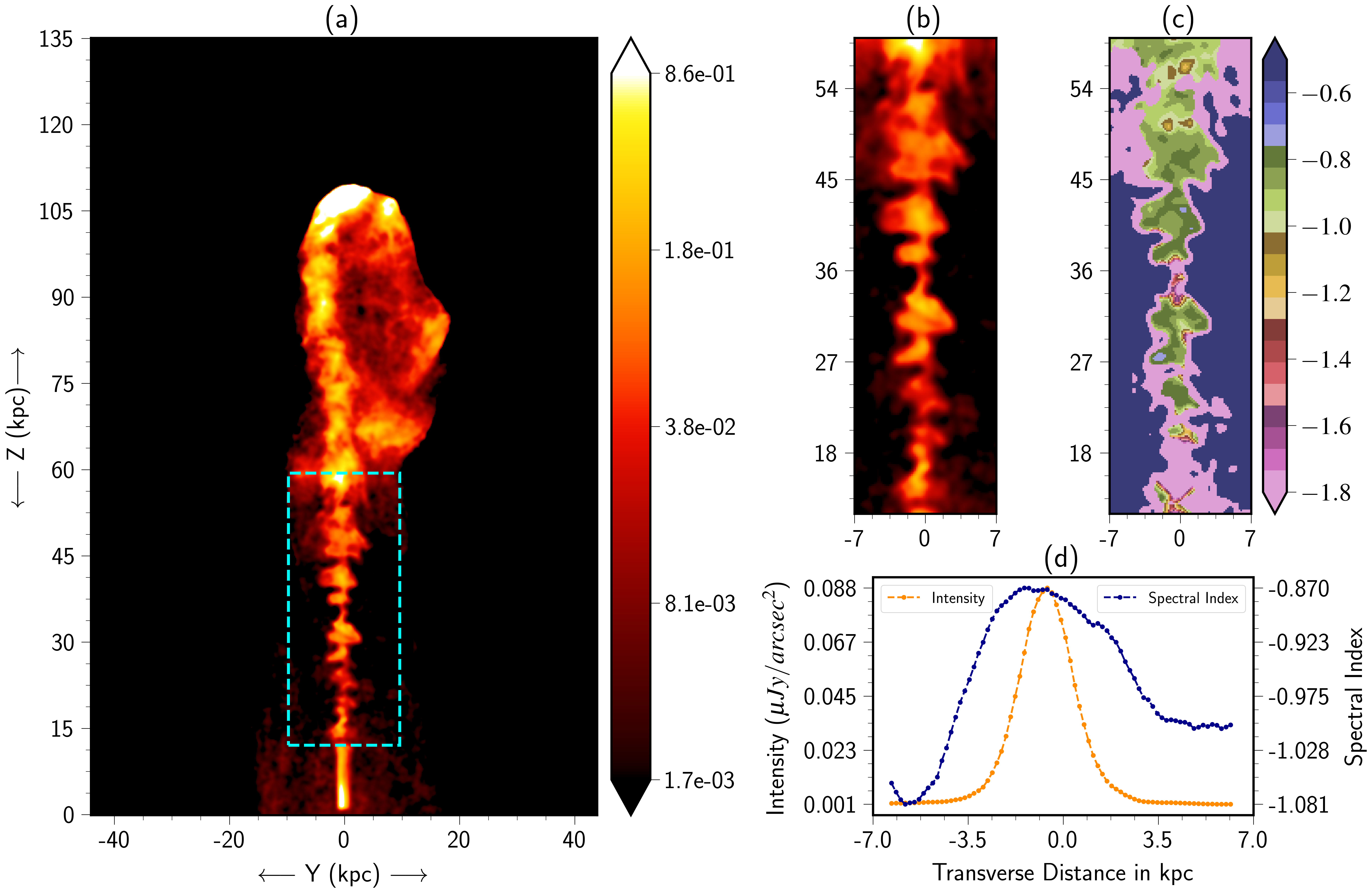}
    \caption{This figure illustrates the spectral morphology and signatures of the jet at time $t/t_0 = 920$, corresponding to 6.528 kyr after the jet injection restarted. In the left panel, Subfigure (a) displays the integrated intensity map at a frequency of 1.285 GHz, with a cyan-colored rectangle highlighting a region within the jet exhibiting transverse extensions resembling a ribs-like structure. The right panel, Subfigure (b), presents a zoomed-in view of the total intensity map, showcasing ribs extending up to 47 kpc in length. Subfigure (c) illustrates the distribution of spectral indices within the ribs, while Subfigure (d) depicts a profile cut in total intensity and spectral index across the ribs patch, obtained by taking the median along the jet axis (Z-axis). }
    \label{fig:92}
\end{figure*}

Additionally, at the $t/t_0 = 920$ time-step also, the intensity map in figure \ref{fig:92}  displays qualitatively similar morphological features and closely resembles the observed characteristics of \textit{MysTail}. The profile cut in total intensity and spectral index shows a similar trend to the observations. However, the cross-sectional profile cut in the spectral index at $ t/t_0 = 920 $ differs slightly from the profile cut at $ t/t_0 = 740 $. At $t/t_0=740$, there is an additional peak on the left side of the spectral index profile cut, which originates from the surrounding cocoon. This suggests that the cocoon still contains sufficiently energetic non-thermal particles. In contrast, by $ t/t_0 = 920 $, the particles in the cocoon have significantly cooled, resulting in steeper spectral index values in the profile cut. The spectral evolution for the restarted jet simulation run can be viewed through this \href{https://drive.google.com/file/d/1lKvT8plbrQUYdgu91Ji2rEptuYb2Uhx3/view?usp=sharing}{movie link}.


\begin{figure}[t]
    \centering
    \includegraphics[width=\columnwidth, keepaspectratio]{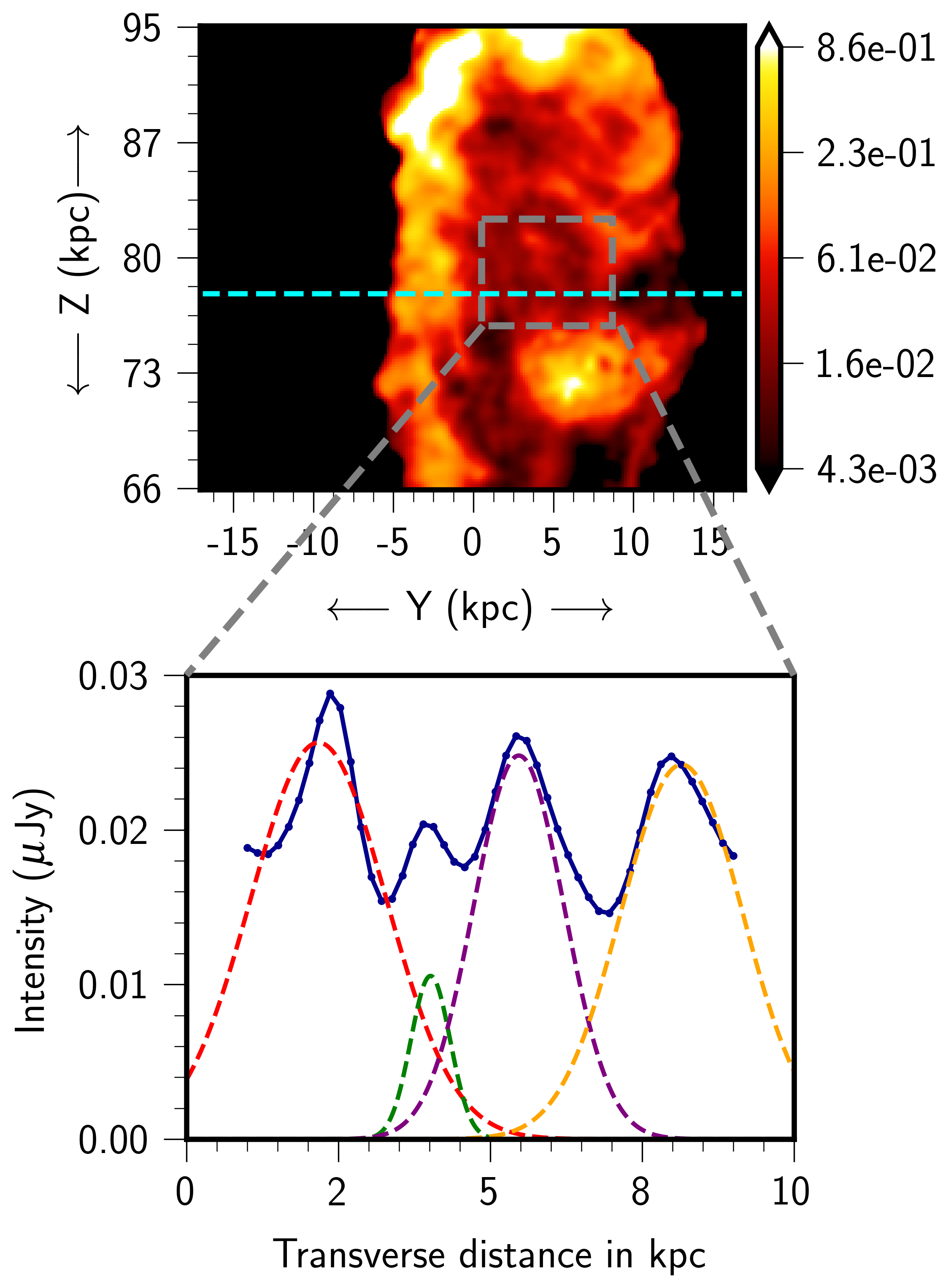}
    \caption{This figure illustrates the spectral signatures of tethers observed within the simulated jet at a frequency of 1.285 GHz, captured at time $t/t_0=740$ during the \textit{restarted} run. In the top panel, an intensity map of the region displaying tether-like structures is presented, with a gray box highlighting these features. The bottom panel displays a cross-sectional profile of total intensity across the tethers obtained along the cyan-colored line shown in the top panel. The distinct peaks observed in total intensity (depicted in blue) are subsequently fitted with independent Gaussian functions, providing evidence for the presence of filamentary structures. }
    \label{fig:tethers_plot}
\end{figure}

For the restarted jet simulation run, we also observed the emergence of longer tether-like structures. Despite the continued back-flow from the forward shock, the cessation of injection into our jet resulted in weaker shocks in that region, leading to fewer and less intense hot spots compared to the steady run. Consequently, longer filamentary structures resembling tethers are formed. Through the cross-sectional profile cut analysis in the tether region, which is outlined by the gray box in figure \ref{fig:tethers_plot}, we identified four separate peaks accounting for the majority of the flux in that region. This profile cut of the tethers can be characterized by four distinct Gaussian components, similar to how observations of tethers were resolved into three separate Gaussian components (refer to Figure 7 in \cite{rudnick}).
Furthermore, observations indicate that the width of individual tethers is approximately 7-8 kpc. The filamentary structures obtained from our restarted simulation run exhibit widths that are of the same order, measuring around 2-3 kpc. This comparison indicates that the structures generated in our restarted simulation run bear a qualitative resemblance to those observed in the \textit{MysTail}.

\subsubsection{Polarisation}
\begin{figure}[t]
    \centering
        \includegraphics[width=\columnwidth]{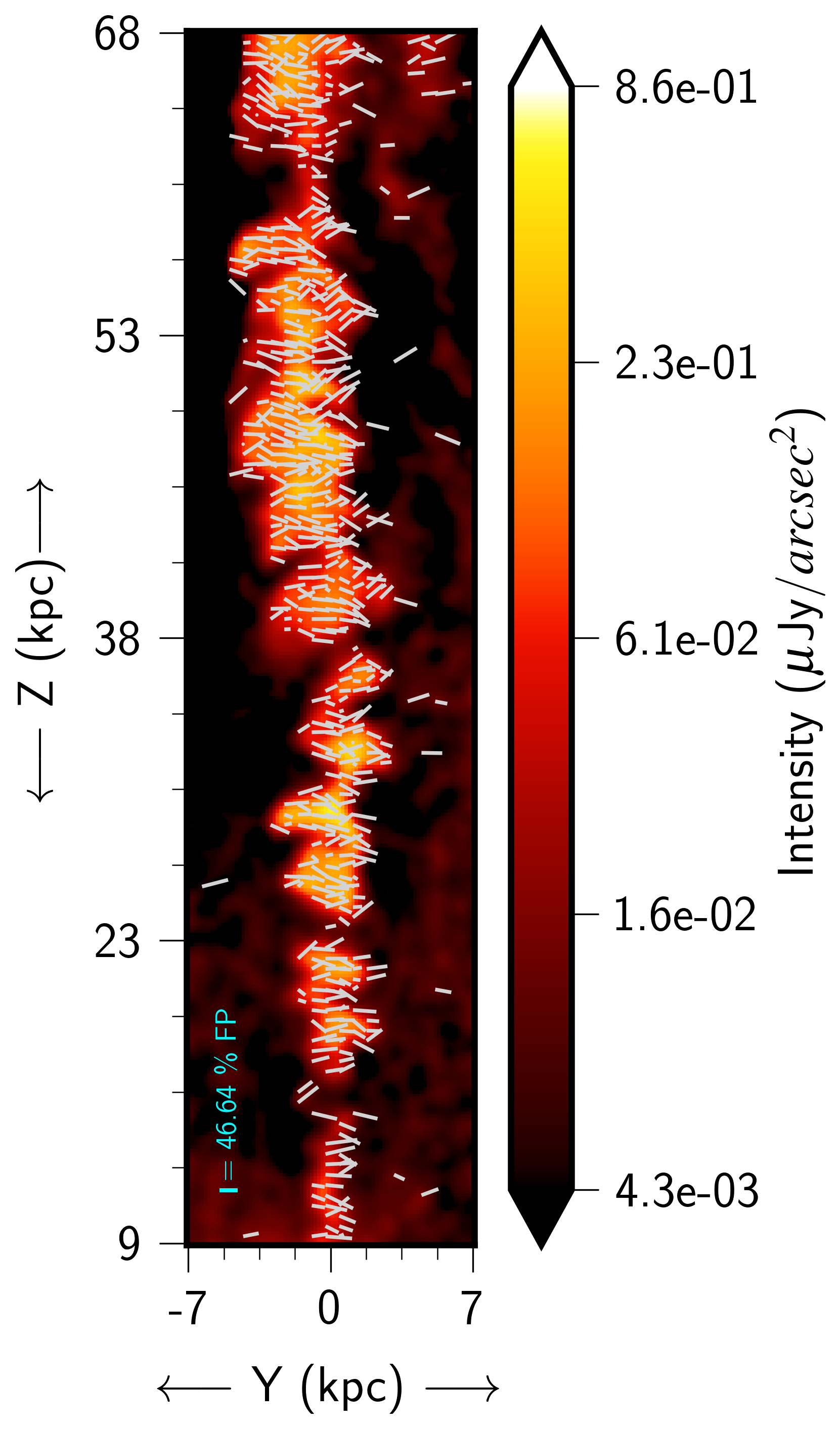}
        \caption{A plot representing the inferred magnetic field vectors in the ribs plotted over the 1.285 GHz convolved synchrotron intensity map. The length of the magnetic field vectors is proportional to the fractional polarisation at that location. }
        \label{fig:ribs_polar}
\end{figure}

\begin{figure}
    \centering
    \includegraphics[width=\columnwidth]{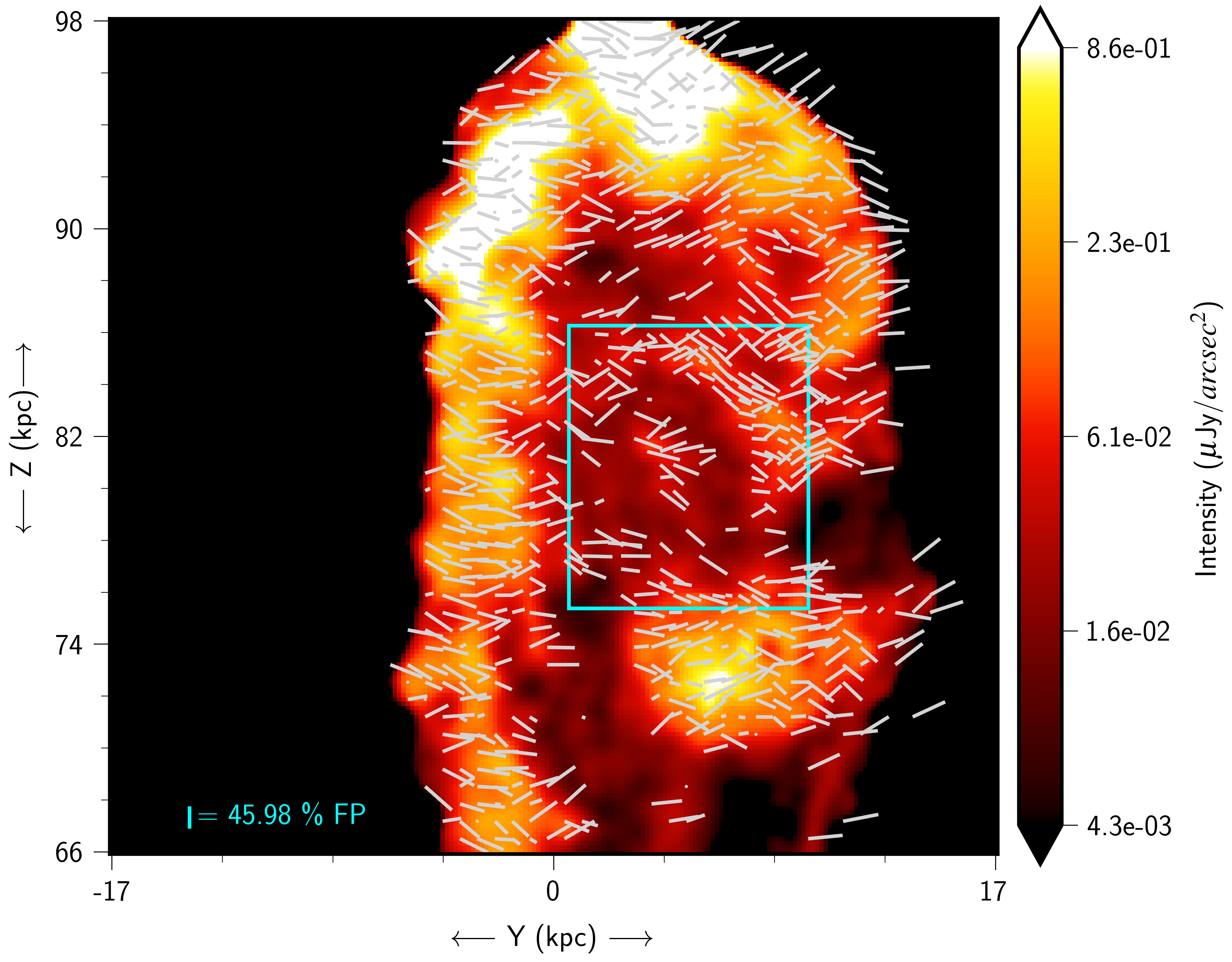}
    \caption{This figure presents the inferred magnetic field vectors within the region near the bow shock, which contains the tethers, plotted over the 1.285 GHz convolved synchrotron intensity map. The length of the magnetic field vectors is proportional to the fractional polarisation at that point.  }
    \label{fig:tethers_polar}
\end{figure}

Furthermore, we have also constructed the polarisation maps for the ribs and tethers observed in the restarted jet activity simulation at time $t/t_0=740$, derived from the linearly polarised synchrotron emission at a frequency of 1.285 GHz. 
Figure \ref{fig:ribs_polar} and \ref{fig:tethers_polar} illustrates the alignment of magnetic field vectors within the ribs and tethers, superimposed on a convolved synchrotron intensity map at the same frequency. We computed the polarisation angles within the jet utilizing the expression $\Theta = 0.5\tan^{-1}(Q_{\nu}/U_{\nu})$, followed by a rotation of $90^\circ$ to determine the orientation of the magnetic field vectors. However, this analysis did not take into account the effects of Faraday rotation. Moreover, the length of these vectors is scaled proportionally to the fractional polarisation at that point, calculated as $\sqrt{Q_{\nu}^2+U_{\nu}^2}/I_{\nu}$.

We observed that the jet exhibits highly linearly polarized regions, with an average fractional polarization of approximately $46 \%$ in both the ribs and tethers. 
In the ribs region, magnetic field vectors are predominantly perpendicular to the jet axis, although, in some areas, these vectors align along the bends in the jet spine. Additionally, in areas near the front shock that contain backflow material, the magnetic field vectors are less ordered. In the bright hotspots or shocked regions, the magnetic field vectors are aligned perpendicular to these areas. Conversely, for the extended filamentary structures, as highlighted in the area marked by the cyan box in Figure \ref{fig:tethers_polar}, the vectors align along the length of these structures. 
Such a patchy polarisation distribution mirrors the observed polarisation properties of \textit{MysTail} and suggests a complex magnetic field configuration within the jet. The polarisation results are also consistent with the magnetic field profiles shown in Figure~\ref{fig:test} for the steady jet case.

\section{Discussion and Summary}
\label{sec4}

We conducted a three-dimensional relativistic magnetohydrodynamics (RMHD) simulation using the hybrid framework of \texttt{PLUTO} code to explore the dynamics of a rotating jet, which was introduced into the computational domain via an injection nozzle. This study is aimed at understanding the mechanisms responsible for the emergence of distinctive morphological features such as ribs and tethers observed in the \textit{MysTail}. This unique morphology might be attributed to fluctuating jets, a phase of restarted AGN activity, or possibly dynamical instabilities within the jet. Our aim is to qualitatively interpret these features through the lens of dynamic instabilities. To achieve this, we ensured that the computational domain was expansive enough to comfortably encompass the entire jet throughout its evolution. The simulation setup involved a simplistic ambient medium characterized by a uniform density and a transverse magnetic field, $B_z$. Within this medium, we introduced an under-dense rotating jet that features a helical magnetic field configuration with Pitch, $P_c=0.01$ at the central jet axis. While jet rotation is naturally linked to magnetic energy extraction from rotating objects, such as accretion disks or black holes, it's important to note that in our simulations, jet dynamics are not solely governed by rotation. Other parameters, such as the magnetic pitch ($P_c$), Alfvénic Mach number ($M_a$), and Lorentz factor ($\gamma_c$), also play crucial roles in determining jet behavior. We conducted two types of simulation runs to investigate the jet dynamics: one featuring continuous jet injection and the other incorporating an interrupted injection. The latter approach involved halting the jet injection process at a time, $t/t_0=500$ and subsequently restarting it at time $t/t_0=900$, thereby mimicking the behavior of intermittent jet activity observed in certain astrophysical contexts. We then calculated the synthetic synchrotron emissivity at radio frequency 1.285 GHz using the particle module of the \texttt{PLUTO} code and integrated it along a line of sight defined by $(\theta, \phi)$=$(90^{\circ}, 0^{\circ})$, thereby observing the jet edge-on. This integration subsequently gave us the 2D intensity maps, which enable us to examine and interpret the spectral morphology of the jet in detail.  

To place our simulation results in context, it is important to consider the arguments provided from observations regarding whether \textit{MysTail} is a single-source or a two-source structure. Observations of \textit{MysTail} at a low radio frequency of $1.285\ \rm{GHz}$  using MeerKAT revealed a complex morphology characterized by ribs, tethers, and a diffused emission at the far region of the source, known as the triple, adjoining the tethers \citep{rudnick}. Apart from its confusing peculiar morphology, \textit{MysTail} appeared to have two cores, both aligning with the extended structure. This raised the question of whether this source was a single, contiguous radio source or if it consisted of overlapping jets from two separate radio galaxies. Spectral analysis of \textit{MysTail} suggested that it is indeed a single, contiguous structure, and the apparent host coinciding with the tail is located at a different redshift \citep{rudnick}.  Additionally, \cite{riseley} also studied this radio galaxy in their multiwavelength observations of the Abell 3266 cluster using ATCA and ASKAP-EMU. Their analysis of the spectral properties corroborated the findings of \cite{rudnick}. Furthermore, they assessed the core prominence of \textit{MysTail} and determined that it was consistent with known samples of remnant radio galaxies. Based on the observed spectral properties,  \cite{riseley} proposed that \textit{MysTail} could represent a phase of restarted AGN activity.

Taking into account the aforementioned arguments and observational characteristics of \textit{MysTail}, we intended to explain its peculiar morphology through simulations. Through our two simulation runs, we generated similar morphological features resembling ribs and tethers observed in \textit{MysTail}. However, in the steady injection scenario, the formed ribs were located far from the nozzle, and the spectral index profile did not quite match the observed trend. This discrepancy in the spectral index profile is attributed to the space between the loosely wound ribs, which contains fainter emission from the less-shocked cocoon, resulting in a dip in the spectral index profile plot. Additionally, the spectral index distribution indicated a steeper spectral index at the bottom of the ribs and a flatter profile towards the top, contrasting with observations. Furthermore, the filamentary structures obtained were not sufficiently elongated to be classified as tethers, which are defined by having lengths that are significantly greater than their widths. In contrast, the restarted run produced ribs that were formed closer to the nozzle and were more closely stacked. The region containing ribs spanned a larger length compared to the steady run. The profile cut in total intensity and spectral index were qualitatively similar to the observed spectral properties of the ribs. Additionally, for the restarted run, the turbulent filaments displayed lengths that were marginally greater than their widths, aligning more closely with the defining characteristics of tethers.

Moreover, we noted that the simulated jet has regions with high linear polarization, exhibiting a mean fractional polarization of approximately $46\%$, which is significantly higher than the observed values in \textit{MysTail} \citep{rudnick}. This overestimation is due to the fact that the values quantified in this study do not account for depolarisation due to Faraday rotation, which is beyond the scope of the current study. 
Nevertheless, such high fractional polarization indicates the presence of a highly ordered magnetic field configuration, likely resulting from moderate to strong shocks. Additionally, the prominent toroidal component of the magnetic field observed in both simulation runs primarily originates from the initially injected helical magnetic field configuration. Observational studies do suggest that magnetic field lines in FRI jets are predominantly toroidal \citep{bridle1,laing}. Furthermore, helical magnetic fields have been documented in several sources \citep{asada, m87_helicalfield, zamaninasab}, supporting the presence of a significant toroidal component in the magnetic fields of launched jets. Moreover, we also noted that in our simulation, these inferred magnetic field vectors tend to align along the elongated filaments.

While these findings are promising, it is important to acknowledge the limitations of our parameter survey. Given the four-dimensional parameter space that governs the jet dynamics, it is challenging and computationally expensive to pinpoint the exact set of parameters needed to accurately simulate such complex morphology. Further exploration of the parameter space is essential for obtaining quantitative estimates, particularly if the objective is to study such peculiar morphological features in other sources. Additionally, the scope of our simulations is also constrained by computational costs, preventing us from making a direct quantitative comparison with \textit{MysTail}. The ribs and tethers of \textit{MysTail} span approximately 400 kpc, while our simulated jet extends only 135 kpc. To match the scales of \textit{MysTail}, we would need to run the simulation for a longer time in a significantly larger domain size, which would increase the computational expense. However, the morphological features observed in the simulation are independent of the size or duration of the simulation; these features would simply be more developed and would span larger lengths with longer simulation runs.

Also, due to the significant computational resources required for simulating a larger domain, our study focused on a one-sided jet injection. For simulations, the counter jet is supposed to be symmetric. However, in the observations of \textit{MysTail}, the jet is visible only on one side. This asymmetry may suggest that \textit{MysTail} could be a one-sided jet. Alternatively, it is possible that the jet initially emerged as a two-sided structure, with the counter jet being decelerated due to the upwind. Another plausible and more complex explanation is that the counter jet might have bent or altered its trajectory, leading to a projection effect that makes it appear one-sided in the observed data. Hence, there are likely other complex phenomena at play. In our study, while we used MysTail as an example to represent the ribs and tethers, our primary goal is to qualitatively correlate these non-axisymmetric peculiar features with kink instability. Therefore, to clearly capture the effects of the kink instability, we set the viewing angle to 90$^{\circ}$. It should also be noted that we have primarily focused on diffusive shock acceleration (DSA) as the acceleration mechanism in this work. However, we acknowledge that other mechanisms, such as magnetic reconnection \citep{striani, Singh2016} and turbulence \citep{sayan2022}, also contribute to particle acceleration. While these processes are indeed important, there are a lot of uncertainties due to their inherently micro-physical nature, which is beyond the scope of the present study. Our pilot study provides a viable explanation for the observed features, but more intricate dynamics could also be contributing to the jet's morphology, which can be explored in further studies through a larger set of parameter runs.

In summary, we have demonstrated the applicability of a hybrid modeling toolkit that combines 3D RMHD simulation with particle acceleration and associated non-thermal emission signatures to produce distinctive features in the complex \textit{MysTail} radio jet. Our runs clearly indicate that observed ribs and tethers resemble synthetic features via the MHD kink instability from a single source, which is currently in the remnant phase, supporting pieces of evidence from radio observation of this complex radio galaxy.

\section*{Acknowledgements}
The authors express their gratitude to the reviewer, Yosuke Mizuno, for his insightful suggestions and comments, which contributed to the significant improvement of this manuscript. NU acknowledges the support provided by DAASE, IIT Indore during the course of the MS(by Research) Degree. BV extends thanks to Christian Fendt and Lawrence Rudnick for their fruitful discussions. BV also acknowledges the support of the Max Planck Partner Group established at IIT Indore. The computations in this work were performed using the facilities at IIT Indore and the MPG supercomputing cluster "Raven."

\appendix
\section{Jet parameter profile}
\label{appendixA}
For the setup of our injection nozzle, we implemented the equilibrium configuration as outlined by \citep{Bodo_2019}.
The magnetized jet for the simulation model is defined as an axisymmetric cylindrical flow defined in a cylindrical system of coordinates, and we seek a steady-state solution for this. The z-direction is assumed to be the direction of propagation of the jet, and the magnetic field is configured such that it has only two components toroidal component ($B_\phi$) and a poloidal(vertical) component ($B_{z}$), both having a dependence on radial distance r. This configuration can be characterized by another parameter called pitch parameter, which is defined as the ratio of these components.

\begin{align}
    &\ P= r\frac{B_z}{B_{\phi}}
\end{align}
Using the steady state condition, continuity equation, and the fact that there is no dependence on z and $\phi$, one can write $v_r$ as :

\begin{align}
   & \mathbf{v} = v_z(r) \mathbf{e}_z + v_\phi(r) \mathbf{e}_\phi = \kappa(r) \mathbf{B} + \Omega(r) r \mathbf{e}_\phi
\end{align}
where $v_\phi$ and $\Omega$ are the toroidal velocity and the angular velocity of field lines, respectively.
As B has no component in the radial direction and there is no dependence of z and $\phi$, we will have an electric field that is directed radially outwards.
Now, the only equation that remains non-trivial is the radial component of the momentum equation, which, in the zero pressure limit, on simplification gives,
\begin{align}
    &\ \rho \gamma^2 v_\phi^2 = \frac{1}{2r} \frac{d(r^2 H^2)}{dr} + \frac{r^2}{2} \frac{d B_z^2}{dr}
\end{align}
here $H^2=B_{\phi}^2-E_{r}^2$ and $E_r=v_zB_{\phi}-v_{\phi}B_z$.

The radial momentum equation presented in the model allows for selecting radial profiles for specific flow variables, which can then be used to determine the remaining profiles. The approach begins with defining the profiles for proper density and velocity, ensuring that the variations within these quantities are contained within the jet radius, denoted as $r_j$. Specifically, the velocity profile for the z component of velocity is described through the Lorentz factor $\gamma_j$, given as
\begin{align}
    &\gamma_z(r)= 1 + \frac{\gamma_c - 1}{\cosh^6 \left(\frac{r}{r_j}\right)}
\end{align}
where $\gamma_c$ is the Lorentz factor at the central axis. Quantities with a subscript c would denote values at r=0.

Density is defined as,
\begin{align}
    &\rho(r)= \eta + \frac{1 - \eta}{\cosh^6 \left(\frac{r}{r_j}\right)} \\
\end{align}
where $\eta$ is the contrast between ambient and jet density.

They assign a profile to H as,
\begin{align}
    &H^2 = H_c^2 r^2 \left[ 1 - \exp \left(-\frac{r^4}{a^4} \right) \right]
\end{align}

and the azimuthal velocity is described as,
\begin{align}
     &\gamma^2 v_\phi^2 = r^2 \gamma_c^2 \Omega_c^2\exp \left(-\frac{r^4}{a^4} \right),
\end{align}
where $\Omega_c$ denotes the angular velocity of the jet rotation at the central axis. The characteristic radius, \textit{a}, is chosen to be 0.6 ($< r_j$) so that the current density would be concentrated inside the jet.  On putting the value of $\gamma$ in the above equation, one obtains the expression for azimuthal velocity as,
\begin{align}
     &v_\phi^2 = \frac{r^2 \gamma_c^2 \Omega_c^2 }{\gamma_z^2}\left[1 + r^2 \gamma_c^2 \beta_c^2 \exp \left(-\frac{r^4}{a^4} \right) \right]^{-1}\exp \left(-\frac{r^4}{a^4} \right)
\end{align}

\begin{figure*}[t]
    \centering
    \includegraphics[width=1\linewidth]{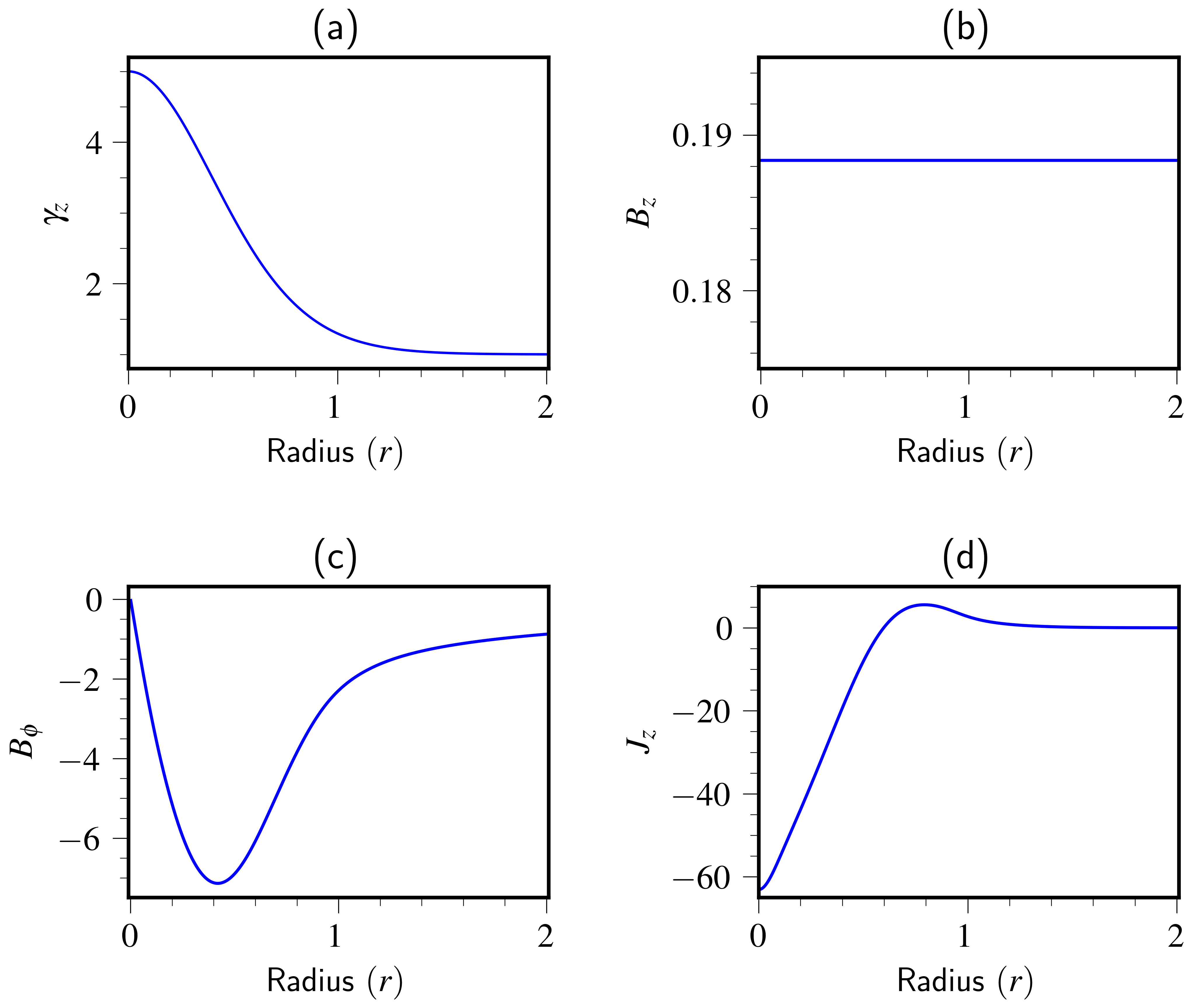}
    \caption{This figure displays the radial profiles for various physical quantities within the jet: the bulk Lorentz factor in the z-direction ($\gamma_z$), the poloidal magnetic field component ($B_z$), the toroidal magnetic field component ($B_{\phi}$), and the axial current density ($J_z$).}
    \label{fig:radial profiles}
\end{figure*}

This equation will ensure that $v_{\phi}$ is always less than unity for any value of $\Omega_c$.
Using the radial momentum equation(2.7), and the expression for H and $v_{\phi}$ one can derive the profile for $B_z$ as, 
 \begin{align}
       &B_z^2 = B_{zc}^2 - \frac{(1 - \alpha) H_c^2}{\sqrt{\pi} a^2} \operatorname{erf} \left(\frac{r^2}{a^2} \right)
 \end{align}
 where erf denotes the error function and $\alpha$ measures the strength of rotation defined as,
 \begin{align}
     &\alpha = \frac{\rho \gamma_c^2 \Omega_c^2 a^4}{2 H_c^2}.
 \end{align}
 $\alpha$=0 depicts no rotation in the jet so the $\Omega_c$ will go to zero and thus the gradient of $r^2H^2 $ gets exactly balanced out by the gradient of $B_z^2$ which means $B_z$ will decrease outwards. However, for $\alpha$=1, $B_z$ becomes constant throughout, and thus, the centrifugal force balances out the $r^2H^2$ gradient. Whereas for $\alpha>1$, $B_z$ will increase radially outwards. For this study, we have considered fully rotating jets ($\alpha$=1).
 The profile for the toroidal magnetic field is obtained by the definitions of $H^2$ and $E_r^2$, which will result in a quadratic equation in $B_{\phi}$. Upon solving the form of $B_{\phi}$ is obtained as,

 \begin{align}
     B_\phi &= \frac{-v_\phi v_z B_z \pm\sqrt{v_\phi^2 B_z^2 + H^2(1 - v_z^2)}}{1 - v_z^2}
     \label{b_phi profile}
\end{align}
As the acceleration models suggest that the direction of azimuthal velocity and toroidal magnetic field should be different \citep{payne}, so here, the negative branch of the solution is considered. Doing so will make sure they both are opposite in direction.
As mentioned, the pitch parameter is used to characterize the configuration of the magnetic field, and it is defined near the jet axis ($r\to0$) as,

\begin{align}
\label{1}
     &P_c \equiv \left|\frac{r B_z}{B_\phi}\right|_{r=0}
\end{align}
Another parameter that is used to characterize the magnetic field configuration is $M_a^2$ (alfvenic mach number), defined as the ratio of the matter-energy density to the magnetic energy density, 
\begin{align}
\label{2}
      &M_a^2 \equiv \frac{\rho \gamma^2_c}{\langle B^2\rangle}
\end{align}
 where $\langle B^2 \rangle$ denotes the average magnetic field in the radial direction
 \begin{align}
 \label{3}
     &\langle B^2 \rangle =\frac{\int_{0}^{r_j} (B_z^2 + B_\phi^2) r dr}{\int_{0}^{r_j} r dr}, 
 \end{align}
 Here $r_j$=1 in this set of units. Now only remaining unknowns are the constant $B_{zc}$ and$H_c$ they can be found out by simultaneously solving the equations \eqref{1}, \eqref{2} and \eqref{3} using the expressions for the magnetic field components.
 After conducting thorough calculations, a relationship between these parameters in the limit $r\to 0$ is obtained as,
 \begin{align}
    & B_{zc}^2 = \frac{H_c^2 P_c^2}{1 - (P_c \Omega_c - v_{zc})^2}.
 \end{align}
 To ensure that these equations yield a meaningful physical solution, there are additional constraints on the magnetic field configuration such that $B_z^2$ and $B_{\phi}^2$ should be positive everywhere. This limits the number of combinations $\Omega_c$, $\gamma_c$,$P_c$, and $M_a$ can have to give a physical solution.
\\
Through these equations, a magnetized rotating relativistic jet can be described, contingent on the selection of parameters such as $\rho_j$, $\gamma_c$, $\Omega_c$, $B_{zc}$, $P_c$, and $p$. The chosen values for these parameters are outlined in Table \ref{tab:Table 1}. In Figure \ref{fig:radial profiles}, the profiles for $\gamma_z$, $B_z$, $B_{phi}$, and $J_z$ are presented, showcasing their variation across the radius r within the stable jet nozzle.

\bibliography{sample631}

\begin{thebibliography}{96}
\expandafter\ifx\csname natexlab\endcsname\relax\def\natexlab#1{#1}\fi
\providecommand{\url}[1]{\texttt{#1}}
\providecommand{\href}[2]{#2}
\providecommand{\path}[1]{#1}
\providecommand{\DOIprefix}{doi:}
\providecommand{\ArXivprefix}{arXiv:}
\providecommand{\URLprefix}{URL: }
\providecommand{\Pubmedprefix}{pmid:}
\providecommand{\doi}[1]{\href{http://dx.doi.org/#1}{\path{#1}}}
\providecommand{\Pubmed}[1]{\href{pmid:#1}{\path{#1}}}
\providecommand{\bibinfo}[2]{#2}
\ifx\xfnm\relax \def\xfnm[#1]{\unskip,\space#1}\fi
\bibitem[{{Acharya} et~al.(2023){Acharya}, {Vaidya}, {Kalpa Dihingia}, {Agarwal} and {Shukla}}]{Acharya2023}
\bibinfo{author}{{Acharya}, S.}, \bibinfo{author}{{Vaidya}, B.}, \bibinfo{author}{{Kalpa Dihingia}, I.}, \bibinfo{author}{{Agarwal}, S.}, \bibinfo{author}{{Shukla}, A.}, \bibinfo{year}{2023}.
\newblock \bibinfo{title}{{A numerical study on the role of instabilities on multi-wavelength emission signatures of blazar jets}}.
\newblock \bibinfo{journal}{Astronomy \& Astrophysics} \bibinfo{volume}{671}, \bibinfo{pages}{A161}.
\newblock \DOIprefix\doi{10.1051/0004-6361/202244256}.
\bibitem[{{Asada} et~al.(2002){Asada}, {Inoue}, {Uchida}, {Kameno}, {Fujisawa}, {Iguchi} and {Mutoh}}]{asada}
\bibinfo{author}{{Asada}, K.}, \bibinfo{author}{{Inoue}, M.}, \bibinfo{author}{{Uchida}, Y.}, \bibinfo{author}{{Kameno}, S.}, \bibinfo{author}{{Fujisawa}, K.}, \bibinfo{author}{{Iguchi}, S.}, \bibinfo{author}{{Mutoh}, M.}, \bibinfo{year}{2002}.
\newblock \bibinfo{title}{{A Helical Magnetic Field in the Jet of 3C 273}}.
\newblock \bibinfo{journal}{Publications of the Astronomical Society of Japan} \bibinfo{volume}{54}, \bibinfo{pages}{L39--L43}.
\newblock \DOIprefix\doi{10.1093/pasj/54.3.L39}.
\bibitem[{Barniol~Duran et~al.(2017)Barniol~Duran, Tchekhovskoy and Giannios}]{barniolduran}
\bibinfo{author}{Barniol~Duran, R.}, \bibinfo{author}{Tchekhovskoy, A.}, \bibinfo{author}{Giannios, D.}, \bibinfo{year}{2017}.
\newblock \bibinfo{title}{{Simulations of AGN jets: magnetic kink instability versus conical shocks}}.
\newblock \bibinfo{journal}{Monthly Notices of the Royal Astronomical Society} \bibinfo{volume}{469}, \bibinfo{pages}{4957--4978}.
\newblock \DOIprefix\doi{10.1093/mnras/stx1165}.
\bibitem[{Bassani et~al.(2016)Bassani, Venturi, Molina, Malizia, Dallacasa, Panessa, Bazzano and Ubertini}]{10.1093/mnras/stw1468}
\bibinfo{author}{Bassani, L.}, \bibinfo{author}{Venturi, T.}, \bibinfo{author}{Molina, M.}, \bibinfo{author}{Malizia, A.}, \bibinfo{author}{Dallacasa, D.}, \bibinfo{author}{Panessa, F.}, \bibinfo{author}{Bazzano, A.}, \bibinfo{author}{Ubertini, P.}, \bibinfo{year}{2016}.
\newblock \bibinfo{title}{{Soft gamma-ray selected radio galaxies: favouring giant size discovery}}.
\newblock \bibinfo{journal}{Monthly Notices of the Royal Astronomical Society} \bibinfo{volume}{461}, \bibinfo{pages}{3165--3171}.
\newblock \DOIprefix\doi{10.1093/mnras/stw1468}.
\bibitem[{{Bateman}(1978)}]{1978mit..book.....B}
\bibinfo{author}{{Bateman}, G.}, \bibinfo{year}{1978}.
\newblock \bibinfo{title}{{MHD instabilities}}.
\bibitem[{{B{\^\i}rzan} et~al.(2008){B{\^\i}rzan}, {McNamara}, {Nulsen}, {Carilli} and {Wise}}]{birzan2008}
\bibinfo{author}{{B{\^\i}rzan}, L.}, \bibinfo{author}{{McNamara}, B.R.}, \bibinfo{author}{{Nulsen}, P.E.J.}, \bibinfo{author}{{Carilli}, C.L.}, \bibinfo{author}{{Wise}, M.W.}, \bibinfo{year}{2008}.
\newblock \bibinfo{title}{{Radiative Efficiency and Content of Extragalactic Radio Sources: Toward a Universal Scaling Relation between Jet Power and Radio Power}}.
\newblock \bibinfo{journal}{The Astrophysical Journal} \bibinfo{volume}{686}, \bibinfo{pages}{859--880}.
\newblock \DOIprefix\doi{10.1086/591416}.
\bibitem[{Blandford et~al.(2019)Blandford, Meier and Readhead}]{Blandford_review}
\bibinfo{author}{Blandford, R.}, \bibinfo{author}{Meier, D.}, \bibinfo{author}{Readhead, A.}, \bibinfo{year}{2019}.
\newblock \bibinfo{title}{Relativistic jets from active galactic nuclei}.
\newblock \bibinfo{journal}{Annual Review of Astronomy and Astrophysics} \bibinfo{volume}{57}, \bibinfo{pages}{467--509}.
\newblock \DOIprefix\doi{10.1146/annurev-astro-081817-051948}.
\bibitem[{{Blandford} and {Payne}(1982)}]{payne}
\bibinfo{author}{{Blandford}, R.D.}, \bibinfo{author}{{Payne}, D.G.}, \bibinfo{year}{1982}.
\newblock \bibinfo{title}{{Hydromagnetic flows from accretion disks and the production of radio jets.}}
\newblock \bibinfo{journal}{Monthly Notices of the Royal Astronomical Society} \bibinfo{volume}{199}, \bibinfo{pages}{883--903}.
\newblock \DOIprefix\doi{10.1093/mnras/199.4.883}.
\bibitem[{Blandford and Znajek(1977)}]{blandford1977electromagnetic}
\bibinfo{author}{Blandford, R.D.}, \bibinfo{author}{Znajek, R.L.}, \bibinfo{year}{1977}.
\newblock \bibinfo{title}{Electromagnetic extraction of energy from kerr black holes}.
\newblock \bibinfo{journal}{Monthly Notices of the Royal Astronomical Society} \bibinfo{volume}{179}, \bibinfo{pages}{433--456}.
\bibitem[{{B{\"o}ckmann} et~al.(2023){B{\"o}ckmann}, {Br{\"u}ggen}, {Koribalski}, {Veronica}, {Reiprich}, {Bulbul}, {Bahar}, {Balzer}, {Comparat}, {Garrel}, {Ghirardini}, {G{\"u}rkan}, {Kluge}, {Leahy}, {Merloni}, {Liu}, {Ramos-Ceja}, {Salvato}, {Sanders}, {Shabala} and {Zhang}}]{erosita}
\bibinfo{author}{{B{\"o}ckmann}, K.}, \bibinfo{author}{{Br{\"u}ggen}, M.}, \bibinfo{author}{{Koribalski}, B.}, \bibinfo{author}{{Veronica}, A.}, \bibinfo{author}{{Reiprich}, T.H.}, \bibinfo{author}{{Bulbul}, E.}, \bibinfo{author}{{Bahar}, Y.E.}, \bibinfo{author}{{Balzer}, F.}, \bibinfo{author}{{Comparat}, J.}, \bibinfo{author}{{Garrel}, C.}, \bibinfo{author}{{Ghirardini}, V.}, \bibinfo{author}{{G{\"u}rkan}, G.}, \bibinfo{author}{{Kluge}, M.}, \bibinfo{author}{{Leahy}, D.}, \bibinfo{author}{{Merloni}, A.}, \bibinfo{author}{{Liu}, A.}, \bibinfo{author}{{Ramos-Ceja}, M.E.}, \bibinfo{author}{{Salvato}, M.}, \bibinfo{author}{{Sanders}, J.}, \bibinfo{author}{{Shabala}, S.}, \bibinfo{author}{{Zhang}, X.}, \bibinfo{year}{2023}.
\newblock \bibinfo{title}{{Central radio galaxies in galaxy clusters: Joint surveys by eROSITA and ASKAP}}.
\newblock \bibinfo{journal}{Astronomy \& Astrophysics} \bibinfo{volume}{677}, \bibinfo{pages}{A188}.
\newblock \DOIprefix\doi{10.1051/0004-6361/202346912}.
\bibitem[{Bodo et~al.(2013)Bodo, Mamatsashvili, Rossi and Mignone}]{bodo2013__}
\bibinfo{author}{Bodo, G.}, \bibinfo{author}{Mamatsashvili, G.}, \bibinfo{author}{Rossi, P.}, \bibinfo{author}{Mignone, A.}, \bibinfo{year}{2013}.
\newblock \bibinfo{title}{{Linear stability analysis of magnetized relativistic jets: the non-rotating case}}.
\newblock \bibinfo{journal}{Monthly Notices of the Royal Astronomical Society} \bibinfo{volume}{434}, \bibinfo{pages}{3030--3046}.
\newblock \DOIprefix\doi{10.1093/mnras/stt1225}.
\bibitem[{{Bodo} et~al.(2019){Bodo}, {Mamatsashvili}, {Rossi} and {Mignone}}]{Bodo_2019}
\bibinfo{author}{{Bodo}, G.}, \bibinfo{author}{{Mamatsashvili}, G.}, \bibinfo{author}{{Rossi}, P.}, \bibinfo{author}{{Mignone}, A.}, \bibinfo{year}{2019}.
\newblock \bibinfo{title}{{Linear stability analysis of magnetized relativistic rotating jets}}.
\newblock \bibinfo{journal}{Monthly Notices of the Royal Astronomical Society} \bibinfo{volume}{485}, \bibinfo{pages}{2909--2921}.
\newblock \DOIprefix\doi{10.1093/mnras/stz591}.
\bibitem[{{Bodo} and {Tavecchio}(2018)}]{bodo_tavechio}
\bibinfo{author}{{Bodo}, G.}, \bibinfo{author}{{Tavecchio}, F.}, \bibinfo{year}{2018}.
\newblock \bibinfo{title}{{Recollimation shocks and radiative losses in extragalactic relativistic jets}}.
\newblock \bibinfo{journal}{Astronomy \& Astrophysics} \bibinfo{volume}{609}, \bibinfo{pages}{A122}.
\newblock \DOIprefix\doi{10.1051/0004-6361/201732000}.
\bibitem[{Bodo et~al.(2020)Bodo, Tavecchio and Sironi}]{bodo2021}
\bibinfo{author}{Bodo, G.}, \bibinfo{author}{Tavecchio, F.}, \bibinfo{author}{Sironi, L.}, \bibinfo{year}{2020}.
\newblock \bibinfo{title}{{Kink-driven magnetic reconnection in relativistic jets: consequences for X-ray polarimetry of BL Lacs}}.
\newblock \bibinfo{journal}{Monthly Notices of the Royal Astronomical Society} \bibinfo{volume}{501}, \bibinfo{pages}{2836--2847}.
\newblock \DOIprefix\doi{10.1093/mnras/staa3620}.
\bibitem[{{Borse, Nikhil} et~al.(2021){Borse, Nikhil}, {Acharya, Sriyasriti}, {Vaidya, Bhargav}, {Mukherjee, Dipanjan}, {Bodo, Gianluigi}, {Rossi, Paola} and {Mignone, Andrea}}]{Nikhil2021}
\bibinfo{author}{{Borse, Nikhil}}, \bibinfo{author}{{Acharya, Sriyasriti}}, \bibinfo{author}{{Vaidya, Bhargav}}, \bibinfo{author}{{Mukherjee, Dipanjan}}, \bibinfo{author}{{Bodo, Gianluigi}}, \bibinfo{author}{{Rossi, Paola}}, \bibinfo{author}{{Mignone, Andrea}}, \bibinfo{year}{2021}.
\newblock \bibinfo{title}{Numerical study of the kelvin-helmholtz instability and its effect on synthetic emission from magnetized jets}.
\newblock \bibinfo{journal}{Astronomy \& Astrophysics} \bibinfo{volume}{649}, \bibinfo{pages}{A150}.
\newblock \DOIprefix\doi{10.1051/0004-6361/202140440}.
\bibitem[{{Bridle} and {Perley}(1984)}]{bridle1}
\bibinfo{author}{{Bridle}, A.H.}, \bibinfo{author}{{Perley}, R.A.}, \bibinfo{year}{1984}.
\newblock \bibinfo{title}{{Extragalactic Radio Jets}}.
\newblock \bibinfo{journal}{Annual Review of Astronomy and Astrophysics} \bibinfo{volume}{22}, \bibinfo{pages}{319--358}.
\newblock \DOIprefix\doi{10.1146/annurev.aa.22.090184.001535}.
\bibitem[{{Brienza} et~al.(2022){Brienza}, {Lovisari}, {Rajpurohit}, {Bonafede}, {Gastaldello}, {Murgia}, {Vazza}, {Bonnassieux}, {Botteon}, {Brunetti}, {Drabent}, {Hardcastle}, {Pasini}, {Riseley}, {R{\"o}ttgering}, {Shimwell}, {Simionescu} and {van Weeren}}]{brienza}
\bibinfo{author}{{Brienza}, M.}, \bibinfo{author}{{Lovisari}, L.}, \bibinfo{author}{{Rajpurohit}, K.}, \bibinfo{author}{{Bonafede}, A.}, \bibinfo{author}{{Gastaldello}, F.}, \bibinfo{author}{{Murgia}, M.}, \bibinfo{author}{{Vazza}, F.}, \bibinfo{author}{{Bonnassieux}, E.}, \bibinfo{author}{{Botteon}, A.}, \bibinfo{author}{{Brunetti}, G.}, \bibinfo{author}{{Drabent}, A.}, \bibinfo{author}{{Hardcastle}, M.J.}, \bibinfo{author}{{Pasini}, T.}, \bibinfo{author}{{Riseley}, C.J.}, \bibinfo{author}{{R{\"o}ttgering}, H.J.A.}, \bibinfo{author}{{Shimwell}, T.}, \bibinfo{author}{{Simionescu}, A.}, \bibinfo{author}{{van Weeren}, R.J.}, \bibinfo{year}{2022}.
\newblock \bibinfo{title}{{The galaxy group NGC 507: Newly detected AGN remnant plasma transported by sloshing}}.
\newblock \bibinfo{journal}{Astronomy \& Astrophysics} \bibinfo{volume}{661}, \bibinfo{pages}{A92}.
\newblock \DOIprefix\doi{10.1051/0004-6361/202142579}.
\bibitem[{Bromberg and Tchekhovskoy(2015)}]{bromberg_tchekov}
\bibinfo{author}{Bromberg, O.}, \bibinfo{author}{Tchekhovskoy, A.}, \bibinfo{year}{2015}.
\newblock \bibinfo{title}{{Relativistic MHD simulations of core-collapse GRB jets: 3D instabilities and magnetic dissipation}}.
\newblock \bibinfo{journal}{Monthly Notices of the Royal Astronomical Society} \bibinfo{volume}{456}, \bibinfo{pages}{1739--1760}.
\newblock \DOIprefix\doi{10.1093/mnras/stv2591}.
\bibitem[{{Capetti} et~al.(2002){Capetti}, {Zamfir}, {Rossi}, {Bodo}, {Zanni} and {Massaglia}}]{capetti}
\bibinfo{author}{{Capetti}, A.}, \bibinfo{author}{{Zamfir}, S.}, \bibinfo{author}{{Rossi}, P.}, \bibinfo{author}{{Bodo}, G.}, \bibinfo{author}{{Zanni}, C.}, \bibinfo{author}{{Massaglia}, S.}, \bibinfo{year}{2002}.
\newblock \bibinfo{title}{{On the origin of X-shaped radio-sources: New insights from the properties of their host galaxies}}.
\newblock \bibinfo{journal}{Astronomy \& Astrophysics} \bibinfo{volume}{394}, \bibinfo{pages}{39--45}.
\newblock \DOIprefix\doi{10.1051/0004-6361:20021070}.
\bibitem[{Carilli and Barthel(1996)}]{carilli1996cygnus}
\bibinfo{author}{Carilli, C.}, \bibinfo{author}{Barthel, P.}, \bibinfo{year}{1996}.
\newblock \bibinfo{title}{Cygnus a}.
\newblock \bibinfo{journal}{The astronomy and astrophysics review} \bibinfo{volume}{7}, \bibinfo{pages}{1--54}.
\bibitem[{{Cavagnolo} et~al.(2010){Cavagnolo}, {McNamara}, {Nulsen}, {Carilli}, {Jones} and {B{\^\i}rzan}}]{cavagnolo}
\bibinfo{author}{{Cavagnolo}, K.W.}, \bibinfo{author}{{McNamara}, B.R.}, \bibinfo{author}{{Nulsen}, P.E.J.}, \bibinfo{author}{{Carilli}, C.L.}, \bibinfo{author}{{Jones}, C.}, \bibinfo{author}{{B{\^\i}rzan}, L.}, \bibinfo{year}{2010}.
\newblock \bibinfo{title}{{A Relationship Between AGN Jet Power and Radio Power}}.
\newblock \bibinfo{journal}{The Astrophysical Journal} \bibinfo{volume}{720}, \bibinfo{pages}{1066--1072}.
\newblock \DOIprefix\doi{10.1088/0004-637X/720/2/1066}.
\bibitem[{{Chen} et~al.(2023){Chen}, {Heinz} and {Hooper}}]{yi-hao_chen}
\bibinfo{author}{{Chen}, Y.H.}, \bibinfo{author}{{Heinz}, S.}, \bibinfo{author}{{Hooper}, E.}, \bibinfo{year}{2023}.
\newblock \bibinfo{title}{{A numerical study of the impact of jet magnetic topology on radio galaxy evolution}}.
\newblock \bibinfo{journal}{Monthly Notices of the Royal Astronomical Society} \bibinfo{volume}{522}, \bibinfo{pages}{2850--2868}.
\newblock \DOIprefix\doi{10.1093/mnras/stad1074}.
\bibitem[{{Colella} and {Woodward}(1984)}]{colella1984}
\bibinfo{author}{{Colella}, P.}, \bibinfo{author}{{Woodward}, P.R.}, \bibinfo{year}{1984}.
\newblock \bibinfo{title}{{The Piecewise Parabolic Method (PPM) for Gas-Dynamical Simulations}}.
\newblock \bibinfo{journal}{Journal of Computational Physics} \bibinfo{volume}{54}, \bibinfo{pages}{174--201}.
\newblock \DOIprefix\doi{10.1016/0021-9991(84)90143-8}.
\bibitem[{Cotton et~al.(2020)Cotton, Thorat, Condon, Frank, Józsa, White, Deane, Oozeer, Atemkeng, Bester, Fanaroff, Kupa, Smirnov, Mauch, Krishnan and Camilo}]{cotton2020}
\bibinfo{author}{Cotton, W.D.}, \bibinfo{author}{Thorat, K.}, \bibinfo{author}{Condon, J.J.}, \bibinfo{author}{Frank, B.S.}, \bibinfo{author}{Józsa, G.I.G.}, \bibinfo{author}{White, S.V.}, \bibinfo{author}{Deane, R.}, \bibinfo{author}{Oozeer, N.}, \bibinfo{author}{Atemkeng, M.}, \bibinfo{author}{Bester, L.}, \bibinfo{author}{Fanaroff, B.}, \bibinfo{author}{Kupa, R.S.}, \bibinfo{author}{Smirnov, O.M.}, \bibinfo{author}{Mauch, T.}, \bibinfo{author}{Krishnan, V.}, \bibinfo{author}{Camilo, F.}, \bibinfo{year}{2020}.
\newblock \bibinfo{title}{{Hydrodynamical backflow in X-shaped radio galaxy PKS 2014-55}}.
\newblock \bibinfo{journal}{Monthly Notices of the Royal Astronomical Society} \bibinfo{volume}{495}, \bibinfo{pages}{1271--1283}.
\newblock \DOIprefix\doi{10.1093/mnras/staa1240}.
\bibitem[{{Dabhade} et~al.(2020){Dabhade}, {Mahato}, {Bagchi}, {Saikia}, {Combes}, {Sankhyayan}, {R{\"o}ttgering}, {Ho}, {Gaikwad}, {Raychaudhury}, {Vaidya} and {Guiderdoni}}]{2020A&A...642A.153D}
\bibinfo{author}{{Dabhade}, P.}, \bibinfo{author}{{Mahato}, M.}, \bibinfo{author}{{Bagchi}, J.}, \bibinfo{author}{{Saikia}, D.J.}, \bibinfo{author}{{Combes}, F.}, \bibinfo{author}{{Sankhyayan}, S.}, \bibinfo{author}{{R{\"o}ttgering}, H.J.A.}, \bibinfo{author}{{Ho}, L.C.}, \bibinfo{author}{{Gaikwad}, M.}, \bibinfo{author}{{Raychaudhury}, S.}, \bibinfo{author}{{Vaidya}, B.}, \bibinfo{author}{{Guiderdoni}, B.}, \bibinfo{year}{2020}.
\newblock \bibinfo{title}{{Search and analysis of giant radio galaxies with associated nuclei (SAGAN). I. New sample and multi-wavelength studies}}.
\newblock \bibinfo{journal}{Astronomy \& Astrophysics} \bibinfo{volume}{642}, \bibinfo{pages}{A153}.
\newblock \DOIprefix\doi{10.1051/0004-6361/202038344}.
\bibitem[{{Dedner} et~al.(2002){Dedner}, {Kemm}, {Kr{\"o}ner}, {Munz}, {Schnitzer} and {Wesenberg}}]{dedner2002}
\bibinfo{author}{{Dedner}, A.}, \bibinfo{author}{{Kemm}, F.}, \bibinfo{author}{{Kr{\"o}ner}, D.}, \bibinfo{author}{{Munz}, C.D.}, \bibinfo{author}{{Schnitzer}, T.}, \bibinfo{author}{{Wesenberg}, M.}, \bibinfo{year}{2002}.
\newblock \bibinfo{title}{{Hyperbolic Divergence Cleaning for the MHD Equations}}.
\newblock \bibinfo{journal}{Journal of Computational Physics} \bibinfo{volume}{175}, \bibinfo{pages}{645--673}.
\newblock \DOIprefix\doi{10.1006/jcph.2001.6961}.
\bibitem[{{Dehghan} et~al.(2017){Dehghan}, {Johnston-Hollitt}, {Colless} and {Miller}}]{dehghan2017}
\bibinfo{author}{{Dehghan}, S.}, \bibinfo{author}{{Johnston-Hollitt}, M.}, \bibinfo{author}{{Colless}, M.}, \bibinfo{author}{{Miller}, R.}, \bibinfo{year}{2017}.
\newblock \bibinfo{title}{{Dynamics of Abell 3266 - I. An optical view of a complex merging cluster}}.
\newblock \bibinfo{journal}{Monthly Notices of the Royal Astronomical Society} \bibinfo{volume}{468}, \bibinfo{pages}{2645--2654}.
\newblock \DOIprefix\doi{10.1093/mnras/stx582}.
\bibitem[{{Del Zanna} et~al.(2003){Del Zanna}, {Bucciantini} and {Londrillo}}]{del2003}
\bibinfo{author}{{Del Zanna}, L.}, \bibinfo{author}{{Bucciantini}, N.}, \bibinfo{author}{{Londrillo}, P.}, \bibinfo{year}{2003}.
\newblock \bibinfo{title}{{An efficient shock-capturing central-type scheme for multidimensional relativistic flows. II. Magnetohydrodynamics}}.
\newblock \bibinfo{journal}{Astronomy \& Astrophysics} \bibinfo{volume}{400}, \bibinfo{pages}{397--413}.
\newblock \DOIprefix\doi{10.1051/0004-6361:20021641}.
\bibitem[{Dennett-Thorpe et~al.(2002)Dennett-Thorpe, Scheuer, Laing, Bridle, Pooley and Reich}]{dennett-thorpe}
\bibinfo{author}{Dennett-Thorpe, J.}, \bibinfo{author}{Scheuer, P.A.G.}, \bibinfo{author}{Laing, R.A.}, \bibinfo{author}{Bridle, A.H.}, \bibinfo{author}{Pooley, G.G.}, \bibinfo{author}{Reich, W.}, \bibinfo{year}{2002}.
\newblock \bibinfo{title}{{Jet reorientation in active galactic nuclei: two winged radio galaxies}}.
\newblock \bibinfo{journal}{Monthly Notices of the Royal Astronomical Society} \bibinfo{volume}{330}, \bibinfo{pages}{609--620}.
\newblock \DOIprefix\doi{10.1046/j.1365-8711.2002.05106.x}.
\bibitem[{{Dubey} et~al.(2023){Dubey}, {Fendt} and {Vaidya}}]{Dubey2023}
\bibinfo{author}{{Dubey}, R.P.}, \bibinfo{author}{{Fendt}, C.}, \bibinfo{author}{{Vaidya}, B.}, \bibinfo{year}{2023}.
\newblock \bibinfo{title}{{Particles in Relativistic MHD Jets. I. Role of Jet Dynamics in Particle Acceleration}}.
\newblock \bibinfo{journal}{The Astrophysical Journal} \bibinfo{volume}{952}, \bibinfo{pages}{1}.
\newblock \DOIprefix\doi{10.3847/1538-4357/ace0bf}.
\bibitem[{{Ekers} et~al.(1978){Ekers}, {Fanti}, {Lari} and {Parma}}]{ekers}
\bibinfo{author}{{Ekers}, R.D.}, \bibinfo{author}{{Fanti}, R.}, \bibinfo{author}{{Lari}, C.}, \bibinfo{author}{{Parma}, P.}, \bibinfo{year}{1978}.
\newblock \bibinfo{title}{{NGC326 - A radio galaxy with a precessing beam}}.
\newblock \bibinfo{journal}{Nature} \bibinfo{volume}{276}, \bibinfo{pages}{588--590}.
\newblock \DOIprefix\doi{10.1038/276588a0}.
\bibitem[{English et~al.(2016)English, Hardcastle and Krause}]{hardcastle}
\bibinfo{author}{English, W.}, \bibinfo{author}{Hardcastle, M.J.}, \bibinfo{author}{Krause, M.G.}, \bibinfo{year}{2016}.
\newblock \bibinfo{title}{Numerical modelling of the lobes of radio galaxies in cluster environments--iii. powerful relativistic and non-relativistic jets}.
\newblock \bibinfo{journal}{Monthly Notices of the Royal Astronomical Society} \bibinfo{volume}{461}, \bibinfo{pages}{2025--2043}.
\bibitem[{Fanaroff and Riley(1974)}]{fanaroff1974morphology}
\bibinfo{author}{Fanaroff, B.L.}, \bibinfo{author}{Riley, J.M.}, \bibinfo{year}{1974}.
\newblock \bibinfo{title}{The morphology of extragalactic radio sources of high and low luminosity}.
\newblock \bibinfo{journal}{Monthly Notices of the Royal Astronomical Society} \bibinfo{volume}{167}, \bibinfo{pages}{31P--36P}.
\bibitem[{Fuentes et~al.(2018)Fuentes, Gómez, Martí and Perucho}]{Fuentes_2018}
\bibinfo{author}{Fuentes, A.}, \bibinfo{author}{Gómez, J.L.}, \bibinfo{author}{Martí, J.M.}, \bibinfo{author}{Perucho, M.}, \bibinfo{year}{2018}.
\newblock \bibinfo{title}{Total and linearly polarized synchrotron emission from overpressured magnetized relativistic jets}.
\newblock \bibinfo{journal}{The Astrophysical Journal} \bibinfo{volume}{860}, \bibinfo{pages}{121}.
\newblock \DOIprefix\doi{10.3847/1538-4357/aac091}.
\bibitem[{Gaibler et~al.(2009)Gaibler, Krause and Camenzind}]{gaiblerr_krause}
\bibinfo{author}{Gaibler, V.}, \bibinfo{author}{Krause, M.}, \bibinfo{author}{Camenzind, M.}, \bibinfo{year}{2009}.
\newblock \bibinfo{title}{{Very light magnetized jets on large scales – I. Evolution and magnetic fields}}.
\newblock \bibinfo{journal}{Monthly Notices of the Royal Astronomical Society} \bibinfo{volume}{400}, \bibinfo{pages}{1785--1802}.
\newblock \DOIprefix\doi{10.1111/j.1365-2966.2009.15625.x}.
\bibitem[{Giri et~al.(2023)Giri, Vaidya and Fendt}]{Giri_2023}
\bibinfo{author}{Giri, G.}, \bibinfo{author}{Vaidya, B.}, \bibinfo{author}{Fendt, C.}, \bibinfo{year}{2023}.
\newblock \bibinfo{title}{Deciphering the morphological origins of x-shaped radio galaxies: Numerical modeling of backflow versus jet reorientation}.
\newblock \bibinfo{journal}{The Astrophysical Journal Supplement Series} \bibinfo{volume}{268}, \bibinfo{pages}{49}.
\newblock \DOIprefix\doi{10.3847/1538-4365/acebca}.
\bibitem[{Giri et~al.(2022)Giri, Vaidya, Rossi, Bodo, Mukherjee and Mignone}]{Giri_2022}
\bibinfo{author}{Giri, G.}, \bibinfo{author}{Vaidya, B.}, \bibinfo{author}{Rossi, P.}, \bibinfo{author}{Bodo, G.}, \bibinfo{author}{Mukherjee, D.}, \bibinfo{author}{Mignone, A.}, \bibinfo{year}{2022}.
\newblock \bibinfo{title}{Modelling x-shaped radio galaxies: Dynamical and emission signatures from the back-flow model}.
\newblock \bibinfo{journal}{Astronomy \& Astrophysics} \bibinfo{volume}{662}, \bibinfo{pages}{A5}.
\newblock \DOIprefix\doi{10.1051/0004-6361/202142546}.
\bibitem[{Gopal-Krishna et~al.(2003)Gopal-Krishna, Biermann and Wiita}]{Gopal-Krishna_2003}
\bibinfo{author}{Gopal-Krishna}, \bibinfo{author}{Biermann, P.L.}, \bibinfo{author}{Wiita, P.J.}, \bibinfo{year}{2003}.
\newblock \bibinfo{title}{The origin of x-shaped radio galaxies: Clues from the z-symmetric secondary lobes}.
\newblock \bibinfo{journal}{The Astrophysical Journal} \bibinfo{volume}{594}, \bibinfo{pages}{L103}.
\newblock \DOIprefix\doi{10.1086/378766}.
\bibitem[{{Gower} et~al.(1982){Gower}, {Gregory}, {Unruh} and {Hutchings}}]{gower1982}
\bibinfo{author}{{Gower}, A.C.}, \bibinfo{author}{{Gregory}, P.C.}, \bibinfo{author}{{Unruh}, W.G.}, \bibinfo{author}{{Hutchings}, J.B.}, \bibinfo{year}{1982}.
\newblock \bibinfo{title}{{Relativistic precessing jets in quasars and radio galaxies : models to fit high resolution data.}}
\newblock \bibinfo{journal}{The Astrophysical Journal} \bibinfo{volume}{262}, \bibinfo{pages}{478--496}.
\newblock \DOIprefix\doi{10.1086/160442}.
\bibitem[{Guan et~al.(2014)Guan, Li and Li}]{guan2014relativistic}
\bibinfo{author}{Guan, X.}, \bibinfo{author}{Li, H.}, \bibinfo{author}{Li, S.}, \bibinfo{year}{2014}.
\newblock \bibinfo{title}{Relativistic mhd simulations of poynting flux-driven jets}.
\newblock \bibinfo{journal}{The Astrophysical Journal} \bibinfo{volume}{781}, \bibinfo{pages}{48}.
\bibitem[{{Hardcastle} et~al.(2019){Hardcastle}, {Croston}, {Shimwell}, {Tasse}, {G{\"u}rkan}, {Morganti}, {Murgia}, {R{\"o}ttgering}, {van Weeren} and {Williams}}]{hardcastle2019}
\bibinfo{author}{{Hardcastle}, M.J.}, \bibinfo{author}{{Croston}, J.H.}, \bibinfo{author}{{Shimwell}, T.W.}, \bibinfo{author}{{Tasse}, C.}, \bibinfo{author}{{G{\"u}rkan}, G.}, \bibinfo{author}{{Morganti}, R.}, \bibinfo{author}{{Murgia}, M.}, \bibinfo{author}{{R{\"o}ttgering}, H.J.A.}, \bibinfo{author}{{van Weeren}, R.J.}, \bibinfo{author}{{Williams}, W.L.}, \bibinfo{year}{2019}.
\newblock \bibinfo{title}{{NGC 326: X-shaped no more}}.
\newblock \bibinfo{journal}{Monthly Notices of the Royal Astronomical Society} \bibinfo{volume}{488}, \bibinfo{pages}{3416--3422}.
\newblock \DOIprefix\doi{10.1093/mnras/stz1910}.
\bibitem[{{Hines} et~al.(1989){Hines}, {Owen} and {Eilek}}]{owen}
\bibinfo{author}{{Hines}, D.C.}, \bibinfo{author}{{Owen}, F.N.}, \bibinfo{author}{{Eilek}, J.A.}, \bibinfo{year}{1989}.
\newblock \bibinfo{title}{{Filaments in the Radio Lobes of M87}}.
\newblock \bibinfo{journal}{The Astrophysical Journal} \bibinfo{volume}{347}, \bibinfo{pages}{713}.
\newblock \DOIprefix\doi{10.1086/168163}.
\bibitem[{{Hodges-Kluck} and {Reynolds}(2012)}]{hodges-kluck}
\bibinfo{author}{{Hodges-Kluck}, E.J.}, \bibinfo{author}{{Reynolds}, C.S.}, \bibinfo{year}{2012}.
\newblock \bibinfo{title}{{A Chandra Study of the Radio Galaxy NGC 326: Wings, Outburst History, and Active Galactic Nucleus Feedback}}.
\newblock \bibinfo{journal}{The Astrophysical Journal} \bibinfo{volume}{746}, \bibinfo{pages}{167}.
\newblock \DOIprefix\doi{10.1088/0004-637X/746/2/167}.
\bibitem[{Hodges-Kluck et~al.(2010)Hodges-Kluck, Reynolds, Miller and Cheung}]{Hodges-Kluck_2010}
\bibinfo{author}{Hodges-Kluck, E.J.}, \bibinfo{author}{Reynolds, C.S.}, \bibinfo{author}{Miller, M.C.}, \bibinfo{author}{Cheung, C.C.}, \bibinfo{year}{2010}.
\newblock \bibinfo{title}{A deep chandra observation of the x-shaped radio galaxy 4c +00.58: A candidate for merger-induced reorientation?}
\newblock \bibinfo{journal}{The Astrophysical Journal Letters} \bibinfo{volume}{717}, \bibinfo{pages}{L37}.
\newblock \DOIprefix\doi{10.1088/2041-8205/717/1/L37}.
\bibitem[{Huarte-Espinosa et~al.(2011)Huarte-Espinosa, Krause and Alexander}]{huarte}
\bibinfo{author}{Huarte-Espinosa, M.}, \bibinfo{author}{Krause, M.}, \bibinfo{author}{Alexander, P.}, \bibinfo{year}{2011}.
\newblock \bibinfo{title}{{3D magnetohydrodynamic simulations of the evolution of magnetic fields in Fanaroff–Riley class II radio sources}}.
\newblock \bibinfo{journal}{Monthly Notices of the Royal Astronomical Society} \bibinfo{volume}{417}, \bibinfo{pages}{382--399}.
\newblock \DOIprefix\doi{10.1111/j.1365-2966.2011.19271.x}.
\bibitem[{{Intema} et~al.(2017){Intema}, {Jagannathan}, {Mooley} and {Frail}}]{gmrt}
\bibinfo{author}{{Intema}, H.T.}, \bibinfo{author}{{Jagannathan}, P.}, \bibinfo{author}{{Mooley}, K.P.}, \bibinfo{author}{{Frail}, D.A.}, \bibinfo{year}{2017}.
\newblock \bibinfo{title}{{The GMRT 150 MHz all-sky radio survey. First alternative data release TGSS ADR1}}.
\newblock \bibinfo{journal}{Astronomy \& Astrophysics} \bibinfo{volume}{598}, \bibinfo{pages}{A78}.
\newblock \DOIprefix\doi{10.1051/0004-6361/201628536}.
\bibitem[{{Jarvis} et~al.(2016){Jarvis}, {Taylor}, {Agudo}, {Allison}, {Deane}, {Frank}, {Gupta}, {Heywood}, {Maddox}, {McAlpine}, {Santos}, {Scaife}, {Vaccari}, {Zwart}, {Adams}, {Bacon}, {Baker}, {Bassett}, {Best}, {Beswick}, {Blyth}, {Brown}, {Bruggen}, {Cluver}, {Colafrancesco}, {Cotter}, {Cress}, {Dav{\'e}}, {Ferrari}, {Hardcastle}, {Hale}, {Harrison}, {Hatfield}, {Klockner}, {Kolwa}, {Malefahlo}, {Marubini}, {Mauch}, {Moodley}, {Morganti}, {Norris}, {Peters}, {Prandoni}, {Prescott}, {Oliver}, {Oozeer}, {Rottgering}, {Seymour}, {Simpson}, {Smirnov} and {Smith}}]{MIGHTEE}
\bibinfo{author}{{Jarvis}, M.}, \bibinfo{author}{{Taylor}, R.}, \bibinfo{author}{{Agudo}, I.}, \bibinfo{author}{{Allison}, J.R.}, \bibinfo{author}{{Deane}, R.P.}, \bibinfo{author}{{Frank}, B.}, \bibinfo{author}{{Gupta}, N.}, \bibinfo{author}{{Heywood}, I.}, \bibinfo{author}{{Maddox}, N.}, \bibinfo{author}{{McAlpine}, K.}, \bibinfo{author}{{Santos}, M.}, \bibinfo{author}{{Scaife}, A.M.M.}, \bibinfo{author}{{Vaccari}, M.}, \bibinfo{author}{{Zwart}, J.T.L.}, \bibinfo{author}{{Adams}, E.}, \bibinfo{author}{{Bacon}, D.J.}, \bibinfo{author}{{Baker}, A.J.}, \bibinfo{author}{{Bassett}, B.A.}, \bibinfo{author}{{Best}, P.N.}, \bibinfo{author}{{Beswick}, R.}, \bibinfo{author}{{Blyth}, S.}, \bibinfo{author}{{Brown}, M.L.}, \bibinfo{author}{{Bruggen}, M.}, \bibinfo{author}{{Cluver}, M.}, \bibinfo{author}{{Colafrancesco}, S.}, \bibinfo{author}{{Cotter}, G.}, \bibinfo{author}{{Cress}, C.}, \bibinfo{author}{{Dav{\'e}}, R.}, \bibinfo{author}{{Ferrari}, C.}, \bibinfo{author}{{Hardcastle}, M.J.}, \bibinfo{author}{{Hale}, C.L.},
  \bibinfo{author}{{Harrison}, I.}, \bibinfo{author}{{Hatfield}, P.W.}, \bibinfo{author}{{Klockner}, H.R.}, \bibinfo{author}{{Kolwa}, S.}, \bibinfo{author}{{Malefahlo}, E.}, \bibinfo{author}{{Marubini}, T.}, \bibinfo{author}{{Mauch}, T.}, \bibinfo{author}{{Moodley}, K.}, \bibinfo{author}{{Morganti}, R.}, \bibinfo{author}{{Norris}, R.P.}, \bibinfo{author}{{Peters}, J.A.}, \bibinfo{author}{{Prandoni}, I.}, \bibinfo{author}{{Prescott}, M.}, \bibinfo{author}{{Oliver}, S.}, \bibinfo{author}{{Oozeer}, N.}, \bibinfo{author}{{Rottgering}, H.J.A.}, \bibinfo{author}{{Seymour}, N.}, \bibinfo{author}{{Simpson}, C.}, \bibinfo{author}{{Smirnov}, O.}, \bibinfo{author}{{Smith}, D.J.B.}, \bibinfo{year}{2016}.
\newblock \bibinfo{title}{{The MeerKAT International GHz Tiered Extragalactic Exploration (MIGHTEE) Survey}}, in: \bibinfo{booktitle}{MeerKAT Science: On the Pathway to the SKA}, p.~\bibinfo{pages}{6}.
\newblock \DOIprefix\doi{10.22323/1.277.0006}, \href{http://arxiv.org/abs/1709.01901}{{\tt arXiv:1709.01901}}.
\bibitem[{{Kadomtsev}(1966)}]{kadomstev}
\bibinfo{author}{{Kadomtsev}, B.B.}, \bibinfo{year}{1966}.
\newblock \bibinfo{title}{{Hydromagnetic Stability of a Plasma}}.
\newblock \bibinfo{journal}{Reviews of Plasma Physics} \bibinfo{volume}{2}, \bibinfo{pages}{153}.
\bibitem[{Kaiser and Alexander(1997)}]{kaiser}
\bibinfo{author}{Kaiser, C.R.}, \bibinfo{author}{Alexander, P.}, \bibinfo{year}{1997}.
\newblock \bibinfo{title}{{A self-similar model for extragalactic radio sources}}.
\newblock \bibinfo{journal}{Monthly Notices of the Royal Astronomical Society} \bibinfo{volume}{286}, \bibinfo{pages}{215--222}.
\newblock \DOIprefix\doi{10.1093/mnras/286.1.215}.
\bibitem[{{Keppens} et~al.(2008){Keppens}, {Meliani}, {van der Holst} and {Casse}}]{keppens}
\bibinfo{author}{{Keppens}, R.}, \bibinfo{author}{{Meliani}, Z.}, \bibinfo{author}{{van der Holst}, B.}, \bibinfo{author}{{Casse}, F.}, \bibinfo{year}{2008}.
\newblock \bibinfo{title}{{Extragalactic jets with helical magnetic fields: relativistic MHD simulations}}.
\newblock \bibinfo{journal}{Astronomy \& Astrophysics} \bibinfo{volume}{486}, \bibinfo{pages}{663--678}.
\newblock \DOIprefix\doi{10.1051/0004-6361:20079174}.
\bibitem[{{Knowles} et~al.(2022){Knowles}, {Cotton}, {Rudnick}, {Camilo}, {Goedhart}, {Deane}, {Ramatsoku}, {Bietenholz}, {Br{\"u}ggen}, {Button}, {Chen}, {Chibueze}, {Clarke}, {de Gasperin}, {Ianjamasimanana}, {J{\'o}zsa}, {Hilton}, {Kesebonye}, {Kolokythas}, {Kraan-Korteweg}, {Lawrie}, {Lochner}, {Loubser}, {Marchegiani}, {Mhlahlo}, {Moodley}, {Murphy}, {Namumba}, {Oozeer}, {Parekh}, {Pillay}, {Passmoor}, {Ramaila}, {Ranchod}, {Retana-Montenegro}, {Sebokolodi}, {Sikhosana}, {Smirnov}, {Thorat}, {Venturi}, {Abbott}, {Adam}, {Adams}, {Aldera}, {Bauermeister}, {Bennett}, {Bode}, {Botha}, {Botha}, {Brederode}, {Buchner}, {Burger}, {Cheetham}, {de Villiers}, {Dikgale-Mahlakoana}, {du Toit}, {Esterhuyse}, {Fadana}, {Fanaroff}, {Fataar}, {Foley}, {Fourie}, {Frank}, {Gamatham}, {Gatsi}, {Geyer}, {Gouws}, {Gumede}, {Heywood}, {Hlakola}, {Hokwana}, {Hoosen}, {Horn}, {Horrell}, {Hugo}, {Isaacson}, {Jonas}, {Jordaan}, {Joubert}, {Julie}, {Kapp}, {Kasper}, {Kenyon}, {Kotz{\'e}}, {Kotze}, {Kriek}, {Kriel}, {Krishnan},
  {Kusel}, {Legodi}, {Lehmensiek}, {Liebenberg}, {Lord}, {Lunsky}, {Madisa}, {Magnus}, {Main}, {Makhaba}, {Makhathini}, {Malan}, {Manley}, {Marais}, {Maree}, {Martens}, {Mauch}, {McAlpine}, {Merry}, {Millenaar}, {Mokone}, {Monama}, {Mphego}, {New}, {Ngcebetsha}, {Ngoasheng}, {Ockards}, {Otto}, {Patel}, {Peens-Hough}, {Perkins}, {Ramanujam}, {Ramudzuli}, {Ratcliffe}, {Renil}, {Robyntjies}, {Rust}, {Salie}, {Sambu}, {Schollar}, {Schwardt}, {Schwartz}, {Serylak}, {Siebrits}, {Sirothia}, {Slabber}, {Sofeya}, {Taljaard}, {Tasse}, {Tiplady}, {Toruvanda}, {Twum}, {van Balla}, {van der Byl}, {van der Merwe}, {van Dyk}, {Van Tonder}, {Van Wyk}, {Venter}, {Venter}, {Welz}, {Williams} and {Xaia}}]{knowles}
\bibinfo{author}{{Knowles}, K.}, \bibinfo{author}{{Cotton}, W.D.}, \bibinfo{author}{{Rudnick}, L.}, \bibinfo{author}{{Camilo}, F.}, \bibinfo{author}{{Goedhart}, S.}, \bibinfo{author}{{Deane}, R.}, \bibinfo{author}{{Ramatsoku}, M.}, \bibinfo{author}{{Bietenholz}, M.F.}, \bibinfo{author}{{Br{\"u}ggen}, M.}, \bibinfo{author}{{Button}, C.}, \bibinfo{author}{{Chen}, H.}, \bibinfo{author}{{Chibueze}, J.O.}, \bibinfo{author}{{Clarke}, T.E.}, \bibinfo{author}{{de Gasperin}, F.}, \bibinfo{author}{{Ianjamasimanana}, R.}, \bibinfo{author}{{J{\'o}zsa}, G.I.G.}, \bibinfo{author}{{Hilton}, M.}, \bibinfo{author}{{Kesebonye}, K.C.}, \bibinfo{author}{{Kolokythas}, K.}, \bibinfo{author}{{Kraan-Korteweg}, R.C.}, \bibinfo{author}{{Lawrie}, G.}, \bibinfo{author}{{Lochner}, M.}, \bibinfo{author}{{Loubser}, S.I.}, \bibinfo{author}{{Marchegiani}, P.}, \bibinfo{author}{{Mhlahlo}, N.}, \bibinfo{author}{{Moodley}, K.}, \bibinfo{author}{{Murphy}, E.}, \bibinfo{author}{{Namumba}, B.}, \bibinfo{author}{{Oozeer}, N.},
  \bibinfo{author}{{Parekh}, V.}, \bibinfo{author}{{Pillay}, D.S.}, \bibinfo{author}{{Passmoor}, S.S.}, \bibinfo{author}{{Ramaila}, A.J.T.}, \bibinfo{author}{{Ranchod}, S.}, \bibinfo{author}{{Retana-Montenegro}, E.}, \bibinfo{author}{{Sebokolodi}, L.}, \bibinfo{author}{{Sikhosana}, S.P.}, \bibinfo{author}{{Smirnov}, O.}, \bibinfo{author}{{Thorat}, K.}, \bibinfo{author}{{Venturi}, T.}, \bibinfo{author}{{Abbott}, T.D.}, \bibinfo{author}{{Adam}, R.M.}, \bibinfo{author}{{Adams}, G.}, \bibinfo{author}{{Aldera}, M.A.}, \bibinfo{author}{{Bauermeister}, E.F.}, \bibinfo{author}{{Bennett}, T.G.H.}, \bibinfo{author}{{Bode}, W.A.}, \bibinfo{author}{{Botha}, D.H.}, \bibinfo{author}{{Botha}, A.G.}, \bibinfo{author}{{Brederode}, L.R.S.}, \bibinfo{author}{{Buchner}, S.}, \bibinfo{author}{{Burger}, J.P.}, \bibinfo{author}{{Cheetham}, T.}, \bibinfo{author}{{de Villiers}, D.I.L.}, \bibinfo{author}{{Dikgale-Mahlakoana}, M.A.}, \bibinfo{author}{{du Toit}, L.J.}, \bibinfo{author}{{Esterhuyse}, S.W.P.}, \bibinfo{author}{{Fadana},
  G.}, \bibinfo{author}{{Fanaroff}, B.L.}, \bibinfo{author}{{Fataar}, S.}, \bibinfo{author}{{Foley}, A.R.}, \bibinfo{author}{{Fourie}, D.J.}, \bibinfo{author}{{Frank}, B.S.}, \bibinfo{author}{{Gamatham}, R.R.G.}, \bibinfo{author}{{Gatsi}, T.G.}, \bibinfo{author}{{Geyer}, M.}, \bibinfo{author}{{Gouws}, M.}, \bibinfo{author}{{Gumede}, S.C.}, \bibinfo{author}{{Heywood}, I.}, \bibinfo{author}{{Hlakola}, M.J.}, \bibinfo{author}{{Hokwana}, A.}, \bibinfo{author}{{Hoosen}, S.W.}, \bibinfo{author}{{Horn}, D.M.}, \bibinfo{author}{{Horrell}, J.M.G.}, \bibinfo{author}{{Hugo}, B.V.}, \bibinfo{author}{{Isaacson}, A.R.}, \bibinfo{author}{{Jonas}, J.L.}, \bibinfo{author}{{Jordaan}, J.D.B.}, \bibinfo{author}{{Joubert}, A.F.}, \bibinfo{author}{{Julie}, R.P.M.}, \bibinfo{author}{{Kapp}, F.B.}, \bibinfo{author}{{Kasper}, V.A.}, \bibinfo{author}{{Kenyon}, J.S.}, \bibinfo{author}{{Kotz{\'e}}, P.P.A.}, \bibinfo{author}{{Kotze}, A.G.}, \bibinfo{author}{{Kriek}, N.}, \bibinfo{author}{{Kriel}, H.}, \bibinfo{author}{{Krishnan}, V.K.},
  \bibinfo{author}{{Kusel}, T.W.}, \bibinfo{author}{{Legodi}, L.S.}, \bibinfo{author}{{Lehmensiek}, R.}, \bibinfo{author}{{Liebenberg}, D.}, \bibinfo{author}{{Lord}, R.T.}, \bibinfo{author}{{Lunsky}, B.M.}, \bibinfo{author}{{Madisa}, K.}, \bibinfo{author}{{Magnus}, L.G.}, \bibinfo{author}{{Main}, J.P.L.}, \bibinfo{author}{{Makhaba}, A.}, \bibinfo{author}{{Makhathini}, S.}, \bibinfo{author}{{Malan}, J.A.}, \bibinfo{author}{{Manley}, J.R.}, \bibinfo{author}{{Marais}, S.J.}, \bibinfo{author}{{Maree}, M.D.J.}, \bibinfo{author}{{Martens}, A.}, \bibinfo{author}{{Mauch}, T.}, \bibinfo{author}{{McAlpine}, K.}, \bibinfo{author}{{Merry}, B.C.}, \bibinfo{author}{{Millenaar}, R.P.}, \bibinfo{author}{{Mokone}, O.J.}, \bibinfo{author}{{Monama}, T.E.}, \bibinfo{author}{{Mphego}, M.C.}, \bibinfo{author}{{New}, W.S.}, \bibinfo{author}{{Ngcebetsha}, B.}, \bibinfo{author}{{Ngoasheng}, K.J.}, \bibinfo{author}{{Ockards}, M.T.}, \bibinfo{author}{{Otto}, A.J.}, \bibinfo{author}{{Patel}, A.A.}, \bibinfo{author}{{Peens-Hough}, A.},
  \bibinfo{author}{{Perkins}, S.J.}, \bibinfo{author}{{Ramanujam}, N.M.}, \bibinfo{author}{{Ramudzuli}, Z.R.}, \bibinfo{author}{{Ratcliffe}, S.M.}, \bibinfo{author}{{Renil}, R.}, \bibinfo{author}{{Robyntjies}, A.}, \bibinfo{author}{{Rust}, A.N.}, \bibinfo{author}{{Salie}, S.}, \bibinfo{author}{{Sambu}, N.}, \bibinfo{author}{{Schollar}, C.T.G.}, \bibinfo{author}{{Schwardt}, L.C.}, \bibinfo{author}{{Schwartz}, R.L.}, \bibinfo{author}{{Serylak}, M.}, \bibinfo{author}{{Siebrits}, R.}, \bibinfo{author}{{Sirothia}, S.K.}, \bibinfo{author}{{Slabber}, M.}, \bibinfo{author}{{Sofeya}, L.}, \bibinfo{author}{{Taljaard}, B.}, \bibinfo{author}{{Tasse}, C.}, \bibinfo{author}{{Tiplady}, A.J.}, \bibinfo{author}{{Toruvanda}, O.}, \bibinfo{author}{{Twum}, S.N.}, \bibinfo{author}{{van Balla}, T.J.}, \bibinfo{author}{{van der Byl}, A.}, \bibinfo{author}{{van der Merwe}, C.}, \bibinfo{author}{{van Dyk}, C.L.}, \bibinfo{author}{{Van Tonder}, V.}, \bibinfo{author}{{Van Wyk}, R.}, \bibinfo{author}{{Venter}, A.J.},
  \bibinfo{author}{{Venter}, M.}, \bibinfo{author}{{Welz}, M.G.}, \bibinfo{author}{{Williams}, L.P.}, \bibinfo{author}{{Xaia}, B.}, \bibinfo{year}{2022}.
\newblock \bibinfo{title}{{The MeerKAT Galaxy Cluster Legacy Survey. I. Survey Overview and Highlights}}.
\newblock \bibinfo{journal}{Astronomy \& Astrophysics} \bibinfo{volume}{657}, \bibinfo{pages}{A56}.
\newblock \DOIprefix\doi{10.1051/0004-6361/202141488}.
\bibitem[{Komissarov and Falle(1998)}]{komissarov1998large}
\bibinfo{author}{Komissarov, S.}, \bibinfo{author}{Falle, S.}, \bibinfo{year}{1998}.
\newblock \bibinfo{title}{The large-scale structure of fr-ii radio sources}.
\newblock \bibinfo{journal}{Monthly Notices of the Royal Astronomical Society} \bibinfo{volume}{297}, \bibinfo{pages}{1087--1108}.
\bibitem[{{Krause} et~al.(2019){Krause}, {Shabala}, {Hardcastle}, {Bicknell}, {B{\"o}hringer}, {Chon}, {Nawaz}, {Sarzi} and {Wagner}}]{krause2019}
\bibinfo{author}{{Krause}, M.G.H.}, \bibinfo{author}{{Shabala}, S.S.}, \bibinfo{author}{{Hardcastle}, M.J.}, \bibinfo{author}{{Bicknell}, G.V.}, \bibinfo{author}{{B{\"o}hringer}, H.}, \bibinfo{author}{{Chon}, G.}, \bibinfo{author}{{Nawaz}, M.A.}, \bibinfo{author}{{Sarzi}, M.}, \bibinfo{author}{{Wagner}, A.Y.}, \bibinfo{year}{2019}.
\newblock \bibinfo{title}{{How frequent are close supermassive binary black holes in powerful jet sources?}}
\newblock \bibinfo{journal}{Monthly Notices of the Royal Astronomical Society} \bibinfo{volume}{482}, \bibinfo{pages}{240--261}.
\newblock \DOIprefix\doi{10.1093/mnras/sty2558}.
\bibitem[{{Kundu} et~al.(2022){Kundu}, {Vaidya}, {Mignone} and {Hardcastle}}]{sayan2022}
\bibinfo{author}{{Kundu}, S.}, \bibinfo{author}{{Vaidya}, B.}, \bibinfo{author}{{Mignone}, A.}, \bibinfo{author}{{Hardcastle}, M.J.}, \bibinfo{year}{2022}.
\newblock \bibinfo{title}{{A numerical study of the interplay between Fermi acceleration mechanisms in radio lobes of FR-II radio galaxies}}.
\newblock \bibinfo{journal}{Astronomy \& Astrophysics} \bibinfo{volume}{667}, \bibinfo{pages}{A138}.
\newblock \DOIprefix\doi{10.1051/0004-6361/202244251}.
\bibitem[{{Laing}(1981)}]{laing}
\bibinfo{author}{{Laing}, R.A.}, \bibinfo{year}{1981}.
\newblock \bibinfo{title}{{Magnetic fields in extragalactic radio sources.}}
\newblock \bibinfo{journal}{The Astrophysical Journal} \bibinfo{volume}{248}, \bibinfo{pages}{87--104}.
\newblock \DOIprefix\doi{10.1086/159132}.
\bibitem[{Leahy and Williams(1984)}]{leahy1984bridges}
\bibinfo{author}{Leahy, J.}, \bibinfo{author}{Williams, A.}, \bibinfo{year}{1984}.
\newblock \bibinfo{title}{The bridges of classical double radio sources}.
\newblock \bibinfo{journal}{Monthly Notices of the Royal Astronomical Society} \bibinfo{volume}{210}, \bibinfo{pages}{929--951}.
\bibitem[{López-Miralles et~al.(2022)López-Miralles, Perucho, Martí, Migliari and Bosch-Ramon}]{L_pez_Miralles_2022}
\bibinfo{author}{López-Miralles, J.}, \bibinfo{author}{Perucho, M.}, \bibinfo{author}{Martí, J.M.}, \bibinfo{author}{Migliari, S.}, \bibinfo{author}{Bosch-Ramon, V.}, \bibinfo{year}{2022}.
\newblock \bibinfo{title}{3d rmhd simulations of jet-wind interactions in high-mass x-ray binaries}.
\newblock \bibinfo{journal}{Astronomy \& Astrophysics} \bibinfo{volume}{661}, \bibinfo{pages}{A117}.
\newblock \DOIprefix\doi{10.1051/0004-6361/202142968}.
\bibitem[{Martí(2019)}]{marti}
\bibinfo{author}{Martí, J.M.}, \bibinfo{year}{2019}.
\newblock \bibinfo{title}{Numerical simulations of jets from active galactic nuclei}.
\newblock \bibinfo{journal}{Galaxies} \bibinfo{volume}{7}.
\newblock \DOIprefix\doi{10.3390/galaxies7010024}.
\bibitem[{{Massaglia} et~al.(2022){Massaglia}, {Bodo}, {Rossi}, {Capetti} and {Mignone}}]{massaglia}
\bibinfo{author}{{Massaglia}, S.}, \bibinfo{author}{{Bodo}, G.}, \bibinfo{author}{{Rossi}, P.}, \bibinfo{author}{{Capetti}, A.}, \bibinfo{author}{{Mignone}, A.}, \bibinfo{year}{2022}.
\newblock \bibinfo{title}{{Making Fanaroff-Riley I radio sources. III. The effects of the magnetic field on relativistic jets' propagation and source morphologies}}.
\newblock \bibinfo{journal}{Astronomy \& Astrophysics} \bibinfo{volume}{659}, \bibinfo{pages}{A139}.
\newblock \DOIprefix\doi{10.1051/0004-6361/202038724}.
\bibitem[{Massaglia et~al.(2019)Massaglia, Bodo, Rossi, Capetti and Mignone}]{Massaglia_2019}
\bibinfo{author}{Massaglia, S.}, \bibinfo{author}{Bodo, G.}, \bibinfo{author}{Rossi, P.}, \bibinfo{author}{Capetti, S.}, \bibinfo{author}{Mignone, A.}, \bibinfo{year}{2019}.
\newblock \bibinfo{title}{Making faranoff-riley i radio sources: Ii. the effects of jet magnetization}.
\newblock \bibinfo{journal}{Astronomy \& Astrophysics} \bibinfo{volume}{621}, \bibinfo{pages}{A132}.
\newblock \DOIprefix\doi{10.1051/0004-6361/201834512}.
\bibitem[{Meenakshi et~al.(2023)Meenakshi, Mukherjee, Bodo and Rossi}]{meenakshi2023polarization}
\bibinfo{author}{Meenakshi, M.}, \bibinfo{author}{Mukherjee, D.}, \bibinfo{author}{Bodo, G.}, \bibinfo{author}{Rossi, P.}, \bibinfo{year}{2023}.
\newblock \bibinfo{title}{A polarization study of jets interacting with turbulent magnetic fields}.
\newblock \href{http://arxiv.org/abs/2310.03139}{{\tt arXiv:2310.03139}}.
\bibitem[{{Mignone} et~al.(2007){Mignone}, {Bodo}, {Massaglia}, {Matsakos}, {Tesileanu}, {Zanni} and {Ferrari}}]{pluto2007}
\bibinfo{author}{{Mignone}, A.}, \bibinfo{author}{{Bodo}, G.}, \bibinfo{author}{{Massaglia}, S.}, \bibinfo{author}{{Matsakos}, T.}, \bibinfo{author}{{Tesileanu}, O.}, \bibinfo{author}{{Zanni}, C.}, \bibinfo{author}{{Ferrari}, A.}, \bibinfo{year}{2007}.
\newblock \bibinfo{title}{{PLUTO: A Numerical Code for Computational Astrophysics}}.
\newblock \bibinfo{journal}{Astrophysical Journal Supplement Series} \bibinfo{volume}{170}, \bibinfo{pages}{228--242}.
\newblock \DOIprefix\doi{10.1086/513316}.
\bibitem[{Mignone and McKinney(2007)}]{mignone_taub}
\bibinfo{author}{Mignone, A.}, \bibinfo{author}{McKinney, J.C.}, \bibinfo{year}{2007}.
\newblock \bibinfo{title}{{Equation of state in relativistic magnetohydrodynamics: variable versus constant adiabatic index}}.
\newblock \bibinfo{journal}{Monthly Notices of the Royal Astronomical Society} \bibinfo{volume}{378}, \bibinfo{pages}{1118--1130}.
\newblock \DOIprefix\doi{10.1111/j.1365-2966.2007.11849.x}.
\bibitem[{Mignone et~al.(2005)Mignone, Plewa and Bodo}]{Mignone_2005}
\bibinfo{author}{Mignone, A.}, \bibinfo{author}{Plewa, T.}, \bibinfo{author}{Bodo, G.}, \bibinfo{year}{2005}.
\newblock \bibinfo{title}{The piecewise parabolic method for multidimensional relativistic fluid dynamics}.
\newblock \bibinfo{journal}{The Astrophysical Journal Supplement Series} \bibinfo{volume}{160}, \bibinfo{pages}{199}.
\newblock \DOIprefix\doi{10.1086/430905}.
\bibitem[{Mignone et~al.(2010)Mignone, Rossi, Bodo, Ferrari and Massaglia}]{mignone_2010}
\bibinfo{author}{Mignone, A.}, \bibinfo{author}{Rossi, P.}, \bibinfo{author}{Bodo, G.}, \bibinfo{author}{Ferrari, A.}, \bibinfo{author}{Massaglia, S.}, \bibinfo{year}{2010}.
\newblock \bibinfo{title}{{High-resolution 3D relativistic MHD simulations of jets}}.
\newblock \bibinfo{journal}{Monthly Notices of the Royal Astronomical Society} \bibinfo{volume}{402}, \bibinfo{pages}{7--12}.
\newblock \DOIprefix\doi{10.1111/j.1365-2966.2009.15642.x}.
\bibitem[{{Mignone} et~al.(2010){Mignone}, {Rossi}, {Bodo}, {Ferrari} and {Massaglia}}]{mignone2010}
\bibinfo{author}{{Mignone}, A.}, \bibinfo{author}{{Rossi}, P.}, \bibinfo{author}{{Bodo}, G.}, \bibinfo{author}{{Ferrari}, A.}, \bibinfo{author}{{Massaglia}, S.}, \bibinfo{year}{2010}.
\newblock \bibinfo{title}{{High-resolution 3D relativistic MHD simulations of jets}}.
\newblock \bibinfo{journal}{Monthly Notices of the Royal Astronomical Society} \bibinfo{volume}{402}, \bibinfo{pages}{7--12}.
\newblock \DOIprefix\doi{10.1111/j.1365-2966.2009.15642.x}.
\bibitem[{Mignone et~al.(2009)Mignone, Ugliano and Bodo}]{mignone2009five}
\bibinfo{author}{Mignone, A.}, \bibinfo{author}{Ugliano, M.}, \bibinfo{author}{Bodo, G.}, \bibinfo{year}{2009}.
\newblock \bibinfo{title}{A five-wave harten--lax--van leer riemann solver for relativistic magnetohydrodynamics}.
\newblock \bibinfo{journal}{Monthly Notices of the Royal Astronomical Society} \bibinfo{volume}{393}, \bibinfo{pages}{1141--1156}.
\bibitem[{{Mizuno} et~al.(2012){Mizuno}, {Lyubarsky}, {Nishikawa} and {Hardee}}]{mizuno2012}
\bibinfo{author}{{Mizuno}, Y.}, \bibinfo{author}{{Lyubarsky}, Y.}, \bibinfo{author}{{Nishikawa}, K.I.}, \bibinfo{author}{{Hardee}, P.E.}, \bibinfo{year}{2012}.
\newblock \bibinfo{title}{{Three-dimensional Relativistic Magnetohydrodynamic Simulations of Current-driven Instability. III. Rotating Relativistic Jets}}.
\newblock \bibinfo{journal}{The Astrophysical Journal} \bibinfo{volume}{757}, \bibinfo{pages}{16}.
\newblock \DOIprefix\doi{10.1088/0004-637X/757/1/16}.
\bibitem[{{Mukherjee} et~al.(2020){Mukherjee}, {Bodo}, {Mignone}, {Rossi} and {Vaidya}}]{mukherjee2020}
\bibinfo{author}{{Mukherjee}, D.}, \bibinfo{author}{{Bodo}, G.}, \bibinfo{author}{{Mignone}, A.}, \bibinfo{author}{{Rossi}, P.}, \bibinfo{author}{{Vaidya}, B.}, \bibinfo{year}{2020}.
\newblock \bibinfo{title}{{Simulating the dynamics and non-thermal emission of relativistic magnetized jets I. Dynamics}}.
\newblock \bibinfo{journal}{Monthly Notices of the Royal Astronomical Society} \bibinfo{volume}{499}, \bibinfo{pages}{681--701}.
\newblock \DOIprefix\doi{10.1093/mnras/staa2934}.
\bibitem[{Mukherjee et~al.(2021)Mukherjee, Bodo, Rossi, Mignone and Vaidya}]{dipanjan2021}
\bibinfo{author}{Mukherjee, D.}, \bibinfo{author}{Bodo, G.}, \bibinfo{author}{Rossi, P.}, \bibinfo{author}{Mignone, A.}, \bibinfo{author}{Vaidya, B.}, \bibinfo{year}{2021}.
\newblock \bibinfo{title}{{Simulating the dynamics and synchrotron emission from relativistic jets – II. Evolution of non-thermal electrons}}.
\newblock \bibinfo{journal}{Monthly Notices of the Royal Astronomical Society} \bibinfo{volume}{505}, \bibinfo{pages}{2267--2284}.
\newblock \DOIprefix\doi{10.1093/mnras/stab1327}.
\bibitem[{Murphy(1999)}]{Murphy1999}
\bibinfo{author}{Murphy, T.}, \bibinfo{year}{1999}.
\newblock \bibinfo{title}{PhD Thesis}.
\newblock Ph.D. thesis. University of Sydney.
\newblock \URLprefix \url{http://www.astrop.physics.usyd.edu.au/RELICS/thesis/thesis.html}.
\bibitem[{Müller et~al.(2021)Müller, Pfrommer, Ignesti, Moretti, Lourenço, Paladino, Jaffé, Gitti, Venturi, Gullieuszik, Poggianti, Vulcani, Biviano, Adebahr and Dettmar}]{muller2021}
\bibinfo{author}{Müller, A.}, \bibinfo{author}{Pfrommer, C.}, \bibinfo{author}{Ignesti, A.}, \bibinfo{author}{Moretti, A.}, \bibinfo{author}{Lourenço, A.}, \bibinfo{author}{Paladino, R.}, \bibinfo{author}{Jaffé, Y.}, \bibinfo{author}{Gitti, M.}, \bibinfo{author}{Venturi, T.}, \bibinfo{author}{Gullieuszik, M.}, \bibinfo{author}{Poggianti, B.}, \bibinfo{author}{Vulcani, B.}, \bibinfo{author}{Biviano, A.}, \bibinfo{author}{Adebahr, B.}, \bibinfo{author}{Dettmar, R.J.}, \bibinfo{year}{2021}.
\newblock \bibinfo{title}{{Two striking head–tail galaxies in the galaxy cluster IIZW108: insights into transition to turbulence, magnetic fields, and particle re-acceleration}}.
\newblock \bibinfo{journal}{Monthly Notices of the Royal Astronomical Society} \bibinfo{volume}{508}, \bibinfo{pages}{5326--5344}.
\newblock \DOIprefix\doi{10.1093/mnras/stab2928}.
\bibitem[{Nakamura et~al.(2007)Nakamura, Li and Li}]{Nakamura_2007}
\bibinfo{author}{Nakamura, M.}, \bibinfo{author}{Li, H.}, \bibinfo{author}{Li, S.}, \bibinfo{year}{2007}.
\newblock \bibinfo{title}{Stability properties of magnetic tower jets}.
\newblock \bibinfo{journal}{The Astrophysical Journal} \bibinfo{volume}{656}, \bibinfo{pages}{721}.
\newblock \DOIprefix\doi{10.1086/510361}.
\bibitem[{{Narayan} et~al.(2009){Narayan}, {Li} and {Tchekhovskoy}}]{narayan2009}
\bibinfo{author}{{Narayan}, R.}, \bibinfo{author}{{Li}, J.}, \bibinfo{author}{{Tchekhovskoy}, A.}, \bibinfo{year}{2009}.
\newblock \bibinfo{title}{{Stability of Relativistic Force-free Jets}}.
\newblock \bibinfo{journal}{The Astrophysical Journal} \bibinfo{volume}{697}, \bibinfo{pages}{1681--1694}.
\newblock \DOIprefix\doi{10.1088/0004-637X/697/2/1681}.
\bibitem[{Nawaz et~al.(2014)Nawaz, Wagner, Bicknell, Sutherland and McNamara}]{nawaz}
\bibinfo{author}{Nawaz, M.A.}, \bibinfo{author}{Wagner, A.Y.}, \bibinfo{author}{Bicknell, G.V.}, \bibinfo{author}{Sutherland, R.S.}, \bibinfo{author}{McNamara, B.R.}, \bibinfo{year}{2014}.
\newblock \bibinfo{title}{{Jet–intracluster medium interaction in Hydra A – I. Estimates of jet velocity from inner knots}}.
\newblock \bibinfo{journal}{Monthly Notices of the Royal Astronomical Society} \bibinfo{volume}{444}, \bibinfo{pages}{1600--1614}.
\newblock \DOIprefix\doi{10.1093/mnras/stu1563}.
\bibitem[{Norman et~al.(1982)Norman, Winkler, Smarr and Smith}]{norman1982structure}
\bibinfo{author}{Norman, M.}, \bibinfo{author}{Winkler, K.H.}, \bibinfo{author}{Smarr, L.}, \bibinfo{author}{Smith, M.}, \bibinfo{year}{1982}.
\newblock \bibinfo{title}{Structure and dynamics of supersonic jets}.
\newblock \bibinfo{journal}{Astronomy and Astrophysics} \bibinfo{volume}{113}, \bibinfo{pages}{285--302}.
\bibitem[{O’Neill et~al.(2019)O’Neill, Jones, Nolting and Mendygral}]{O_Neill_2019}
\bibinfo{author}{O’Neill, B.J.}, \bibinfo{author}{Jones, T.W.}, \bibinfo{author}{Nolting, C.}, \bibinfo{author}{Mendygral, P.J.}, \bibinfo{year}{2019}.
\newblock \bibinfo{title}{Shocked narrow-angle tail radio galaxies: Simulations and emissions}.
\newblock \bibinfo{journal}{The Astrophysical Journal} \bibinfo{volume}{887}, \bibinfo{pages}{26}.
\newblock \DOIprefix\doi{10.3847/1538-4357/ab4efa}.
\bibitem[{Pandge et~al.(2021)Pandge, Kale, Dabhade, Mahato and Raychaudhury}]{pandge2022}
\bibinfo{author}{Pandge, M.B.}, \bibinfo{author}{Kale, R.}, \bibinfo{author}{Dabhade, P.}, \bibinfo{author}{Mahato, M.}, \bibinfo{author}{Raychaudhury, S.}, \bibinfo{year}{2021}.
\newblock \bibinfo{title}{{Giant Metrewave Radio Telescope unveils steep-spectrum antique filaments in the galaxy cluster Abell 725}}.
\newblock \bibinfo{journal}{Monthly Notices of the Royal Astronomical Society} \bibinfo{volume}{509}, \bibinfo{pages}{1837--1847}.
\newblock \DOIprefix\doi{10.1093/mnras/stab2945}.
\bibitem[{{Pasetto} et~al.(2021){Pasetto}, {Carrasco-Gonz{\'a}lez}, {G{\'o}mez}, {Mart{\'\i}}, {Perucho}, {O'Sullivan}, {Anderson}, {D{\'\i}az-Gonz{\'a}lez}, {Fuentes} and {Wardle}}]{m87_helicalfield}
\bibinfo{author}{{Pasetto}, A.}, \bibinfo{author}{{Carrasco-Gonz{\'a}lez}, C.}, \bibinfo{author}{{G{\'o}mez}, J.L.}, \bibinfo{author}{{Mart{\'\i}}, J.M.}, \bibinfo{author}{{Perucho}, M.}, \bibinfo{author}{{O'Sullivan}, S.P.}, \bibinfo{author}{{Anderson}, C.}, \bibinfo{author}{{D{\'\i}az-Gonz{\'a}lez}, D.J.}, \bibinfo{author}{{Fuentes}, A.}, \bibinfo{author}{{Wardle}, J.}, \bibinfo{year}{2021}.
\newblock \bibinfo{title}{{Reading M87's DNA: A Double Helix Revealing a Large-scale Helical Magnetic Field}}.
\newblock \bibinfo{journal}{Astrophysical Journal Letters} \bibinfo{volume}{923}, \bibinfo{pages}{L5}.
\newblock \DOIprefix\doi{10.3847/2041-8213/ac3a88}.
\bibitem[{Perucho and López-Miralles(2023)}]{Perucho_López-Miralles_2023}
\bibinfo{author}{Perucho, M.}, \bibinfo{author}{López-Miralles, J.}, \bibinfo{year}{2023}.
\newblock \bibinfo{title}{Numerical simulations of relativistic jets}.
\newblock \bibinfo{journal}{Journal of Plasma Physics} \bibinfo{volume}{89}, \bibinfo{pages}{915890501}.
\newblock \DOIprefix\doi{10.1017/S0022377823000892}.
\bibitem[{{Rajpurohit} et~al.(2022){Rajpurohit}, {van Weeren}, {Hoeft}, {Vazza}, {Brienza}, {Forman}, {Wittor}, {Dom{\'\i}nguez-Fern{\'a}ndez}, {Rajpurohit}, {Riseley}, {Botteon}, {Osinga}, {Brunetti}, {Bonnassieux}, {Bonafede}, {Rajpurohit}, {Stuardi}, {Drabent}, {Br{\"u}ggen}, {Dallacasa}, {Shimwell}, {R{\"o}ttgering}, {de Gasperin}, {Miley} and {Rossetti}}]{rajpurohit2022}
\bibinfo{author}{{Rajpurohit}, K.}, \bibinfo{author}{{van Weeren}, R.J.}, \bibinfo{author}{{Hoeft}, M.}, \bibinfo{author}{{Vazza}, F.}, \bibinfo{author}{{Brienza}, M.}, \bibinfo{author}{{Forman}, W.}, \bibinfo{author}{{Wittor}, D.}, \bibinfo{author}{{Dom{\'\i}nguez-Fern{\'a}ndez}, P.}, \bibinfo{author}{{Rajpurohit}, S.}, \bibinfo{author}{{Riseley}, C.J.}, \bibinfo{author}{{Botteon}, A.}, \bibinfo{author}{{Osinga}, E.}, \bibinfo{author}{{Brunetti}, G.}, \bibinfo{author}{{Bonnassieux}, E.}, \bibinfo{author}{{Bonafede}, A.}, \bibinfo{author}{{Rajpurohit}, A.S.}, \bibinfo{author}{{Stuardi}, C.}, \bibinfo{author}{{Drabent}, A.}, \bibinfo{author}{{Br{\"u}ggen}, M.}, \bibinfo{author}{{Dallacasa}, D.}, \bibinfo{author}{{Shimwell}, T.W.}, \bibinfo{author}{{R{\"o}ttgering}, H.J.A.}, \bibinfo{author}{{de Gasperin}, F.}, \bibinfo{author}{{Miley}, G.K.}, \bibinfo{author}{{Rossetti}, M.}, \bibinfo{year}{2022}.
\newblock \bibinfo{title}{{Deep Low-frequency Radio Observations of A2256. I. The Filamentary Radio Relic}}.
\newblock \bibinfo{journal}{The Astrophysical Journal} \bibinfo{volume}{927}, \bibinfo{pages}{80}.
\newblock \DOIprefix\doi{10.3847/1538-4357/ac4708}.
\bibitem[{Riseley et~al.(2022)Riseley, Bonnassieux, Vernstrom, Galvin, Chokshi, Botteon, Rajpurohit, Duchesne, Bonafede, Rudnick, Hoeft, Quici, Eckert, Brienza, Tasse, Carretti, Collier, Diego, Di Mascolo, Hopkins, Johnston-Hollitt, Keel, Koribalski and Reiprich}]{riseley}
\bibinfo{author}{Riseley, C.J.}, \bibinfo{author}{Bonnassieux, E.}, \bibinfo{author}{Vernstrom, T.}, \bibinfo{author}{Galvin, T.J.}, \bibinfo{author}{Chokshi, A.}, \bibinfo{author}{Botteon, A.}, \bibinfo{author}{Rajpurohit, K.}, \bibinfo{author}{Duchesne, S.W.}, \bibinfo{author}{Bonafede, A.}, \bibinfo{author}{Rudnick, L.}, \bibinfo{author}{Hoeft, M.}, \bibinfo{author}{Quici, B.}, \bibinfo{author}{Eckert, D.}, \bibinfo{author}{Brienza, M.}, \bibinfo{author}{Tasse, C.}, \bibinfo{author}{Carretti, E.}, \bibinfo{author}{Collier, J.D.}, \bibinfo{author}{Diego, J.M.}, \bibinfo{author}{Di Mascolo, L.}, \bibinfo{author}{Hopkins, A.M.}, \bibinfo{author}{Johnston-Hollitt, M.}, \bibinfo{author}{Keel, R.R.}, \bibinfo{author}{Koribalski, B.S.}, \bibinfo{author}{Reiprich, T.H.}, \bibinfo{year}{2022}.
\newblock \bibinfo{title}{{Radio fossils, relics, and haloes in Abell 3266: cluster archaeology with ASKAP-EMU and the ATCA}}.
\newblock \bibinfo{journal}{Monthly Notices of the Royal Astronomical Society} \bibinfo{volume}{515}, \bibinfo{pages}{1871--1896}.
\newblock \DOIprefix\doi{10.1093/mnras/stac1771}.
\bibitem[{{Rossi} et~al.(2017){Rossi}, {Bodo}, {Capetti} and {Massaglia}}]{rossi2017}
\bibinfo{author}{{Rossi}, P.}, \bibinfo{author}{{Bodo}, G.}, \bibinfo{author}{{Capetti}, A.}, \bibinfo{author}{{Massaglia}, S.}, \bibinfo{year}{2017}.
\newblock \bibinfo{title}{{3D relativistic MHD numerical simulations of X-shaped radio sources}}.
\newblock \bibinfo{journal}{Astronomy \& Astrophysics} \bibinfo{volume}{606}, \bibinfo{pages}{A57}.
\newblock \DOIprefix\doi{10.1051/0004-6361/201730594}.
\bibitem[{Rottmann(2001)}]{rottmann2001jet}
\bibinfo{author}{Rottmann, H.}, \bibinfo{year}{2001}.
\newblock \bibinfo{title}{Jet-reorientation in x-shaped radio galaxies}.
\newblock \bibinfo{journal}{Ph. D. Thesis} .
\bibitem[{{Rudnick} et~al.(2022){Rudnick}, {Br{\"u}ggen}, {Brunetti}, {Cotton}, {Forman}, {Jones}, {Nolting}, {Schellenberger} and {van Weeren}}]{magnetic_filament}
\bibinfo{author}{{Rudnick}, L.}, \bibinfo{author}{{Br{\"u}ggen}, M.}, \bibinfo{author}{{Brunetti}, G.}, \bibinfo{author}{{Cotton}, W.D.}, \bibinfo{author}{{Forman}, W.}, \bibinfo{author}{{Jones}, T.W.}, \bibinfo{author}{{Nolting}, C.}, \bibinfo{author}{{Schellenberger}, G.}, \bibinfo{author}{{van Weeren}, R.}, \bibinfo{year}{2022}.
\newblock \bibinfo{title}{{Intracluster Magnetic Filaments and an Encounter with a Radio Jet}}.
\newblock \bibinfo{journal}{The Astrophysical Journal} \bibinfo{volume}{935}, \bibinfo{pages}{168}.
\newblock \DOIprefix\doi{10.3847/1538-4357/ac7c76}.
\bibitem[{Rudnick et~al.(2021)Rudnick, Cotton, Knowles and Kolokythas}]{rudnick}
\bibinfo{author}{Rudnick, L.}, \bibinfo{author}{Cotton, W.}, \bibinfo{author}{Knowles, K.}, \bibinfo{author}{Kolokythas, K.}, \bibinfo{year}{2021}.
\newblock \bibinfo{title}{One source, two source(s): Ribs and tethers}.
\newblock \bibinfo{journal}{Galaxies} \bibinfo{volume}{9}, \bibinfo{pages}{81}.
\newblock \DOIprefix\doi{10.3390/galaxies9040081}.
\bibitem[{Sarazin(1986)}]{sarazin}
\bibinfo{author}{Sarazin, C.L.}, \bibinfo{year}{1986}.
\newblock \bibinfo{title}{X-ray emission from clusters of galaxies}.
\newblock \bibinfo{journal}{Rev. Mod. Phys.} \bibinfo{volume}{58}, \bibinfo{pages}{1--115}.
\newblock \DOIprefix\doi{10.1103/RevModPhys.58.1}.
\bibitem[{Shafranov(1970)}]{shafranov1970hydromagnetic}
\bibinfo{author}{Shafranov, V.}, \bibinfo{year}{1970}.
\newblock \bibinfo{title}{Hydromagnetic stability of a current-carrying plasma filament in a strong longitudinal magnetic field(hydromagnetic flute and surface wave instabilities of current-carrying plasma filament in strong longitudinal field, showing radial density stabilization)}.
\newblock \bibinfo{journal}{Zhurnal Tekhnicheskoi Fiziki} \bibinfo{volume}{40}, \bibinfo{pages}{241--253}.
\bibitem[{{Shimwell} et~al.(2017){Shimwell}, {R{\"o}ttgering}, {Best}, {Williams}, {Dijkema}, {de Gasperin}, {Hardcastle}, {Heald}, {Hoang}, {Horneffer}, {Intema}, {Mahony}, {Mandal}, {Mechev}, {Morabito}, {Oonk}, {Rafferty}, {Retana-Montenegro}, {Sabater}, {Tasse}, {van Weeren}, {Br{\"u}ggen}, {Brunetti}, {Chy{\.z}y}, {Conway}, {Haverkorn}, {Jackson}, {Jarvis}, {McKean}, {Miley}, {Morganti}, {White}, {Wise}, {van Bemmel}, {Beck}, {Brienza}, {Bonafede}, {Calistro Rivera}, {Cassano}, {Clarke}, {Cseh}, {Deller}, {Drabent}, {van Driel}, {Engels}, {Falcke}, {Ferrari}, {Fr{\"o}hlich}, {Garrett}, {Harwood}, {Heesen}, {Hoeft}, {Horellou}, {Israel}, {Kapi{\'n}ska}, {Kunert-Bajraszewska}, {McKay}, {Mohan}, {Orr{\'u}}, {Pizzo}, {Prandoni}, {Schwarz}, {Shulevski}, {Sipior}, {Smith}, {Sridhar}, {Steinmetz}, {Stroe}, {Varenius}, {van der Werf}, {Zensus} and {Zwart}}]{LoTSS}
\bibinfo{author}{{Shimwell}, T.W.}, \bibinfo{author}{{R{\"o}ttgering}, H.J.A.}, \bibinfo{author}{{Best}, P.N.}, \bibinfo{author}{{Williams}, W.L.}, \bibinfo{author}{{Dijkema}, T.J.}, \bibinfo{author}{{de Gasperin}, F.}, \bibinfo{author}{{Hardcastle}, M.J.}, \bibinfo{author}{{Heald}, G.H.}, \bibinfo{author}{{Hoang}, D.N.}, \bibinfo{author}{{Horneffer}, A.}, \bibinfo{author}{{Intema}, H.}, \bibinfo{author}{{Mahony}, E.K.}, \bibinfo{author}{{Mandal}, S.}, \bibinfo{author}{{Mechev}, A.P.}, \bibinfo{author}{{Morabito}, L.}, \bibinfo{author}{{Oonk}, J.B.R.}, \bibinfo{author}{{Rafferty}, D.}, \bibinfo{author}{{Retana-Montenegro}, E.}, \bibinfo{author}{{Sabater}, J.}, \bibinfo{author}{{Tasse}, C.}, \bibinfo{author}{{van Weeren}, R.J.}, \bibinfo{author}{{Br{\"u}ggen}, M.}, \bibinfo{author}{{Brunetti}, G.}, \bibinfo{author}{{Chy{\.z}y}, K.T.}, \bibinfo{author}{{Conway}, J.E.}, \bibinfo{author}{{Haverkorn}, M.}, \bibinfo{author}{{Jackson}, N.}, \bibinfo{author}{{Jarvis}, M.J.}, \bibinfo{author}{{McKean}, J.P.},
  \bibinfo{author}{{Miley}, G.K.}, \bibinfo{author}{{Morganti}, R.}, \bibinfo{author}{{White}, G.J.}, \bibinfo{author}{{Wise}, M.W.}, \bibinfo{author}{{van Bemmel}, I.M.}, \bibinfo{author}{{Beck}, R.}, \bibinfo{author}{{Brienza}, M.}, \bibinfo{author}{{Bonafede}, A.}, \bibinfo{author}{{Calistro Rivera}, G.}, \bibinfo{author}{{Cassano}, R.}, \bibinfo{author}{{Clarke}, A.O.}, \bibinfo{author}{{Cseh}, D.}, \bibinfo{author}{{Deller}, A.}, \bibinfo{author}{{Drabent}, A.}, \bibinfo{author}{{van Driel}, W.}, \bibinfo{author}{{Engels}, D.}, \bibinfo{author}{{Falcke}, H.}, \bibinfo{author}{{Ferrari}, C.}, \bibinfo{author}{{Fr{\"o}hlich}, S.}, \bibinfo{author}{{Garrett}, M.A.}, \bibinfo{author}{{Harwood}, J.J.}, \bibinfo{author}{{Heesen}, V.}, \bibinfo{author}{{Hoeft}, M.}, \bibinfo{author}{{Horellou}, C.}, \bibinfo{author}{{Israel}, F.P.}, \bibinfo{author}{{Kapi{\'n}ska}, A.D.}, \bibinfo{author}{{Kunert-Bajraszewska}, M.}, \bibinfo{author}{{McKay}, D.J.}, \bibinfo{author}{{Mohan}, N.R.}, \bibinfo{author}{{Orr{\'u}},
  E.}, \bibinfo{author}{{Pizzo}, R.F.}, \bibinfo{author}{{Prandoni}, I.}, \bibinfo{author}{{Schwarz}, D.J.}, \bibinfo{author}{{Shulevski}, A.}, \bibinfo{author}{{Sipior}, M.}, \bibinfo{author}{{Smith}, D.J.B.}, \bibinfo{author}{{Sridhar}, S.S.}, \bibinfo{author}{{Steinmetz}, M.}, \bibinfo{author}{{Stroe}, A.}, \bibinfo{author}{{Varenius}, E.}, \bibinfo{author}{{van der Werf}, P.P.}, \bibinfo{author}{{Zensus}, J.A.}, \bibinfo{author}{{Zwart}, J.T.L.}, \bibinfo{year}{2017}.
\newblock \bibinfo{title}{{The LOFAR Two-metre Sky Survey. I. Survey description and preliminary data release}}.
\newblock \bibinfo{journal}{Astronomy \& Astrophysics} \bibinfo{volume}{598}, \bibinfo{pages}{A104}.
\newblock \DOIprefix\doi{10.1051/0004-6361/201629313}.
\bibitem[{{Singh} et~al.(2016){Singh}, {Mizuno} and {de Gouveia Dal Pino}}]{Singh2016}
\bibinfo{author}{{Singh}, C.B.}, \bibinfo{author}{{Mizuno}, Y.}, \bibinfo{author}{{de Gouveia Dal Pino}, E.M.}, \bibinfo{year}{2016}.
\newblock \bibinfo{title}{{Spatial Growth of Current-driven Instability in Relativistic Rotating Jets and the Search for Magnetic Reconnection}}.
\newblock \bibinfo{journal}{The Astrophysical Journal} \bibinfo{volume}{824}, \bibinfo{pages}{48}.
\newblock \DOIprefix\doi{10.3847/0004-637X/824/1/48}.
\bibitem[{Striani et~al.(2016)Striani, Mignone, Vaidya, Bodo and Ferrari}]{striani}
\bibinfo{author}{Striani, E.}, \bibinfo{author}{Mignone, A.}, \bibinfo{author}{Vaidya, B.}, \bibinfo{author}{Bodo, G.}, \bibinfo{author}{Ferrari, A.}, \bibinfo{year}{2016}.
\newblock \bibinfo{title}{{MHD simulations of three-dimensional resistive reconnection in a cylindrical plasma column}}.
\newblock \bibinfo{journal}{Monthly Notices of the Royal Astronomical Society} \bibinfo{volume}{462}, \bibinfo{pages}{2970--2979}.
\newblock \DOIprefix\doi{10.1093/mnras/stw1848}.
\bibitem[{Tchekhovskoy(2015)}]{Tchekhovskoy2015}
\bibinfo{author}{Tchekhovskoy, A.}, \bibinfo{year}{2015}.
\newblock \bibinfo{title}{Launching of Active Galactic Nuclei Jets}. \bibinfo{publisher}{Springer International Publishing}, \bibinfo{address}{Cham}.
\newblock pp. \bibinfo{pages}{45--82}.
\newblock \DOIprefix\doi{10.1007/978-3-319-10356-3_3}.
\bibitem[{Tchekhovskoy and Bromberg(2016)}]{tchekov_2016}
\bibinfo{author}{Tchekhovskoy, A.}, \bibinfo{author}{Bromberg, O.}, \bibinfo{year}{2016}.
\newblock \bibinfo{title}{{Three-dimensional relativistic MHD simulations of active galactic nuclei jets: magnetic kink instability and Fanaroff–Riley dichotomy}}.
\newblock \bibinfo{journal}{Monthly Notices of the Royal Astronomical Society: Letters} \bibinfo{volume}{461}, \bibinfo{pages}{L46--L50}.
\newblock \DOIprefix\doi{10.1093/mnrasl/slw064}.
\bibitem[{{Vaidya} et~al.(2018){Vaidya}, {Mignone}, {Bodo}, {Rossi} and {Massaglia}}]{Vaidya_2018}
\bibinfo{author}{{Vaidya}, B.}, \bibinfo{author}{{Mignone}, A.}, \bibinfo{author}{{Bodo}, G.}, \bibinfo{author}{{Rossi}, P.}, \bibinfo{author}{{Massaglia}, S.}, \bibinfo{year}{2018}.
\newblock \bibinfo{title}{{A Particle Module for the PLUTO Code. II. Hybrid Framework for Modeling Nonthermal Emission from Relativistic Magnetized Flows}}.
\newblock \bibinfo{journal}{The Astrophysical Journal} \bibinfo{volume}{865}, \bibinfo{pages}{144}.
\newblock \DOIprefix\doi{10.3847/1538-4357/aadd17}.
\bibitem[{Velović et~al.(2023)Velović, Cotton, Filipović, Norris, Barnes and Condon}]{Velovic}
\bibinfo{author}{Velović, V.}, \bibinfo{author}{Cotton, W.D.}, \bibinfo{author}{Filipović, M.D.}, \bibinfo{author}{Norris, R.P.}, \bibinfo{author}{Barnes, L.A.}, \bibinfo{author}{Condon, J.J.}, \bibinfo{year}{2023}.
\newblock \bibinfo{title}{{MeerKAT view of the dancing ghosts – peculiar galaxy pair PKS 2130-538 in Abell 3785}}.
\newblock \bibinfo{journal}{Monthly Notices of the Royal Astronomical Society} \bibinfo{volume}{523}, \bibinfo{pages}{1933--1945}.
\newblock \DOIprefix\doi{10.1093/mnras/stad1307}.
\bibitem[{Zamaninasab et~al.(2013)Zamaninasab, Savolainen, Clausen-Brown, Hovatta, Lister, Krichbaum, Kovalev and Pushkarev}]{zamaninasab}
\bibinfo{author}{Zamaninasab, M.}, \bibinfo{author}{Savolainen, T.}, \bibinfo{author}{Clausen-Brown, E.}, \bibinfo{author}{Hovatta, T.}, \bibinfo{author}{Lister, M.L.}, \bibinfo{author}{Krichbaum, T.P.}, \bibinfo{author}{Kovalev, Y.Y.}, \bibinfo{author}{Pushkarev, A.B.}, \bibinfo{year}{2013}.
\newblock \bibinfo{title}{{Evidence for a large-scale helical magnetic field in the quasar 3C 454.3}}.
\newblock \bibinfo{journal}{Monthly Notices of the Royal Astronomical Society} \bibinfo{volume}{436}, \bibinfo{pages}{3341--3356}.
\newblock \DOIprefix\doi{10.1093/mnras/stt1816}.

\end{thebibliography}
\bibliographystyle{elsarticle-harv} 



\end{document}